\documentclass[11pt]{article}
\usepackage{jheppubmod}
\usepackage{amssymb, amsmath, amsopn, amsthm}
\usepackage{epstopdf}
\usepackage{epsfig}
 \usepackage[parsep]{collref}
\usepackage{graphicx}
\usepackage{caption}
\usepackage{subcaption}

\usepackage{graphicx}

\title{Conformal Invariance and the Four Point Scalar Correlator in Slow-Roll Inflation} 

\author[1]{Archisman Ghosh,}
\author[2]{Nilay Kundu,}
\author[1]{Suvrat Raju,}
\author[2]{and Sandip P.  Trivedi}
\affiliation[1]{International Centre for Theoretical Sciences, \\
Tata Institute of Fundamental Research, \\ IISc Campus, \\
  Bangalore 560012, India.}
\affiliation[2]{Tata Institute of  Fundamental Research, \\
1 Homi Bhabha  Road, Colaba, \\
Mumbai 400 005, India.\\}

\emailAdd{archisman.ghosh@icts.res.in}
\emailAdd{nilay.tifr@gmail.com}
\emailAdd{suvrat@icts.res.in}
\emailAdd{trivedi.sp@gmail.com}

\abstract{We calculate the four point correlation function for scalar 
perturbations in the canonical model of slow-roll inflation. We work in the leading 
slow-roll approximation where the calculation can be done in de Sitter space. 
Our calculation uses techniques drawn from  the  AdS/CFT correspondence
 to find  the wave function at late times and then  calculate the four point function from it. 
The answer we get agrees with an earlier  result in the literature, obtained using  different methods. 
Our  analysis reveals a subtlety with regard to the Ward identities for conformal invariance,
which arises in de Sitter space and has no analogue in AdS space. This subtlety arises because  in  de Sitter space
the metric at late times is a genuine degree of freedom, and hence to calculate correlation functions from the wave function of the 
Universe at late times, one must fix gauge completely. The resulting correlators are 
then invariant under  a conformal transformation accompanied by  a compensating  coordinate transformation which
restores the gauge.

}

\preprint{\parbox{3.5cm}{ICTS/2013/23 \\ TIFR/TH/13-31 }}

\def\be{\begin{equation}}

\def\ee{\end{equation}}

\def\bea{\begin{eqnarray}}

\def\eea{\end{eqnarray}}

\def\vk{\vect{k_1}}
\def\vkk{\vect{k_2}}
\def\vkkk{\vect{k_3}}
\def\vkfour{\vect{k_4}}

\newcommand{\norm}[1]{#1}

\def\eepsilon{b}

\newcommand{\vect}[1]{{\boldsymbol{#1}}}

\def\Or[#1]{{\text{O}}\left({#1}\right)}
\def\dotl[#1,#2]{\left\langle #1,\, #2 \right\rangle}
\def\dotlb[#1,#2]{\left\langle #1,\, #2 \right\rangle}
\def\dotlm[#1,#2]{\left[ #1,\, #2 \right]}
\def\dotp[#1,#2]{(\vect{#1} \cdot\vect{#2})}
\begin{document}

\maketitle

\section{Introduction}

Inflation is an attractive idea which explains the approximate 
 homogeneity and isotropy of the early Universe, while also   providing a mechanism for the production of 
 perturbations  which lead to a   small breaking of these symmetries. 
The simplest model of inflation involves a scalar field, called the inflaton, with a potential which is 
positive and slowly varying  during the inflationary era. The positive and approximately constant 
 potential gives rise to a spacetime which is well described,
upto small corrections,  by four dimensional de Sitter space ($dS_4$). This spacetime is  
 homogeneous and isotropic, in fact highly symmetric,  with symmetry group $SO(4,1)$.  
Quantum effects, due to the rapid expansion of the Universe during inflation, give rise to 
small perturbations in this spacetime.
These perturbations  are of two types:  tensor perturbations,  or gravity waves, and scalar perturbations,
 which owe their origin to the presence of the inflaton. 

The past decade or so has seen   impressive advances in observational cosmology. 
These advances, for example in the measurement of the cosmic microwave background, 
increasingly constrain some of the parameters which appear in the inflationary dynamics. 
This includes a 
determination of the amplitude of the  two-point correlator for 
 scalar perturbations, and more recently, 
 a measurement of the tilt in this correlator and a 
 bound on magnitude of the scalar three point function \cite{Ade:2013uln}, \cite{Ade:2013ydc}. 
In fact, observations are  now able to rule out several models of inflation, see for e.g. \cite{Ade:2013uln}. 

These observational advances provide a motivation for  
 a more detailed theoretical  study of the  higher point 
correlation functions for perturbations produced during inflation. 
There are theoretical developments also  which make this an opportune time to carry out such a study.  Assuming that  the  $SO(4,1)$ symmetries of de Sitter space are shared by the  full inflationary dynamics, 
including the scalar field, to good approximation,  
 this  symmetry group   can be used to characterize and in some cases
significantly constrain the 
correlation functions of perturbations produced during inflation. Although the idea of inflation is quite
 old,  such a symmetry based analysis, which can sometimes lead to interesting  model independent consequences, has received 
relatively little attention, until recently. 

Another related theoretical development comes from the recent  intensive study of the AdS/CFT correspondence
in string theory and gravity. Four dimensional AdS space, $AdS_4$,  is related, by analytic continuation to $dS_4$,
and its symmetry group $SO(3,2)$ on continuation becomes the $SO(4,1)$  symmetry  of $dS_4$. As a result,  
 many of the techniques which have been developed to study correlators in  AdS space can 
  be adapted to the study of correlators in de Sitter space.  It is well known that the 
 $SO(3,2)$ symmetries of $AdS_4$ are 
also those of a $2+1$ dimensional  conformal field theory. 
It then follows that the $SO(4,1)$ symmetries of $dS_4$ are the same as those of a $3$ dimensional Euclidean Conformal Field Theory. This 
connection, between symmetries of $dS_4$ and a $3$ dimensional CFT, is often a useful guide in organizing 
the discussion of de Sitter correlation functions. 
A deeper connection between 
de Sitter space and CFTs, analogous to the AdS/CFT correspondence,  is much more tentative at the moment.
We will therefore  not assume that any such deeper connection exists in the discussion below.
 Instead, our analysis  will only use the property that $dS_4$ and $CFT_3$ share the same symmetry group.

More specifically, in this paper, we will use some of  more recent theoretical developments referred to above, 
to calculate the four point 
 correlator for scalar perturbations produced during inflation. We will work in the 
simplest model of inflation mentioned above, consisting of a slowly varying scalar coupled to two-derivative gravity,  which is often referred to as the slow-roll model of inflation. 
This model is characterized by three parameters, the Hubble constant during inflation, $H$,
and the two slow-roll parameters, denoted by $\epsilon$, $\eta$, which are  a
 measure of the deviation from de Sitter invariance. These parameters are defined in eq.\eqref{defeps}, eq.\eqref{defeta}. 
In our calculation, which is already quite complicated, we will work to 
leading order in $\epsilon, \eta$. In this limit the effects of the slow variation of the potential can be neglected and the calculation reduces to one in dS space.  
The tilt of the two-point scalar correlator, as measured for example by the 
Planck experiment, 
suggest that $\epsilon, \eta$ are of order a 
few percent, and thus that the deviations from de Sitter invariance are small, so that 
our approximation should be a good one. 

In the slow-roll model of inflation we consider, one knows before hand, from straightforward estimates,
 that the magnitude of the four point scalar  correlator is very  small. The 
calculation we carry out is therefore not motivated by the 
 hope of any immediate contact with observations. Rather, it is motivated by more theoretical considerations 
mentioned above, 
namely, to explore the connection with calculations in AdS space and investigate the role that conformal symmetry can play in constraining the inflationary correlators. 

In fact, the  calculation of the four point function in this model of inflation has already been carried out in
 \cite{Seery:2008ax}, using the so called ``in-in'' or Schwinger-Keldysh formalism. 
Quite surprisingly, it turns out that 
the result obtained in \cite{Seery:2008ax}  does not seem to satisfy the Ward identities of conformal invariance. 
This is a very puzzling feature of the result.\footnote{We thank P. Creminelli and M. Simonovic for bringing this puzzle to our attention and for sharing
their Mathematica code where the Ward identities are checked with us.} It seemed to us
 that it was clearly important to understand this puzzle further  since  doing so 
 would have implications for other correlation functions as well, and this in fact 
provided one of the main motivations for our work. 

The result we obtain for the four point function using, as we mentioned above,
 techniques motivated by the AdS/CFT correspondence, 
agrees with that obtained in \cite{Seery:2008ax}.
 Since the final answer  is quite complicated, this 
agreement between two calculations using quite different methods is a useful check on the literature. 

But, more importantly, the insights from AdS/CFT also help us resolve the puzzle regarding conformal invariance mentioned above. 
In fact techniques motivated from AdS/CFT are well suited for the study of symmetry related questions in general, 
since these techniques  naturally lead  to the wave function of the Universe
 which is related to the partition function in the CFT.

We find that the wave function, calculated upto the required order for the four point function calculation,
is indeed invariant under conformal transformations. However our calculation also reveals a subtlety, which is present in the de Sitter case and which does not have an analogue in the AdS case.
This subtlety, which holds the key to the resolution,  arises because 
one needs to proceed differently in calculating a correlation function from a wave function as opposed to 
  the partition function (in the presence of sources). Given a wave function, as in the de Sitter case,  
 one must  carry out a further sum over all configurations 
weighting them with the square of  the wave function, as per the standard rules of quantum mechanics, 
to obtain the correlation functions. 

This sum also runs over possible values of the metric. This is a sign of the fact that the metric is itself a dynamical variable on the late-time surface on which we are evaluating the wave function. We emphasize that this is not in contradiction with the fact that the metric perturbation becomes time independent at late times. Rather, the point is that there is also a non-zero amplitude for this
time-independent value to be non-trivial. In contrast, in the AdS case, where the boundary value of the metric 
(the non-normalizable mode) is a source, one does not carry out this further sum; instead correlation functions are calculated by taking derivatives with respect to the boundary metric. 

From a calculational perspective, this further sum over all configurations in the de Sitter case requires a more 
complete fixing of gauge for the metric. This is not surprising since even 
defining local correlators in a theory without a fixed background metric requires
a choice of gauge. This resulting  gauge is not preserved in general by a conformal transformation. As a result, a conformal transformation must be accompanied by a suitable coordinate  parameterization   before it becomes a symmetry in  the chosen gauge. Once this additional parameterization is included, we find  that the  four point function does indeed meet the resulting Ward identities of conformal invariance. We expect this to be true for other correlation functions as well. 

There is another way to state the fact above. The correlation functions that are commonly
computed in the AdS/CFT correspondence can be understood to be limits of bulk correlation functions, where only normalizable modes are turned on \cite{Banks:1998dd}. However, as emphasized in \cite{Harlow:2011ke}, the expectation values of de-Sitter perturbations that we are interested in cannot be obtained in this way as a limit of bulk correlation functions. As a result, they do not directly satisfy the Ward identities of conformal invariance, although a signature of 
this symmetry remains in the wave function of the Universe from which they originate.

The Ward identities of conformal invariance, once they have been appropriately understood, 
serve as a highly non-trivial test on the result especially when the correlation
function is a complicated one, as in the case of the scalar four point function considered here.
The AdS/CFT point of view also suggests other   tests,  including the flat-space limit where we check that
the AdS correlator reduces to the flat space scattering amplitude of four scalars in the appropriate limit. In a third series of checks,
we test the behavior of the correlator in suitable limits that  are related to the operator product expansion in a conformal field theory. 
Our result meets all these checks.

Before proceeding let us discuss some of the other related literature on the subject. For a current review on the present status of inflation and future planned experiments, see \cite{Abazajian:2013vfg}. Two  reviews which discuss  non-Gaussianity from the CMB and from large scale structure are, \cite{Komatsu:2010hc}, and \cite{Desjacques:2010jw} respectively.
The scalar four point function in single field inflation has been discussed in \cite{Seery:2008ax, Seery:2006vu, PhysRevD.77.083517, PhysRevD.80.043527, chen}. 
The general approach we adopt is along the
 lines of the seminal work of Maldacena \cite{Maldacena:2002vr}. (See also \cite{Maldacena:2011nz}.)
Some other references which 
contain a discussion of conformal invariance and its implications for correlators in cosmology are \cite{Antoniadis:1996dj, Larsen:2002et,  Larsen:2003pf, McFadden:2010vh, Antoniadis:2011ib, Creminelli:2011mw, Bzowski:2011ab, McFadden:2011kk, Kehagias:2012pd, Kehagias:2012td, Schalm:2012pi, Bzowski:2012ih, MRT, Garriga:2013rpa}.
Some  discussion of  consistency conditions which arise in the squeezed limit can be found in \cite{Maldacena:2011nz, Creminelli:2004yq, Cheung:2007sv, Senatore:2012wy, Creminelli:2011sq, Bartolo:2011wb, Creminelli:2012ed, Creminelli:2012qr, Hinterbichler:2012nm, Assassi:2012zq, Goldberger:2013rsa, Hinterbichler:2013dpa, Creminelli:2013cga, Berezhiani:2013ewa}.  An approach towards holography in inflationary backgrounds is given in \cite{Banks:2011qf, Banks:2013qra}.

This paper is structured as follows. 
Some basic concepts which are useful in the calculation are discussed in section \ref{prelim}, 
including  the connection between the wave function in dS space and the partition function 
in AdS space. Issues related to conformal invariance are discussed in  section \ref{conformalinv}. 
A term in the wave function  needed for the four point correlator is then calculated in 
section \ref{4ptcorrinads}, leading to the final result for the correlator in section \ref{finaldscalc}. Important 
tests of the result are carried out in section \ref{Testsofresult} including a discussion of  the Ward identities
of conformal invariance.  Finally, we end with a discussion
in section \ref{discussion}. There are six important appendices which contain useful supplementary 
material. 

\section{Basic Concepts}
\label{prelim}

We will consider a theory of gravity coupled to a scalar field, the inflaton, with action
\be
\label{action1}
S=\int d^4x\sqrt{-g}M_{Pl}^2  \bigg[{1\over 2} R-{1\over 2} (\nabla \phi)^2   -V(\phi) \bigg].
\ee
Note we are using conventions in which $\phi$ is dimensionless. 
Also note that in our conventions the relation between the Planck mass and the gravitational constant is
\be
\label{defmpl}
M_{Pl}^2={1\over 8 \pi G_N}.
\ee
If the potential is slowly varying, so that the slow-roll parameters are small,
\be
\label{defeps}
\epsilon \equiv \bigg({V'\over 2 V}\bigg)^2 \ll 1
\ee
and 
\be
\label{defeta}
\eta \equiv {V''\over V} \ll 1,
\ee
then the system has a solution which is approximately de Sitter space, with metric,
\begin{align}
ds^2 &=-  dt^2 + a^{2}(t) \sum_{i=1}^3 dx^i dx^i, \label{dsmetric} \\
a^2 (t) &=  e^{2 Ht}, \label{scaleds}
\end{align}
where the Hubble constant is, 
\be
\label{relhv}
H=\sqrt{V \over 3 }.
\ee
This solution describes the exponentially expanding  inflationary Universe. 

The slow-roll parameters introduced above can be  related to time derivatives of  the Hubble constant,
in the slow-roll approximation, as follows,
\be
\label{defeps2}
\epsilon=-{\dot{H}\over H^2},
\ee
while $\eta$ is given by,
\be
\label{defeta2}
\eta=\epsilon-{\ddot{H}\over 2H \dot{H}}.
\ee

Using the slow-roll approximation we can also express $\epsilon$ in terms of the rate of change of the
 scalar as,
\be
\label{releps3}
\epsilon={1\over 2}{\dot{\phi}^2\over H^2}.
\ee

de Sitter space is well known to be a highly symmetric space with symmetry group
$SO(1,4)$. We will refer to this group as the conformal group because it is also the symmetry group of a
conformal field theory in $3$ dimensions. 
This  group is ten dimensional. It  consists of  $3$ rotations and  $3$ translations in the $x^i$ 
directions, which are obviously symmetries of the metric, eq.\eqref{dsmetric};  a  scale transformations  of the form,
\be
\label{scinv}
x^i \rightarrow \lambda x^i, t \rightarrow t - {1\over H } \log(\lambda);
\ee
and $3$ special conformal transformations whose infinitesimal form is 
\begin{align}
x^i   \rightarrow   x^i -2 (\eepsilon_j x^j) x^i + & \eepsilon^i \bigg(\sum_j (x^j)^2 - {e^{-2Ht}\over H^2}\bigg) \label{spcon}, \\
t  \rightarrow   t + &2{\eepsilon_j x^j \over H} \label{tspcon}, 
\end{align} 
where $b^i, i=1,2,3$ are infinitesimal parameters. 
This symmetry group will play an important role in our discussion below.

As mentioned above during the inflationary epoch the Hubble constant varies with time and 
de Sitter space is only an approximation to the space-time metric. 
The time varying Hubble constant also breaks some of the symmetries of de Sitter space. 
  While translations and
 rotations in the $x^i$ directions are left  unbroken,  the scaling and special conformal symmetries
are broken.
However, as long as the  slow-roll parameters $\epsilon, \eta,$ are
small this breaking is small and the resulting inflationary spacetime
is still approximately conformally invariant.

\subsection{The  Perturbations}
\label{perturbations}

Next we turn to describing perturbation in the inflationary spacetime. 
Following, \cite{Maldacena:2002vr}, we write the metric in the ADM form, 
\be
\label{admmetric}
ds^2=-N^2 dt^2 + h_{i j} (dx^i + N^i dt) (dx^j+ N^j dt).
\ee
By suitable coordinate transformations we can set the lapse function 
\be
\label{lapse}
N=1
\ee
and the shift functions to vanish,
\be
\label{shiftc}
N_i=0.
\ee
The metric then takes the form,
\be
\label{fmetric}
ds^2=-dt^2+h_{i j} dx^i dx^j.
\ee
We will work in this gauge throughout in the following discussion. 

The metric of dS space can be put in this form, eq.\eqref{dsmetric} with 
\be
\label{defgij}
h_{i j}=e^{2Ht} \delta_{i j}.
\ee

Perturbations about dS space take the form 
\be
\label{gfversion1}
h_{i j}=e^{2 H t} g_{i j},
\ee
with 
\begin{align}
\label{defgij2}
g_{ij} & =  \delta_{ij} + \gamma_{ij},  \\
\label{expgamma} \gamma_{ij} & =  2 \zeta \delta_{ij} + \widehat{\gamma}_{ij} .
\end{align}
By definition the metric perturbation $\widehat{\gamma}_{ij}$ meets the condition,
\be
\label{condtrace}
\widehat{\gamma}_{ii}=0.
\ee
The tensor modes are given by  $\widehat{\gamma}_{ij}$. 
Let us note here  that the expansion in eq.\eqref{expgamma} is true to lowest order in the perturbations, 
higher order corrections will be discussed in appendix \ref{addterm} and will be shown to be unimportant to the order we work. 

Besides perturbations in the metric there are also perturbations in the inflaton,
\be
\label{delphi}
\phi={\bar \phi}(t) + \delta \phi
\ee
where ${\bar \phi}(t)$ is the background value of the inflaton. 

The metric of dS space, eq.\eqref{fmetric}, eq.\eqref{defgij} is rotationally invariant with $SO(3)$ symmetry 
in the $x^i$ directions.   This  invariance  can be used to classify the perturbations.
There are two types of perturbations, scalar and tensor, which transform as spin $0$ and spin $2$ under
the rotation group respectively.
The tensor perturbations arise from the metric, $\widehat{\gamma}_{ij}$. 
The scalar perturbation physically arises due to fluctuations in the inflaton field. 

We turn to describing these perturbations more precisely next. 
\subsubsection{Gauge 1}
\label{gauge1}
We will be especially interested in understanding the perturbations at sufficiently late time, when their 
wavelength becomes  bigger than the Horizon scale $H$. At such late times the perturbations  become essentially time independent.  It turns out that the coordinate transformations used to bring the metric in the form eq.\eqref{admmetric}
 meeting conditions, eq.\eqref{lapse}, eq.\eqref{shiftc},
does not exhaust all the gauge invariance in the system for describing such time independent perturbations. 
Additional spatial reparameterizations of the kind
\be
\label{sprepara}
x^i \rightarrow x^i + v^i({\vect{x}})
\ee
can be carried out which keep the form of the metric fixed. These can be used to impose the condition 
\be
\label{condtrans}
\partial_i\widehat{\gamma}_{ij}=0.
\ee
From eq.\eqref{condtrace}, eq.\eqref{condtrans} we see that $\widehat{\gamma}_{ij}$ is now 
both transverse and traceless,
as one would   expect for the tensor perturbations. 

In addition,  a further  coordinate transformation can  also be carried out which  is a time parameterization 
of the form, 
\be
\label{timerepara}
t\rightarrow t + \epsilon({\vect{x}}). 
\ee
Strictly speaking to stay  in the gauge eq.\eqref{lapse}, eq.\eqref{shiftc},
 this time parameterization must be accompanied by a  
 spatial parameterization
\be
\label{spre2}
x^i\rightarrow x^i+v^i(t,{\vect{x}}),
\ee
where to leading order in the perturbations 
\be
\label{spre3}
v^i=-{1\over 2H}(\partial_i\epsilon) e^{-2Ht}.
\ee
However, at  late time we see that $v^i\rightarrow 0$ and thus the spatial parameterization 
 vanishes. 
As a result this additional coordinate transformation does not change $\widehat{\gamma}_{ij}$ which continues
to be transverse, upto exponentially small corrections.  

By suitably choosing the parameter $\epsilon$ in eq.\eqref{timerepara} 
one can set the perturbation in the inflaton to vanish, 
\be
\label{conddelphi}
\delta \phi=0.
\ee
This choice will be called gauge $1$ in the subsequent discussion. 
 The  value of  $\zeta$, defined in eq.\eqref{expgamma}, in this gauge,   then 
 corresponds to the scalar perturbation.  It gives rise to   fluctuations of the spatial curvature.

\subsubsection{Gauge 2}
\label{gauge2}
Alternatively, having fixed the spatial reparameterizations so that $\widehat{\gamma}_{ij}$ is transverse, 
eq.\eqref{condtrans}, we can then choose the time parameterization, $\epsilon$, defined in  eq.\eqref{timerepara} 
  differently, so that 
\be
\label{seth}
\zeta=0,
\ee
and  it is  the scalar component of the metric perturbation,  instead of $\delta \phi$, that   vanishes. 
This choice will be referred to as gauge $2$. 
The scalar  perturbations in this coordinate system  are then  given by fluctuations in the 
inflaton $\delta \phi$.

This second gauge is obtained by starting with the coordinates in which the perturbations take the form 
given in gauge 1, where they are described by    $\zeta, \widehat{\gamma}_{ij}$, and  
carrying  out a time reparameterization 
\be
\label{tre}
t\rightarrow t+ {\zeta  \over H},
\ee
to meet  the condition eq.\eqref{seth}.  The tensor perturbation $\widehat{\gamma}_{ij}$ is unchanged by this coordinate transformation.
If the background value of the inflaton in the inflationary solution is 
\be
\label{backinfl}
\phi=\bar{\phi}(t),
\ee
the resulting value for the perturbation $\delta \phi$ this gives rise to is \footnote{There are  corrections 
involving higher powers of the perturbation in this relation, but these will not be important in our calculation
of the four point function.}
\be
\label{infper}
\delta \phi= -{\dot{\bar{\phi}}  \zeta \over H}. 
\ee

Using eq.\eqref{releps3}  we can   express this relation as 
\be
\label{infper2}
\delta \phi=-\sqrt{2 \epsilon} \zeta.
\ee

For purposes of  calculating the $4$-pt scalar correlator  at late time, once the modes have crossed the horizon,
 it will be most convenient to first use gauge 2, where the perturbation is 
described by fluctuations in the scalar, $\delta \phi$, and   then  
  transform the resulting answer to gauge 1, where the perturbation is given in terms of  fluctuation in 
the metric component, $\zeta$.   This turns out to be a convenient thing to do 
for tracing the subsequent evolution of  scalar perturbations,  since a general argument,  following essentially from gauge invariance, says that $\zeta$ must be a constant once the mode crosses the horizon. This fact is discussed in \cite{Bardeen:1980kt, Bardeen:1983qw, Lyth:1984gv, Salopek:1990jq, Weinberg:2003sw, Weinberg:2003ur}, for a review see section [5.4] of \cite{Weinbergcosmo}.

\subsection{Basic Aspects of the Calculation}
\label{basicaspects}
Let us now turn to describing some basic aspects of the calculation.  Our approach is based on that of \cite{Maldacena:2002vr}. We will calculate the wave function of the Universe as a functional of the scalar and tensor perturbations. Once this wave function is known correlation functions can be calculated from it in a straightforward manner. 

In particular we will be interested in the wave function at late times, when the modes of interest have crossed the horizon so that their wavelength $\lambda \gg H$. 
At such late times the Hubble damping results in the correlation functions acquiring a time independent form. 
Since the correlation functions become time independent the wave function also becomes time independent at these late enough times. 

The perturbations produced during inflation in the slow-roll model are known to be approximately Gaussian. 
This allows the wave function, which is a functional of the perturbations in general, 
 to be written as a power series expansion of the form, 
\be
\label{wf1}
\begin{split}
\psi[\chi({\vect{x}})]  = & \exp\bigg[-{1\over 2} \int d^3 x d^3 y \chi ({\vect{x}}) \chi({\vect{y}}) \langle O({\vect{x}}) O({\vect{y}})\rangle  \\
&+ {1\over 6} \int d^3 x d^3 y d^3 z\, \chi({\vect{x}}) \chi({\vect{y}})\chi({\vect{z}})
\langle O({\vect{x}}) O({\vect{y}}) O({\vect{z}})\rangle + \cdots \bigg].
\end{split}
\ee
This expression is schematic, with  $\chi$ standing for a generic perturbation which could be a scalar or a 
tensor mode, and the coefficients
$\langle O({\vect{x}})O({\vect{y}})\rangle, \langle O({\vect{x}})O({\vect{y}})O({\vect{z}})\rangle$ being functions which determine the 
two-point three point etc correlators. Let us also note, before proceeding, that the coefficient functions
will transform under the $SO(1,4)$ symmetries like correlation functions of appropriate  operators in a Euclidean Conformal Field Theory, and we have denoted them in this suggestive manner to emphasize this feature.

For our situation, we have the tensor perturbation, $\gamma_{ij}$, and the scalar perturbation, which 
 in gauge 1 is given by    $\delta \phi$.  
 With a suitable choice of normalization
the wave function  takes the form\footnote{The coefficient functions include contact terms, which are analytic in some or all of  the momenta.}
\be
\label{wf2old}
\begin{split}
\psi[\delta \phi, \gamma_{ij}]  =  \exp\bigg[{M_{Pl}^2 \over H^2} \bigg(&-{1\over 2} \int  d^3x \sqrt{g({\vect{x}})} \ d^3y \sqrt{g({\vect{y}})}  \ \delta \phi({\vect{x}}) \delta \phi({\vect{y}}) \langle O({\vect{x}}) O({\vect{y}})\rangle  \\
- {1\over 2} \int & d^3 x \sqrt{g({\vect{x}})}  \ d^ 3 y \sqrt{g({\vect{y}})} \ \gamma_{ij} ({\vect{x}}) \gamma_{kl}({\vect{y}}) \langle T^{ij}({\vect{x}}) T^{kl}({\vect{y}})\rangle  \\  
 -{1\over 4} \int & d^3 x \sqrt{g({\vect{x}})} \ d^3 y \sqrt{g({\vect{y}})} \ d^3 z \sqrt{g({\vect{z}})}\\ 
& \delta \phi ({\vect{x}}) \delta \phi({\vect{y}}) \gamma_{ij}({\vect{z}}) \langle O({\vect{x}}) O({\vect{y}}) T^{ij}({\vect{z}})\rangle  \\
+ {1\over 4!}\int & d^3 x \sqrt{g({\vect{x}})} \ d^3 y \sqrt{g({\vect{y}})} \ d^3 z \sqrt{g({\vect{z}})} \ d^3 w  \sqrt{g({\vect{w}})}\\ 
& \delta\phi({\vect{x}})\delta\phi({\vect{y}})\delta\phi({\vect{z}})\delta \phi({\vect{w}}) \langle O({\vect{x}})O({\vect{y}})O({\vect{z}})O({\vect{w}}) \rangle + \cdots \bigg)\bigg].
\end{split}
\ee
Where $g({\vect{x}})=\text{det}[g_{ij}({\vect{x}})]$ and $g_{ij}$ is given in eq.\eqref{defgij2}. 

The terms which appear explicitly on the RHS of eq.\eqref{wf2old} are all the ones needed for calculating
 the four point scalar correlator of interest in this paper. The ellipses indicate additional terms 
which will not enter the calculation of this correlation function, in the leading order approximation
 in ${M_{pl}^2 \over H^2}$, where loop effects can be neglected.  
The graviton two-point correlator and the graviton-scalar-scalar three point function
are relevant because they contribute to the scalar four point correlator after integrating out the graviton 
at tree level as we will see below in more detail in sec \ref{finaldscalc}. 

In fact only a subset of terms in eq.\eqref{wf2old} are relevant for calculating the $4$-pt scalar correlator. 
As was mentioned in subsection \ref{gauge2} we will first calculate the result in gauge 2. Working in this gauge,
where $\zeta=0$, and expanding the metric $g_{ij}$ in terms of the perturbation $\gamma_{ij}$, eq.\eqref{defgij2},
one finds that the terms which are relevant are 
\begin{align}
\label{wf2}
\begin{split}
\psi[\delta \phi({\vect{k}})&,\gamma^s({\vect{k}})]= \exp\bigg[{M_{Pl}^2 \over H^2} \bigg(-{1\over 2} \int {d^3 \vk \over (2 \pi)^3} {d^3 \vkk \over (2 \pi)^3} \delta \phi( \vk) \delta \phi( \vkk) \langle O(- \vk) O( -\vkk)\rangle  \\
&-{1\over 2} \int {d^3 \vk \over (2 \pi)^3} {d^3 \vkk \over (2 \pi)^3} \gamma^s( \vk) \gamma^{s'}( \vkk) \langle T^s(- \vk) T^{s'}( -\vkk)\rangle  \\
&-{1\over 4} \int {d^3 \vk \over (2 \pi)^3} {d^3 \vkk \over (2 \pi)^3} {d^3 \vkkk \over (2 \pi)^3} \delta \phi( \vk) \delta \phi( \vkk)\gamma^s( \vkkk) \langle O(- \vk) O( -\vkk)T^s(- \vkkk)\rangle  \\
&+ {1\over 4!}\int \prod_{J=1}^4 \bigg\{ {d^3 {\vect{k}}_J \over (2 \pi)^3} \delta\phi({\vect{k}}_J) \bigg\} \langle O(-\vk) O(-\vkk) O(-\vkkk) O(-\vkfour) \rangle\bigg)\bigg].
\end{split}
\end{align}
In eq.\eqref{wf2} we have shifted to momentum space, with
\be
\label{ftrel}
\delta\phi({\vect{x}})=\int {d^3k \over (2\pi)^3} \delta \phi({\vect{k}}) e^{i{\vect{k}}\cdot{\vect{x}}}
\ee
and similarly for  $\gamma_{ij}$ and all the coefficient functions appearing in eq.\eqref{wf2}.
Also, since $\gamma_{ij}$ is transverse we can write 
\be
\label{defgammas}
\gamma_{ij}(\vect{k}) =\sum_{s=1}^2 \gamma_s(\vect{k}) \epsilon^s_{ij}(\vect{k}),
\ee
where $\epsilon^s_{ij}(k), \ s=1,2$, is a basis of polarization tensors which are transverse and traceless. 
Some additional conventions pertaining to  our definition for $\epsilon_{ij}^s$ etc are given in Appendix \ref{coefffnnorm}.

Of the four coefficient functions which appear explicitly on  the RHS of eq.\eqref{wf2}, two, 
 the coefficient functions $\langle O(\vk)O(\vkk)\rangle $ and $\langle T^s(\vk) T^{s'}(\vkk)\rangle$  are well known. 
The function $\langle O(\vk)O(\vkk)T^s(\vkkk)\rangle $ was obtained in \cite{Maldacena:2002vr}, for the slow-roll model of inflation being considered here,
 and also obtained  from more general  considerations in \cite{MRT}, see also \cite{Bzowski:2011ab}. These coefficient functions are also summarized in Appendix \ref{coefffnnorm}. 
This only leaves the $\langle O(\vk)O(\vkk) O(\vkkk)O(\vkfour)\rangle$ coefficient function.
Calculating it will be one of the major tasks in this paper. 

{\bf \underline{Conventions}:}
Before proceeding it is worth summarizing some of our conventions. 
Vectors with components in the $x^i$ directions will be denoted as boldface, e.g., ${\vect{x}}, {\vect{k}}$,
while their magnitude will be denoted without the boldface, e.g., $x=|{\vect{x}}|, k=|{\vect{k}}|$.
Components of such vectors will be denoted without bold face, e.g., $k^i$.
The Latin indices on these components will be raised and lowered using the flat space metric, so that
$k^i=k_i$, $x^i=x_i$, and also ${\vect{k}} \cdot {\vect{x}}=k_ix_i$.

\subsection{The Wave Function}
\label{wavefunction}
The wave function as a functional of the late time perturbations can be calculated by doing a path integral, 
\be
\label{wf3}
\psi[\chi({\vect{x}})]=\int^{\chi({\vect{x}})} D \chi e^{i S},
\ee
where $S$ is the action and $\chi$  stands for the value a generic perturbation takes at late time. 

To make the path integral well defined one needs to also specify the boundary conditions in the far past.
In this paper we take these boundary conditions to correspond to the standard 
Bunch Davies boundary conditions.  
In the far past,  the perturbations 
had a wavelength much shorter than the Hubble scale,   the short wavelengths of  the modes
makes them  insensitive to the geometry of de Sitter space and   they essentially propagate as if  in Minkowski spacetime.
The Bunch Davies vacuum corresponds to taking the modes to be in the Minkowski vacuum at early enough time.

An elegant way to impose this boundary condition in the path integral above, as discussed in  
\cite{Maldacena:2002vr}, is as follows.
Consider de Sitter space in conformal  coordinates,
\be
\label{poincoordds}
ds^2={1\over H^2\eta^2} (-d\eta^2 + \sum_{i=1}^3(dx^i)^2),
\ee
with the far past being $\eta \rightarrow -\infty$, and late time being $\eta \rightarrow 0$. 
Continue $\eta$ so that it acquires a small imaginary part $\eta \rightarrow \eta (1- i \epsilon), \epsilon > 0$.
Then the Bunch Davies boundary condition is correctly imposed if the path integral is done over configurations 
which vanish at early times when $\eta \rightarrow -\infty(1- i \epsilon)$. 
Note that in general the resulting path integral is over complex field configurations. 

As an  example, for a free scalar field with equation, 
\be
\label{eqff}
\nabla^2 \phi=0,
\ee
a  mode with momentum ${\vect{k}}$, $\phi=f_{{\vect{k}}}(\eta) e^{i{\vect{k}} \cdot {\vect{x}}}$ which meets the required boundary condition is 
\be
\label{fsol}
f_{{\vect{k}}}=c_1({\vect{k}}) (1-i k\eta) e^{i k\eta}.
\ee
The second solution, 
\be
\label{ssol}
f_{{\vect{k}}}=c_2({\vect{k}}) (1+i k\eta) e^{-i k\eta}
\ee
is not allowed. 
Since $f_{\vec{k}}\ne f_{\vec{k}}^*$ the resulting configuration which dominates the saddle point 
is complex. 

With the  Bunch Davies boundary conditions the path integral is well defined as a functional of the boundary values of the fields at late time. 

We will evaluate the path integral in the leading saddle-point approximation. Corrections corresponding 
to quantum loop effects are  suppressed by powers of $H/M_{Pl}$ and are  small as long as 
$H/M_{Pl} \ll 1$. 
In this leading approximation the procedure to be followed is simple. 
We expand the action about the zeroth order inflationary background solution.  
Next, extremize the  resulting corrections to the action as a function of the perturbations, to get the 
 equations which must be satisfied by the perturbations. Solve these equations  subject to the
 Bunch Davies boundary conditions, in the far past, and the given boundary values of the perturbations 
at late times. And finally evaluate   the correction terms in the action  on-shell, on the resulting solution
for the perturbations,  to obtain the action as a functional of the late time boundary values of the perturbations. 
This gives, from eq.\eqref{wf3}
\be
\label{onswf}
\psi[\chi(x)]=e^{i S^{dS}_{\text{on-shell}}[\chi(x)]}.
\ee

This procedure is further simplified by working in the leading slow-roll approximation, as we will do. 
In this approximation, as was mentioned above, the metric becomes that of de Sitter space, eq.\eqref{dsmetric} with constant $H$. Since the slow-roll parameters, eq.\eqref{defeps}, eq.\eqref{defeta}
  are  put to zero, the potential $V$, eq.\eqref{action1}, can be taken to be a constant, related to the Hubble constant by eq.\eqref{relhv}. The  resulting action for the small perturbations is then given by 
\be
\label{resact}
S=\int d^4x \sqrt{-\text{det}(\bar{g}_{\mu\nu}+ \delta g_{\mu\nu})} M_{Pl}^2\bigg[{1\over 2} R(\bar{g}_{\mu\nu}+\delta g_{\mu\nu})-V -{1\over 2}
(\nabla \delta \phi)^2 \bigg].
\ee
Here $\bar{g}_{\mu\nu}$ denotes the background value for the metric in de Sitter space, eq.\eqref{dsmetric},
 and $V$ is constant, as mentioned above. 
$\delta g_{\mu\nu}$ is the metric perturbation, and $\delta \phi$ is the  perturbation for the scalar
 field, eq.\eqref{delphi}. 

Notice that the action for the perturbation of the scalar, is simply that of a minimally coupled scalar field in de Sitter space. In particular self interaction terms coming from expanding the potential, for example
a  $(\delta \phi)^4$ term which would   be of relevance for the four-point function, can
 be neglected in the leading slow-roll  approximation.  
One important consequence of this observation is that the  correlation functions
to leading order in the slow-roll parameters must obey the symmetries of de Sitter space. In particular,
 this must be true for the scalar 4-point function. 

\subsection{The Partition Function in AdS and the Wave Function in dS}
\label{pfwf}

The procedure  described above for calculating the wave function in de Sitter space is very 
analogous to what is adopted for calculating the partition function  AdS space. In 
fact this connection allows us to conveniently obtain the wave function  in de Sitter space from the partition function in  AdS space, after suitable analytic continuation, as we now explain. 

Euclidean $AdS_4$ space has the metric (in Poincare coordinates):
\be
\label{eadsmet}
ds^2=R_{\text{AdS}}^2 {1\over z^2} (dz^2+ \sum_{i=1}^3 (dx^i)^2)
\ee
with $z\in [0,\infty]$. $R$ is the radius of $AdS$ space.

After  continuing $z,H$ to imaginary values, 
\be
\label{contadsdsa}
z=-i \eta, 
\ee
and,
\be
\label{contadsdsb}
R_{\text{AdS}}={i\over H}
\ee
where $\eta \in [-\infty, 0]$ and $H$ is real, 
this metric becomes that of de Sitter space given in eq.\eqref{poincoordds}.  

The partition function in AdS space is defined as a functional of the boundary values that  fields take
as $z\rightarrow 0$. In the leading semi-classical approximation it is given by 
\be
\label{pfads}
Z[\chi(x)]=e^{-S^{\text{AdS}}_{\text{on-shell}}}
\ee
where $S^{\text{AdS}}_{\text{on-shell}}$ is  the on shell action which is obtained by substituting the 
 classical solution for fields which take the required boundary values,  $z\rightarrow 0$, 
into the action. 
We denote these boundary values generically as $\chi(x)$ in eq.\eqref{pfads}. 

Besides the boundary conditions at $z\rightarrow 0$ one also needs to impose boundary conditions 
as $z\rightarrow \infty$ to make the calculation  well defined. This boundary condition is imposed by requiring regularity for fields as $z\rightarrow \infty$. 

For example, for  a free scalar field with momentum ${\vect{k}}$, $\phi=f_{{\vect{k}}} e^{i{\vect{k}}
\cdot {\vect{x}}}$, the solution to the wave equation, eq.\eqref{eqff} is, 
\be
\label{solads}
f_{{\vect{k}}}=c_1({\vect{k}}) (1+k z) e^{-k z}+c_2({\vect{k}}) (1-k z) e^{k z}.
\ee
Regularity requires that $c_2$ must vanish, and  the solution must be 
\be
\label{fsolads}
f_{{\vect{k}}}=c_1({\vect{k}}) (1+k z) e^{-k z}.
\ee
At $z\rightarrow 0$ the solution above goes to a $z$ independent constant
\be
\label{soladsb}
f_{{\vect{k}}}=c_1({\vect{k}}).
\ee

More generally a solution is obtained by summing over modes of this type,
\be
\label{gensol}
\phi(z,{\vect{x}})=\int {d^3 k \over (2\pi)^3} \phi({\vect{k}})(1+ k z) e^{-k z} e^{i{\vect{k}}\cdot {\vect{x}}}.
\ee
Towards the boundary, as $z\rightarrow 0$, this becomes,
\be
\label{bgen}
\phi({\vect{x}})=\int {d^3 k \over (2\pi)^3} \phi({\vect{k}})e^{i{\vect{k}}\cdot{\vect{x}}}.
\ee
The AdS on-shell action is then a functional of $\phi({\vect{k}})$.  

The reader will notice that the 
 relation between the partition function and \text{on-shell} action in AdS space, eq.\eqref{pfads}, is 
quite analogous to 
that between the wave function and on shell action in dS space eq.\eqref{onswf}. 
We saw above that after the analytic continuation, eq.\eqref{contadsdsa}, eq.\eqref{contadsdsb},
 the AdS metric goes over to the  metric in dS space. 
It is easy to see that this analytic  continuation also takes the solutions for fields
in AdS space which meet the regularity condition as $z\rightarrow \infty$, to   those in dS space meeting the Bunch Davies boundary conditions. 
For example,   the free scalar which meets the
 regularity condition, as $z\rightarrow 0$, in AdS, is given in  eq.\eqref{fsolads}, and this goes
over to  the solution meeting the Bunch 
Davies boundary condition in dS space,   eq.\eqref{fsol}. 
Also after the analytic continuation the boundary value of a field as $z\rightarrow 0$ in AdS, becomes
the boundary value at late times, as $\eta\rightarrow 0$ in dS, as is clear from comparing
eq.\eqref{soladsb} with the behavior of the solution in eq.\eqref{fsol} at $\eta \rightarrow 0$.

These facts imply that the on-shell action in AdS space when analytically continued gives the on-shell 
action in  dS space. For example,  for a massless scalar field the action in AdS space
 is a functional of the boundary value for the field $\phi({\vect{k}})$, eq.\eqref{bgen},
 and also on the AdS radius $R_{\text{AdS}}$. We denote
\be
\label{acads3}
S^{\text{AdS}}_{\text{on-shell}}= S^{\text{AdS}}_{\text{on-shell}}[\phi({\vect{k}}),R_{\text{AdS}}].
\ee
to make this dependence explicit.  
The on shell action in dS space is then obtained by  taking $R_{\text{AdS}} \rightarrow i/H$,
\be
\label{onsdsads}
S^{ds}_{\text{on-shell}}[\phi({\vect{k}}),H] = - i \ S^{\text{AdS}}_{\text{on-shell}}\big[\phi({\vect{k}}),{i\over H}\big].
\ee
Note that on the LHS in this equation $\phi({\vect{k}})$  refers
 to the late time value of the scalar field in dS space. The factor of $i$ on the RHS arises
 because the analytic continuation eq.\eqref{contadsdsa} leads to an
 extra factor  of $i$ when the $z$ integral involved in evaluating the AdS  action is continued to the $\eta$ integral in dS space. 

Two more comments are worth making here. 
First, it is worth being more explicit about the analytic continuation, eq.\eqref{contadsdsa}, eq.\eqref{contadsdsb}. To arrive at dS space with the  Bunch Davies conditions correctly imposed one must 
 start  with regular boundary condition in AdS space, $z\rightarrow \infty$, and then continue $z$ from the positive real axis to the negative imaginary axis by setting
\be
\label{analz}
z=|z| e^{i\phi}
\ee
and taking $\phi$ to go from  $0$ to $\pi/2$.  
In particular when $\phi=\pi/2-\epsilon$, $\eta=i z$ is given by 
\be
\label{analeta}
\eta=- |z| (1 - i \epsilon)=-|\eta|(1 -i \epsilon),
\ee
so that $\eta$ has a small positive imaginary part. Imposing the regularity condition then 
 implies that fields vanish when $\eta \rightarrow \infty$,  this 
 is exactly the condition required to impose
the Bunch Davies boundary condition. 

Second, one subtlety we have not discussed is that the resulting answers for correlation functions
can  sometimes have divergences and needs
 to be regulated by introducing a suitable cutoff in the infra-red. Physical
 answers do not depend on the choice of cut-off procedure and in any case  this issue will not arise
for the calculation of interest here, which is to obtain the scalar four-point correlator. 

\subsection{Feynman-Witten Diagrams in AdS}
As we mentioned above, the wave function of the Universe helps to elucidate the role of various symmetries, such as conformal invariance. This itself makes the on-shell action in AdS a useful quantity to consider. However, there is another
advantage in first doing the calculation in anti-de Sitter space. 

The various coefficient functions in the wave function of the Universe can
be computed by a set of simple diagrammatic rules. These Feynman diagrams in AdS
are sometimes called ``Witten diagrams''. They are closely related to flat-space Feynman diagrams, except that flat space propagators must be replaced by the appropriate Green functions in AdS. Taking the limit where the external
points of the correlators reach the boundary, we obtain correlators in one-lower dimension, which are conformally invariant. 

These correlators have been extensively studied in the AdS/CFT literature,
where several powerful techniques have been devised to calculate them, and check their consistency. We will bring some of these techniques to bear upon this calculation below. In fact, the four point scalar correlator that 
we are interested in, has been computed in position space in \cite{D'Hoker:1999pj} and in \cite{Arutyunov:2000py}. Here we will compute this quantity in momentum space. 

Although, in principle, we could have obtained this answer by Fourier transforming the position space answers, it is much more convenient to do the calculation directly using momentum space Feynman-Witten diagrams. The use of momentum space is particularly convenient in odd boundary dimensions, since the propagators simplify greatly, and exchange interactions can be evaluated by a straightforward algebraic procedure of computing residues of a complex function, as we will see below. 

\subsection{Basic Strategy for the Calculation}
\label{basicstrategy}
Now that we have discussed all the required preliminaries in some detail, we are ready to spell out  
the basic strategy that we will adopt in our calculation of the scalar four point function. 
First, we will calculate the coefficient function $\langle O({\vect{x}}_1) O({\vect{x}}_2) O({\vect{x}}_3) O({\vect{x}}_4)\rangle $, which, as was discussed in section \ref{basicaspects},
 is the one coefficient function which is not already known. This correlator is
given by a simple set of Feynman-Witten diagrams that we can evaluate in momentum space. With this coefficient function in hand, all the relevant terms in the on-shell AdS action are known, and we can analytically continue the Euclidean AdS result of section \ref{adscalc} to de Sitter space in section \ref{dsres}. 
We then proceed to calculate the four point scalar 
correlator from it as discussed in section \ref{finaldscalc}. Before embarking on the
 calculation though we let us  first  pause to discuss some general issues pertaining to  conformal invariance in 
the next section. 
 
\section{Conformal Invariance}
\label{conformalinv}

Working in the ADM formalism, with metric, eq.\eqref{admmetric}, 
the action, eq.\eqref{action1},   is given by  
\be
\label{actadm2}
\begin{split}
S={M_{Pl}^2\over 2}\int \sqrt{-h} \bigg[ & N R^{(3)} - 2 N V +{1 \over N} (E_{ij}E^{ij}-E^2)+ {1 \over N} (\partial_t \phi -N^i \partial_i \phi)^2  \\ & -N h^{ij} \partial_i \phi \partial_j \phi\bigg]
\end{split}
\ee
where 
\be
\label{defE}
\begin{split}
E_{ij}= & {1 \over 2}(\partial_t h_{ij}-\nabla_i N_j-\nabla_j N_i), \\
E= & E^i_i.
\end{split}
\ee

The equations of motion of $N_i, N$ give rise to the constraints
\begin{align}
\label{spreparac}
& \nabla_i[N^{-1} (E^i_j-\delta^i_j E)] = 0, \\ \label{wdw}
R^{(3)}-2V -{1\over N^2}(E_{ij}E^{ij}-&E^2) -{1\over N^2} (\partial_t \phi -N^i \partial_i \phi)^2-h^{ij} \partial_i \phi \partial_j \phi=0
\end{align}
respectively.
 
The constraints obtained by varying the shift functions, $N_i$, leads to invariance under spatial 
reparameterizations, eq.\eqref{sprepara}. The constraint imposed by varying the lapse function, $N$, leads to invariance under time reparameterizations, eq.\eqref{timerepara}. Physical states must meet these constraints. In the quantum theory this implies that the wave function must be invariant under the spatial reparameterizations and the time reparameterization, eq.\eqref{sprepara},eq.\eqref{timerepara},
for a pedagogical introduction see \cite{hartle}. 
 We will see that these conditions give rise to the Ward identities of  interest. 

Under the spatial parameterization, eq.\eqref{sprepara} the metric and scalar perturbations transform as 
\begin{align}
\label{metrtsp}
\gamma_{ij}  \rightarrow  \gamma_{ij}+ \delta \gamma_{ij} & =    \gamma_{ij}-\nabla_i v_j - \nabla_j v_i,  \\ \label{scalartrsp}
\delta \phi  \rightarrow  \delta \phi + \delta (\delta \phi) & =  \delta \phi  - v^k \partial_k \delta \phi .
\end{align}
Requiring that the wave function is invariant under the spatial parameterization imposes the condition that 
\be
\label{wfinvsp}
\psi[\gamma_{ij} + \delta \gamma_{ij}, \delta \phi+ \delta (\delta \phi)] = \psi[\gamma_{ij}, \delta\phi].
\ee
For the wave function in eq.\eqref{wf2old} imposing this condition in turn leads to constraints on the coefficient functions.
For example, it is straightforward to see, as discussed in Appendix \ref{appwardident} that we get the condition, 
\be
\label{spreparaward}
\partial_i \langle T_{ij}({\vect{x}}) O({\vect{y_1}}) O({\vect{y_2}}\rangle = \delta^3({\vect{x}}-{\vect{y_1}}) \langle \partial_{y_1^j}O({\vect{y_1}})
O({\vect{y_2}})\rangle + \delta^3({\vect{x}}-{\vect{y_2}}) \langle O({\vect{y_1}})\partial_{y_2^j}
O({\vect{y_2}})\rangle.
\ee
Similar conditions   rise for other correlation functions. 
These conditions are exactly the  Ward identities due to translational invariance  in the conformal field theory. 

Under the time reparameterizations, eq.\eqref{timerepara}, the metric transforms as 
\be
\label{mettrepara}
\gamma_{ij}\rightarrow \gamma_{ij}+ 2 H \epsilon ({\vect{x}}) \delta_{ij}.
\ee
The scalar perturbation, $\delta \phi$,  at late times is independent of $t$ and thus is invariant under time parameterization. 
The invariance of the wave function then gives rise to the condition
\be
\label{wftrepara}
\psi[\gamma_{ij}+ 2 H \epsilon({\vect{x}}), \delta \phi]=\psi[\gamma_{ij}, \delta \phi]
\ee
which also imposes conditions on the coefficient functions. 
For example from eq.\eqref{wf2}  we get that  the condition
\be
\label{extreparaw}
\langle T_{ii}({\vect{x}})O({\vect{y}}_1) O({\vect{y}}_2)\rangle = - 3 \delta^3({\vect{x}}-{\vect{y}}_1) \langle O({\vect{y}}_1) 
O({\vect{y}}_2) \rangle  - 3 \delta^3({\vect{x}}-{\vect{y}}_2) \langle O({\vect{y}}_1) 
O({\vect{y}}_2) \rangle
\ee
must be true, as shown in Appendix \ref{appwardident}. Similarly other conditions also arise;  these are all 
 exactly the analogue of the Ward identities in the CFT
due to Weyl invariance,  with $O$ being an operator of dimension $3$. 

The isometries corresponding to  special conformal transformations were discussed in eq.\eqref{spcon}, eq.\eqref{tspcon}. 
We see that at late times when $e^{-2Ht} \rightarrow 0$ these are given by
\begin{align}
 x^i & \rightarrow  x^i -2 (\eepsilon_j x^j) x^i + \eepsilon^i \sum_j (x^j)^2,  \label{cftrans1}\\
t & \rightarrow  t + 2 {b_j x^j \over H} \label{cftrans2}.
\end{align}
We see that this is a 
combination of a 
spatial parameterization, eq.\eqref{sprepara} and   a time parameterization, eq.\eqref{timerepara}.  The invariance of the 
wave function under the special conformal transformations then follows from our discussion above. It 
is easy to see, as discussed in Appendix \ref{appwardident} that the invariance of the wave function under conformal transformations leads to the condition that the  coefficient functions must be invariant under the transformations,
\begin{align}
 \label{opstr}
\begin{split}
 &O({\vect{x}})   \rightarrow O({\vect{x}}) + \delta O({\vect{x}}), \\
&\delta O({\vect{x}})  = - 6 ( {\vect{x}} \cdot \vect{\eepsilon}) O({\vect{x}}) + D O({\vect{x}}),\\
&D \equiv   x^2 \dotp[\eepsilon,\partial] - 2 \dotp[\eepsilon,x] \dotp[x,\partial].
\end{split}
\end{align}
and
\begin{align}\label{opttr}
 \begin{split}
&T_{ij}({\vect{x}})  \rightarrow T_{ij}+\delta T_{ij},\\
&\delta T_{ij}  = -6 ({\vect{x}}\cdot \vect{\eepsilon}) T_{ij} - 2 {M^k}_i T_{k j} - 2 {M^k}_j T_{i k}+ D T_{ij}, \\
&{M^k}_i  \equiv   (x^k \eepsilon^i - x^i \eepsilon^k),\\
&D \equiv   x^2 \dotp[\eepsilon,\partial] - 2 \dotp[\eepsilon,x] \dotp[x,\partial].
\end{split}
\end{align}
The resulting conditions on the coefficient functions agree exactly with the Ward identities for conformal invariance which must be satisfied by correlation function in the conformal field theory. 

Specifically for the scalar four point function of interest here, the relevant terms in the wave 
function are given in eq.\eqref{wf2} in momentum space. The momentum space versions of 
eq.\eqref{opstr}, eq.\eqref{opttr} are given in the appendix \ref{appsubtrans} in  
 eq.\eqref{appa2}, eq.\eqref{appa4},  eq.\eqref{osctmom}, eq.\eqref{gamsctmom}.
It is easy to check that  
the two point functions, $\langle O(\vk) O(\vkk)\rangle$ and $\langle T_{ij} (\vk) T_{kl}(\vkk)\rangle $ 
are both invariant under these transformations. The invariance of 
$\langle O(\vk) O(\vkk) T^s(\vkkk)\rangle$ was discussed \emph{e.g.} in \cite{Bzowski:2011ab, MRT}. 
In order to establish conformal invariance for the wave function it is then enough to prove that the 
coefficient function $\langle O(\vk) O(\vkk) O(\vkkk) O(\vkfour)\rangle$ is invariant under the transformation eq.\eqref{osctmom}.   We will see that the answer we calculate in section \ref{4ptcorrinads} does indeed  have this property. 

\subsection{Further Gauge Fixing and Conformal Invariance}
\label{fgaugefix}
We now come to  an interesting subtlety which arises when we consider the conformal invariance of 
 correlation functions, as opposed to the wave function, in  the  de Sitter case. 
This subtlety arises because one needs to integrate over the  metric and scalar perturbations, 
to calculate the correlation functions from the wave function. In order to do so the gauge symmetry 
 needs to be fixed more completely,   as we will see   in the subsequent discussion. 
 However, once this additional gauge fixing is done a general conformal transformation does not preserve the choice of gauge. 
Thus, to test for conformal invariance of the resulting correlation functions, the conformal transformation must be accompanied by  a compensating coordinate transformation which restores the choice of gauge. As we describe below, this compensating
transformation is itself field-dependent. 
The invariance of the correlation functions under the  combined
conformal transformation and compensating coordinate transformation is  then the signature of the 
underlying conformal invariance. 

  Let us note here that   
this subtlety does not have  a corresponding analogue in the AdS case, where one computes the  partition function,
and the boundary value of the metric is a source which is non-dynamical. 
It is also worth emphasizing, before we go further, that due to these complications it is in fact easier  to 
test for the symmetries in the wave function itself rather than in the correlators which are calculated from it.
Calculating the wave function by itself does not require the additional gauge fixing mentioned above. Thus the wave function should be invariant separately under conformal transformations and spatial reparameterizations. 
Once this is shown to be true the invariance of  the probability distribution function $P[\delta \phi]$
and all correlation functions
under the combined conformal transformation and gauge restoring parameterization then  follows.     

We will now discuss this issue in more detail. 
Let us  begin by noting, as was discussed in section \ref{prelim}, that 
the conditions, eq.\eqref{lapse}, eq.\eqref{shiftc}, do not fix the gauge completely.
One has the freedom to do spatial reparameterizations of the form eq.\eqref{sprepara}, and at late times, 
also a time parameterization of the form, eq.\eqref{timerepara}. 
Using this freedom one can then fix the gauge  further,  for example, leading to gauge 1 or gauge 
2 in section \ref{prelim}. In fact it is necessary to do so in order to calculate correlation functions from 
the wave function, otherwise one would end up summing over an infinite set of copies of the same physical configuration.

As a concrete example, consider the case where we make the choice of gauge 2 of subsection \ref{gauge2}. 
In this gauge $\zeta=0$ and the metric $\gamma_{ij}$ is both traceless and transverse. 
On carrying out a conformal transformation, the coordinates $x^i,t$ transform as given in eq.\eqref{cftrans1}
and  eq.\eqref{cftrans2} respectively. As shown in appendix \ref{appwardident} eq.\eqref{gammachange},
\be
\label{gammatr}
\delta \gamma_{ij}(x)=2 {M^m}_j \gamma_{i m} + 2 {M^m}_i \gamma_{m j}-(x^2 \eepsilon^m - 2 x^m \dotp[x, \eepsilon]){\partial \gamma_{ij}({\vect{x}}) \over \partial x^m},
\ee
where $\delta \gamma_{ij}=\gamma'_{ij}(x)-\gamma_{ij}(x) $ is the change in $\gamma_{ij}$ and ${M^m}_j = x^m b^j- x^j b^m$.

Since $\delta \gamma_{ii}=0$, $\gamma'_{ij}$  remains traceless and $\zeta$ continues to vanish. However
\be
\label{transdelg}
\partial_i \delta \gamma_{ij}=-6 b^k \gamma_{k j}\ne 0,
\ee
so we see that  $\gamma'_{ij}(x)$ is not transverse anymore. 

Now, upon carrying out a further coordinate transformation
\be
\label{furthct}
x^i \rightarrow x^i + v^i({\vect{x}})
\ee
$\gamma_{ij}$ transforms as 
\begin{align}
\label{reparagamchan}
\begin{split}
 \gamma_{ij}({\vect{x}}) &\rightarrow \gamma_{ij}({\vect{x}}) +\delta \gamma_{ij},\\
\delta \gamma_{ij} &= - \partial_i v_j -\partial_j v_i.
\end{split}
\end{align}
Choosing
\be
\label{valcompv}
v^j({\vect{x}})={-6 b^k \gamma_{k j}({\vect{x}})\over \partial^2}
\ee
it is easy to see that transformed metric perturbation $\gamma_{ij}$ continues to be traceless and also now becomes transverse.    
 The combination of the conformal
transformation, eq.\eqref{cftrans1} and the compensating spatial parameterization 
eq.\eqref{furthct}, eq.\eqref{valcompv}, thus keep one in gauge 2. 
Let us note here that we will work with perturbation with non-zero momentum, thus ${1\over \partial^2}=-
{1 \over  k^2}$ will be well defined.  

The scalar perturbation $\delta \phi$ transforms like a scalar under both the conformal transformation, eq.\eqref{cftrans1}  and the compensation parameterization eq.\eqref{furthct} with eq.\eqref{valcompv}. It then follows  that under the combined transformation which leaves one in gauge 2 it transforms as follows:
\begin{align}
 \label{comdphi1}
\delta \phi & \rightarrow   \delta \phi + \delta(\delta \phi ), \\ \label{comdphi2}
\delta(\delta \phi ) & =   \delta^C(\delta \phi )+ \delta^R(\delta \phi ),
\end{align}
where $\delta^C(\delta \phi )$ is the change in $\delta \phi$ due to conformal transformation eq.\eqref{cftrans1},
\be
\label{delcdelphi}
\delta^C\big(\delta \phi ({\vect{x}})\big) = - (x^2 \eepsilon^i - 2 x^i \dotp[x, \eepsilon]) {\partial \over \partial x^i} \delta \phi({\vect{x}})
\ee
and $\delta^R(\delta \phi )$ is the change in $\delta \phi$ under spatial parameterization eq.\eqref{furthct} with eq.\eqref{valcompv}
\be
\label{delrdelphi}
\delta^R\big(\delta \phi({\vect{x}}) \big)= -v^i({\vect{x}}) \partial_i \delta \phi({\vect{x}}).
\ee
It is important to note that the coordinate transformation parameter $v^i$, eq.\eqref{reparagamchan} is itself dependent on the metric perturbation, $\gamma_{ij}$, eq.\eqref{valcompv}. As a result the change in $\delta\phi$ under the spatial parameterization is non-linear in the perturbations, $\gamma_{ij}, \delta \phi$. 
This is in contrast to $\delta^C(\delta \phi)$ which is linear in $\delta \phi$. As we will see in section \ref{testconf} 
when we discuss the four point function in more detail, a consequence of this non-linearity  is that terms 
in the probability distribution function which are quadratic in $\delta \phi$ will mix with those which are quartic, thereby 
ensuring invariance under the combined transformation, eq.\eqref{comdphi2}.

The momentum space expression for $\delta^C\big(\delta \phi ({\vect{x}})\big)$ is given in eq.\eqref{delphsctmom} of Appendix \ref{appsubtrans}. We write here the momentum space expression for $v^i$ and $\delta^R\big(\delta \phi ({\vect{x}})\big)$
\begin{align}
 \label{valcompvmom} 
 v_i ({\vect{k}}) &= {6 b^k \gamma_{k i}({\vect{k}}) \over k^2}, \\ 
\label{delphrmom}
\delta^R\big(\delta \phi({\vect{k}}) \big) &= i 6b^k k^i \int {d^3 k_2 \over (2 \pi)^3} {\gamma_{k i}({\vect{k}}-\vkk)\over |{\vect{k}}-\vkk|^2} \delta\phi(\vkk).
\end{align}

\subsection{Conformal Invariance of the Four Point Correlator}
\label{confinv4pt}
Now consider the four point scalar correlator in gauge 2. It can be calculated from 
$\psi[\delta\phi, \gamma_{ij}]$ by evaluating the functional integral:
\be
\label{fourptfi}
\langle \delta \phi({\vect{x}}_1) \delta \phi({\vect{x}}_2) \delta \phi({\vect{x}}_3) \delta \phi({\vect{x}}_4)\rangle=  \mathcal{N} 
\int \mathcal{D}[ \delta \phi] \mathcal{D}[ \gamma_{ij}] \prod_{i=1}^4 \delta \phi({\vect{x}}_i) \  |\psi[\delta \phi, \gamma_{ij}]|^2.
\ee
The normalization $\mathcal{N} $ is given by 
\be
\label{defN}
\mathcal{N} ^{-1}=\int \mathcal{D}[\delta \phi]  \mathcal{D}[\gamma_{ij}] \  |\psi[\delta \phi, \gamma_{ij}]|^2.
\ee
The integral over the field configurations in eq.\eqref{fourptfi}
 can be done in two steps. We can first integrate out the metric
to obtain a probability  distribution which is a functional of $\delta\phi$ alone,
\be
\label{defP}
P[\delta \phi({\vect{x}})]=  \mathcal{N}  \int \mathcal{D}[\gamma_{ij}] \ |\psi[\delta \phi , \gamma_{ij}]|^2,
\ee
and then use $P[\delta \phi(x)]$ to compute correlations of $\delta \phi$, in particular  the correlator,
\be
\label{expcorr}
\langle \delta \phi({\vect{x}}_1) \delta \phi({\vect{x}}_2) \delta \phi({\vect{x}}_3) \delta \phi({\vect{x}}_4)\rangle= \int \mathcal{D} [\delta\phi] \prod_{i=1}^4 \delta \phi({\vect{x}}_i) P[\delta \phi].
\ee
Note that the integral over the metric $\gamma_{ij}$ is well defined only because of the further gauge fixing which was done leading  to gauge 2. 

The invariance of the wave function under conformal transformations and compensating spatial reparameterizations implies that 
the probability distribution $P[\delta \phi]$ must be invariant under the combined transformation generated by 
the conformal transformation and compensating parameterization which leaves one in the gauge 2.
This gives rise to the condition
\be
\label{condpphia}
P[\delta \phi + \delta (\delta \phi)]=P[\delta \phi]
\ee
where $\delta (\delta \phi)$ is given in eq.\eqref{comdphi2} with eq.\eqref{delcdelphi} and eq.\eqref{delrdelphi}. 
We will see in section \ref{testconf} that our final answer for $P[\delta \phi]$ does indeed meet this condition.

\section{The $\langle O({\vect{x}}_1) O({\vect{x}}_2) O({\vect{x}}_3) O({\vect{x}}_4)\rangle  $ Coefficient Function}
\label{4ptcorrinads}
We now compute the coefficient of the quartic term in the wave function of the Universe. This coefficient is the same as the four point correlation function
of marginal scalar operators in anti-de Sitter space. %This calculation is done in two parts. First, we compute the answer in anti-de Sitter space. 
As explained above, this calculation has the advantage that it can be done by standard Feynman-Witten diagram techniques. In the next section, we put
this correlator together with other known correlators to obtain  the wave function of the Universe at late times. This can then easily be used to compute
the expectation value in
de Sitter space that we are interested in.

{\bf \underline {Additional Conventions}:} Some additional conventions we will use are worth stating here. 
The Greek indices $\mu, \nu, \cdots,$ take  $4$ values in the $z, x^i, i=1,2,3,$ directions. 
The inverse of the  back ground metric ${\bar g}_{\mu\nu}$ is denoted by ${\bar g}^{\mu\nu}$, while indices for
the metric perturbation $\delta g_{\mu\nu}$ are   raised or lowered using the flat space metric, so that, e.g.,
$\delta g^{\mu\nu}=\eta^{\mu\rho} \delta g_{\rho\kappa} \eta^{\kappa \nu}$.

\subsection{The Calculation in AdS Space}
\label{adscalc}
We are now ready to begin our calculation of the $\langle O({\vect{x}}_1) O({\vect{x}}_2) O({\vect{x}}_3) O({\vect{x}}_4)\rangle  $ coefficient function.   As discussed in subsection \ref{basicstrategy} we will first calculate the relevant term in the partition function in Euclidean AdS space and then continue the answer to obtain this coefficient function in dS space. 
 This will allow us to readily 
use some of the features  recently employed in  AdS space calculations. However, it  is worth emphasizing at the outset itself that it is not necessary to do the calculation in this way. The problem of interest  is   well posed in de Sitter space and if the reader prefers, the calculation can be  directly done  in de Sitter space, using only  minor modifications in the AdS calculation. 

The perturbations in dS space we are interested in can be studied with the action given in eq.\eqref{resact}. 
For the analogous problem in AdS space we start with the action
\be
\label{eucads}
S={M_{Pl}^2 \over 2}\int d^4 x \sqrt{g} \bigg[R-2 \Lambda - (\nabla \delta \phi)^2\bigg]
\ee
where $g_{\mu\nu}$ here is a Euclidean signature metric, and $\Lambda$, the cosmological constant.
AdS space arises as the solution of this system with metric, eq.\eqref{eadsmet}, and with the scalar 
$\delta\phi=0$. 
The radius $R_{\text{AdS}}$ in eq.\eqref{eadsmet} is related to $\Lambda$ \footnote{Note that in our conventions $\Lambda<0$ corresponds to $AdS$ space.} by  
\be
\Lambda = - {3 \over R_{\text{AdS}}^2}.
\ee

 To simplify the analysis it is convenient
 to set $R_{\text{AdS}}=1$,  the dependence on $R_{\text{AdS}}$ can be restored by  
noting that the action is dimensionless, so that the prefactor which multiples the action
 must appear in the combination ${M_{Pl}^2 R_{\text{AdS}}^2\over 2}$. 
The metric in eq.\eqref{eadsmet} then  becomes 
\be
\label{eadsr}
ds^2={dz^2+(dx^i)^2 \over z^2}
\ee
where the index $i$ takes values,  $i=1,2,3$. 

For studying the small perturbations we expand the metric by writing 
\be
\label{expmetads}
g_{\mu\nu}=\bar{g}_{\mu\nu}+ \delta g_{\mu\nu}
\ee where $\bar{g}_{\mu\nu}$ is the AdS metric given in eq.\eqref{eadsr}
 and $\delta g_{\mu\nu}$ is the metric  perturbation.
Expanding the action, eq.\eqref{eucads}  in powers of the perturbations $\delta g_{\mu\nu}$ and $\delta \phi$ then gives, 
\be
\label{expact}
\begin{split}
S= & S_0+ S_{\text{grav}}^{(2)}
- {M_{Pl}^2 \over 2} \int d^4 x \sqrt{\bar{g}}  \ 
 \bar{g}^{\mu\nu} \ \partial_\mu (\delta \phi) \partial_\nu (\delta \phi) + S_{\text{\text{int}}}.
\end{split}
 \ee
$S_0$ in eq.\eqref{expact} is the action for the background AdS space with metric eq.\eqref{eadsr}. $S_{\text{grav}}^{(2)}$ is the quadratic part of the metric perturbation. Using the action given in \cite{Christensen:1979iy} (see also eq.(98) in \cite{veltman}), and using the first order equations of motion the quadratic
action for the graviton can be simplified to \cite{Raju:2011mp}
\begin{align}
\label{metpertquad}
\begin{split}
S_{\text{grav}}^{(2)} = {M_{Pl}^2 \over 8}\int {d^4 \vect{x} \sqrt{\bar{g}}}  \left(\widetilde{\delta g}^{\mu \nu} \Box \delta g_{\mu \nu} + 2 \widetilde{\delta g}^{\mu \nu} R_{\mu \rho \nu \sigma} \delta g^{\rho \sigma}  + 2 \nabla^{\rho} \widetilde{\delta g}_{\rho \mu} \nabla^{\sigma} \widetilde{\delta g}_{\sigma}^{\mu} \right),
\end{split}
\end{align}
with $\widetilde{\delta g}^{\mu \nu}  = \delta g^{\mu \nu} - {1 \over 2} \bar{g}^{\mu \nu} \delta g^{\alpha}_{\alpha}$.
We also expand 
$S_{\text{int}}$
  to linear order
 in $\delta g_{\mu\nu}$, since higher order terms are not relevant to our calculation. 

\be
\label{defint}
\begin{split}
S_{\text{int}}=&{M_{Pl}^2 \over 2} \int d^4 x \sqrt{\bar{g}} \ {1\over 2} \ \delta g^{\mu\nu} T_{\mu\nu},
\end{split}
\ee
where the scalar stress energy  is 
\be
\label{defstressa}
T_{\mu\nu}= 2\partial_\mu (\delta \phi) \partial_\nu (\delta \phi)  -\bar{g}_{\mu\nu}
\bar{g}^{\alpha\beta} \partial_\alpha (\delta \phi)\partial_\beta (\delta \phi).
\ee
Let us note that the quadratic term, eq.(\ref{metpertquad}), can also be written as, see eq.(98) in \cite{veltman},
\begin{align}
\label{metpertquadvelt}
\begin{split}
S_{grav}^{(2)} =& {M_{Pl}^2 \over 2} \int d^4 x \sqrt{\bar{g}} \bigg[-2\Lambda\bigg\{ -{1\over4}\delta g_{\alpha \beta} \delta g^{\alpha \beta} +{1\over8}(\delta g^{\alpha}_{\alpha})^2\bigg\}
+\bigg\{-R \bigg({1\over8}(\delta g^{\alpha}_{\alpha})^2\\ & -{1\over4}\delta g_{\alpha \beta} \delta g^{\alpha \beta} \bigg) - \delta g^{\nu \beta} \delta g_{\beta \alpha}R^{\alpha \nu} 
+{1 \over 2} \delta g^{\alpha}_{\alpha} \delta g^{\nu \beta}R_{\nu \beta}  -{1\over4}\nabla_{\mu}\delta g_{\alpha \beta}\nabla^{\mu} \delta g^{\alpha \beta} \\ & +{1\over4} \nabla_{\mu}\delta g^{\alpha}_{\alpha}\nabla^{\mu}\delta g^{\beta}_{\beta} -{1\over2}\nabla_{\beta} \delta g^{\alpha}_{\alpha}\nabla_{\mu}\delta g^{\beta \mu} +{1\over2} \nabla^{\alpha}\delta g^{\nu \beta}\nabla_{\nu}\delta g_{\alpha \beta}
\bigg\} \bigg].
\end{split}
\end{align}

From eq.(\ref{expact})  we see that the scalar perturbation is a free field with only gravitational interactions. 
The four point function arises from $S_{\text{int}}$ due to single  graviton exchange. 
The scalar perturbation gives rise to a stress energy which  sources a metric perturbation. 
Using the action eq.\eqref{expact}   we can solve for $\delta g_{\mu\nu}$ in terms of this source and a suitable
Green's function. Then substitute the solution for the metric perturbation back into the action to  
obtain the on-shell action as a function of $\phi_{{\vect{k}}_i}$.

From eq.\eqref{expact} we see that to leading order the scalar perturbation, $\delta \phi$,
 satisfies the free equation in AdS space. 
For a mode with momentum dependence $e^{i {\vect{k}} \cdot {\vect{x}}}$ the solution, which is regular as $z \rightarrow \infty$ is given by  
\be
\label{adsscaa}
\delta \phi=(1+ k z) e^{-k z} e^{i {\vect{k}} \cdot {\vect{x}}}.
\ee
A general solution is obtained by linearly superposing solutions of this type. 
For calculating the four point scalar correlator  we  take 
\be
\label{delphidd}
\delta \phi=\sum_{i=1}^4 \phi({{\vect{k}}_i}) (1+k_i z) e^{-k_i z}e^{i {\vect{k}}_i \cdot {\vect{x}}}
\ee
so that it is a sum of four modes with momenta $\vk, \cdots \vkfour$, with coefficients $\phi({\vect{k}}_i)$. 

Notice that towards the boundary of AdS space,  as $z\rightarrow 0$, 
\be
\label{limdelphi}
\delta \phi = \sum_{i=1}^4 \phi({{\vect{k}}_i}) e^{i {\vect{k}}\cdot {\vect{x}}}.
\ee
Thus the procedure above yields the partition function in AdS space as a function of the external scalar
 source, eq.\eqref{limdelphi}. On suitably analytically continuing this answer we will then obtain the 
wave function in 
de Sitter space as a functional of the boundary value of the scalar field given in eq.\eqref{limdelphi}
from where the four point correlator can be obtained.

To proceed  we must  fix a gauge for the metric perturbations,   because it is only after doing so we can  solve for the metric  uniquely in terms of the matter stress tensor. Alternately stated, the Feynman diagram for graviton exchange involves the graviton propagator, which
is well defined only after a choice of gauge for the graviton. We will choose the gauge 
\be
\label{gmetp}
\delta g_{z z}=0, \ \ \delta g_{z i}=0, 
\ee
with  $i=1,2,3$,  taking values over  the  $x^i$ directions.  We emphasize that, at this stage, our final answer for the correlation
function in anti-de Sitter space, or the on shell action is gauge invariant and independent of our choice of gauge above.
After the analytic continuation, eq.\eqref{contadsdsa}, 
this gauge goes over to the gauge eq.\eqref{lapse}, eq.\eqref{shiftc} 
discussed in section \ref{perturbations} in the context of dS space. 
\footnote{The conformal time $\eta$ in eq.\eqref{contadsdsa} is related to $t$
in eq.\eqref{admmetric} by $\eta=e^{-Ht}$.}

\begin{figure}[!h]
        \centering
        \begin{subfigure}[b]{0.32\textwidth}
                \includegraphics[width=\textwidth]{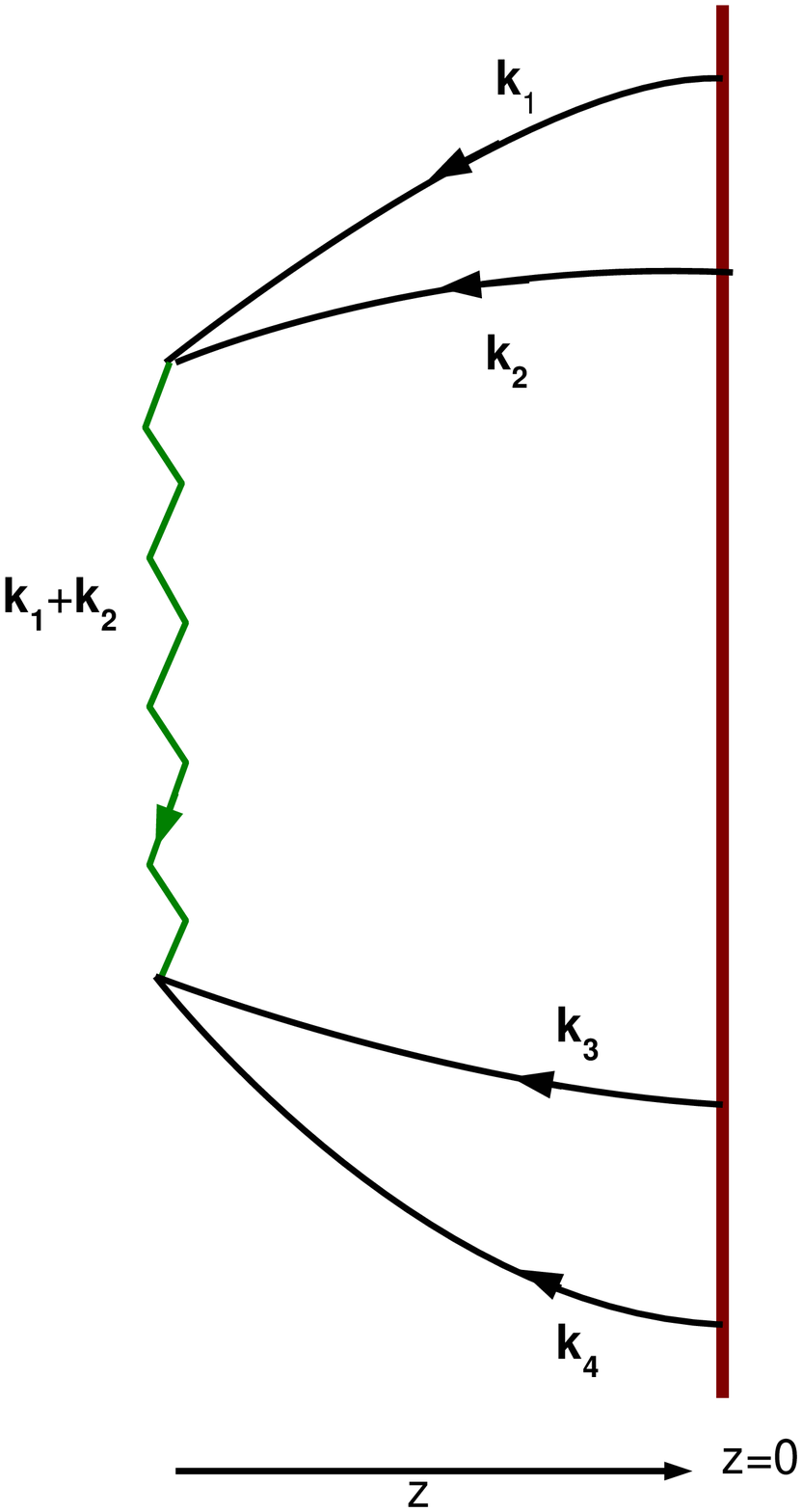}
                \caption{S-Channel}
        \end{subfigure}%
        ~ %add desired spacing between images, e. g. ~, \quad, \qquad etc.
        \begin{subfigure}[b]{0.32\textwidth}
                \includegraphics[width=\textwidth]{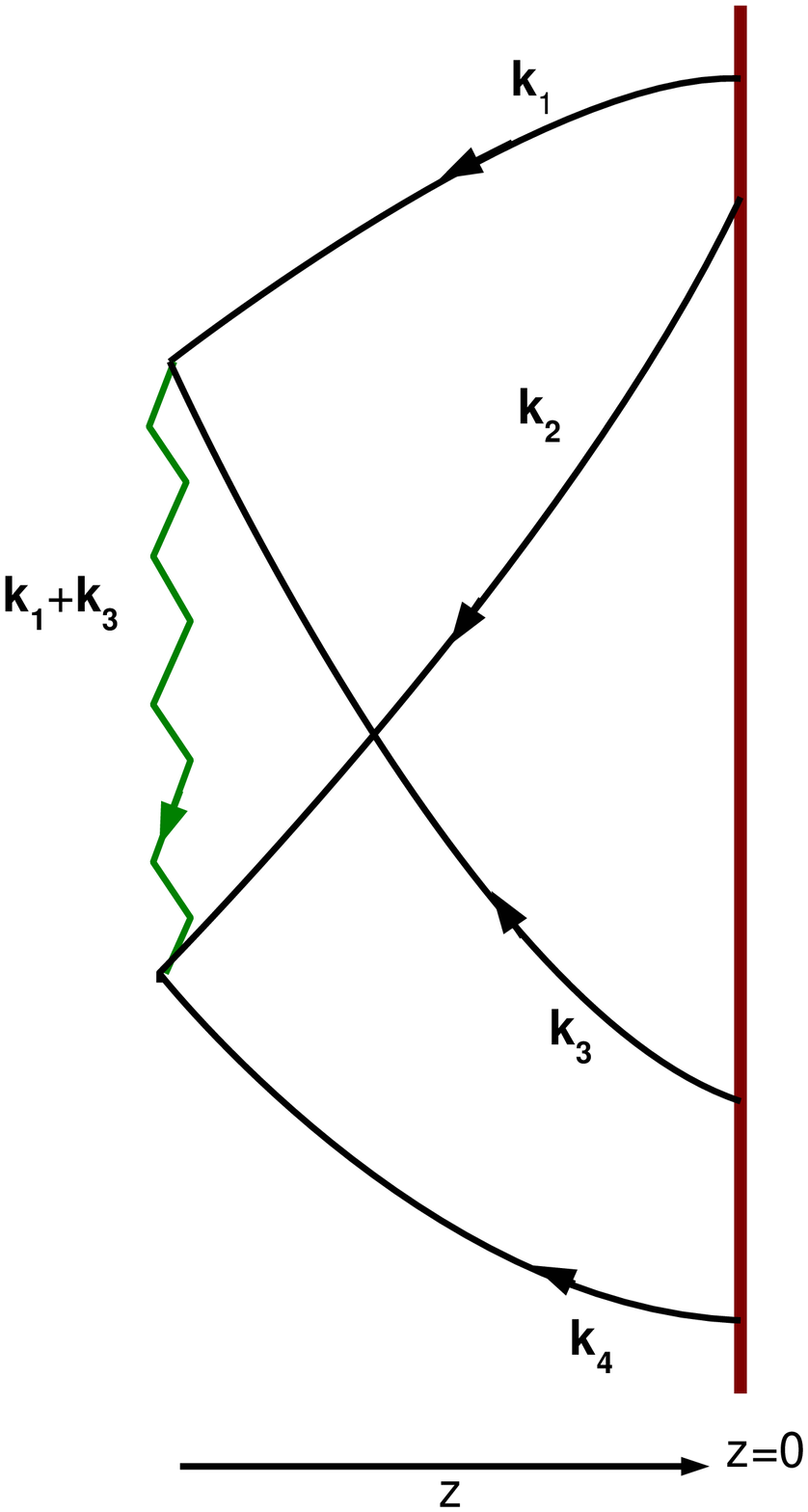}
                \caption{T-Channel}
        \end{subfigure}
        ~ %add desired spacing between images, e. g. ~, \quad, \qquad etc.
        \begin{subfigure}[b]{0.32\textwidth}
                \includegraphics[width=\textwidth]{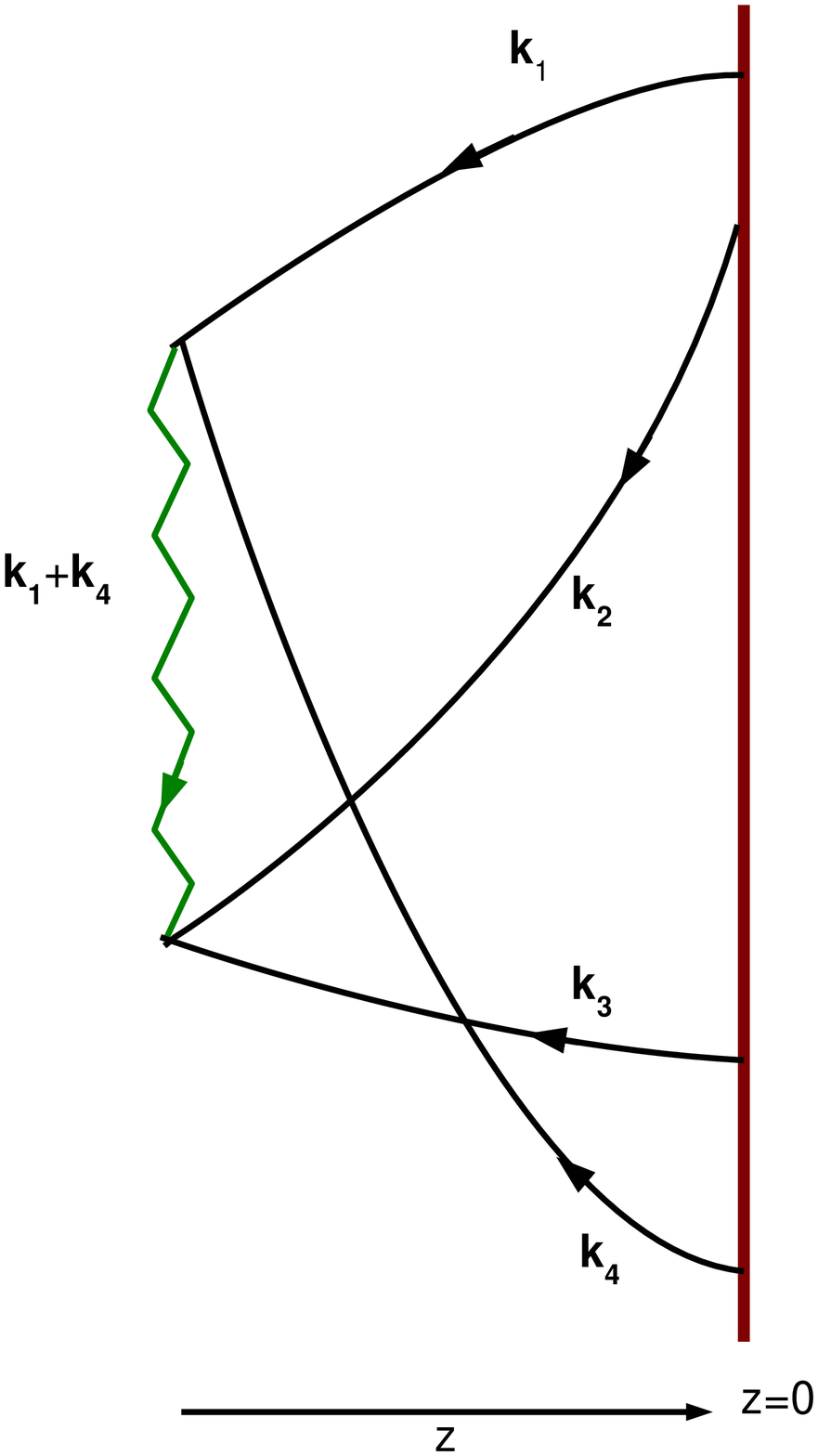}
                \caption{U-Channel}
        \end{subfigure}
        \caption{Three different contribution corresponding to S,T and U-channel to the scalar four point correlator are shown in the three figures. The brown solid vertical line represents the 3-dimensional boundary of $AdS_4$ at $z=0$, the black solid lines are boundary to bulk scalar propagators whereas the green wavy lines are graviton propagators in the bulk. \label{threewittendiag}}
\end{figure}

The on-shell action, with boundary values set for the various perturbations, has an expansion precisely analogous to \eqref{wf2}. As we mentioned there, the only unknown coefficient is the
four-point correlation function $\langle O(\vect{x_1}) O(\vect{x_2}) O(\vect{x_3}) O(\vect{x_4}) \rangle$. Although, at tree-level
this correlator can be computed by solving the classical equations of motion, it is more convenient to simply  evaluate the Feynman-Witten diagrams shown in Fig. \ref{threewittendiag}. The answer is then simply
\be
\label{wittendiagorig}
\begin{split}
S^{\text{AdS}}_{\text{on-shell}} = {M_{\text{pl}}^2 R_{\text{\text{AdS}}}^2 \over 8} \int {d z_1 \over z_1^4} {d z_2 \over z_2^4}  d^3 x_1 d^3 x_2  &\bar{g}^{i_1 i_2} \bar{g}^{j_1 j_2} T_{i_1 j_1} (x_1, z_1)   G^{{\text{grav}}}_{i_2 j_2, k_2 l_2} (x_1, z_1, x_2, z_2) \\ &\bar{g}^{k_1 k_2} \bar{g}^{l_1 l_2} T_{k_1 l_1}(x_2, z_2) .
\end{split}
\ee
In this equation the scalar stress-tensor $T_{i j}$ is given in \eqref{defstressa}, and the graviton propagator $G^{\text{grav}}$ is given by \cite{Raju:2010by, Raju:2011mp}
\begin{equation}
\label{gravitypropagator}
\begin{split}
G^{{\text{grav}}}_{i j, k l} =  \int {d^3 {\vect{k}} \over (2\pi)^3} e^{i{\vect{k}}\cdot
({\vect{x}}_1-{\vect{x}}_2)}\int_0^{\infty} {d p^2\over2} \bigg[{ J_{{3 \over 2}}(p z_1) J_{{3 \over 2}} (p z_2) \over  
 \sqrt{z_1 z_2}\left({\vect{k}}^2 + p^2 \right)}   {1 \over 2}\left({{\cal T}}_{i k} {{\cal T}}_{j l} + {{\cal T}}_{i l} {{\cal T}}_{j k} - 
 {{\cal T}}_{i j} {{\cal T}}_{k l} \right) \bigg],
\end{split}
\end{equation}
where
\be
\label{defprojector}
{\cal T}_{i j} = \delta_{i j} + {k_{i} k_{j} \over p^2}.
\ee
Since the $x$-integrals in \eqref{wittendiagorig} just impose momentum conservation in the boundary directions, the entire four-point function calculation boils down to doing a simple integral in the radial ($z$) direction. Here, the factors of ${1 \over z^4}$ come from the
determinant of the metric to give the appropriate volume factor.
 
Note that the projector that appears in the graviton propagator is not transverse and traceless. As we also discuss in greater detail below in section \ref{subsecflatspace}, this is the well known analogue of the fact that the axial gauge propagator in flat space also has a longitudinal component.  For calculational purposes it is convenient to 
break up our answer into the contribution from the transverse graviton propagator, and the longitudinal propagator. 
This leads us to write the graviton propagator in a form that was analyzed in  \cite{Liu:1998ty} (see eq. 4.14 of that paper),
and we find that the four point correlation function
\be
\label{onshellact}
S^{\text{AdS}}_{\text{on-shell}} ={M_{Pl}^2 R_{\text{AdS}}^2 \over 2} {1 \over 4}[\widetilde{W} + 2 R], 
\ee
where, 
$\widetilde{W}$ is obtained from the scalar stress tensor, eq.\eqref{defstressa}, and the transverse graviton Greens function,
$\widetilde{G}_{ij,kl}(z_1,{\vect{x}}_1;z_2,{\vect{x}}_2)$. 
\begin{equation}
\label{wittendiag}
\widetilde{W} = \int d z_1 d^3 {\vect{x}}_1  d z_2 d^3 {\vect{x}}_2   
T_{i_1 j_1} (z_1,{\vect{x}}_1)  \delta^{i_1 i_2} \delta^{j_1 j_2} 
\widetilde{G}_{i_2 j_2, k_2 l_2} (z_1, {\vect{x}}_1; z_2,{\vect{x}}_2) 
\delta^{k_1 k_2} \delta^{l_1 l_2} T_{k_1 l_1}(z_2,{\vect{x}}_2) \,.
\end{equation}
In the expression above, we have also canceled off the volume factors of ${1 \over z^4}$ in \eqref{wittendiagorig} with the two factors of $z^2$ in the raised metric. 
The transverse graviton Green function is almost the same as \eqref{gravitypropagator}  
\begin{align}
\label{tilwx1x2}
 \begin{split}
  \widetilde{G}_{i j, k l} (z_1, {\vect{x}}_1; z_2,{\vect{x}}_2) = & \int {d^3 {\vect{k}} \over (2\pi)^3} e^{i{\vect{k}}\cdot
({\vect{x}}_1-{\vect{x}}_2)}\int_0^{\infty} {d p^2\over2} \bigg[{ J_{{3 \over 2}}(p z_1) J_{{3 \over 2}} (p z_2) \over  
 \sqrt{z_1 z_2}\left({\vect{k}}^2 + p^2 \right)}   {1 \over 2} \\ & \left({\widetilde{\cal T}}_{i k} {\widetilde{\cal T}}_{j l} + {\widetilde{\cal T}}_{i l} {\widetilde{\cal T}}_{j k} - 
 {\widetilde{\cal T}}_{i j} {\widetilde{\cal T}}_{k l} \right) \bigg],
 \end{split}
\end{align}
except that $\widetilde{\cal T}_{ij}$, which appears here, is a projector onto directions perpendicular to ${\vect{k}}$,
\be
\label{defprojec1}
\widetilde{\cal T}_{ij}= \delta_{ij}-{k_i k_j\over k^2}.
\ee
After momentum conservation is imposed on the intermediate graviton, we have $\vect{k} = \vect{k_1} + \vect{k_2}$.
Details leading to eq.\eqref{wittendiag} are discussed in appendix \ref{detailsadsos}.  

The other term on the RHS of eq.\eqref{onshellact}, $R$, arises from the longitudinal graviton contribution (which is just the difference between \eqref{gravitypropagator} and \eqref{tilwx1x2}) and  it is convenient for us to write it
as a sum of three terms, 
\be
\label{valR}
R=R_1+R_2+R_3,
\ee
with,
\begin{align}
\label{remainder1}
\begin{split}
R_1 &=-  \int {d^3 {\vect{x}}_1 dz_1 \over z_1^{2}} T_{z j}({\vect{x}}_1, z_1) {1 \over \partial^2} T_{z j}({\vect{x}}_1, z_1),\\
R_2 &= -{1 \over 2} \int {d^3 {\vect{x}}_1 dz_1 \over z_1}  \ \partial_j T_{z j}({\vect{x}}_1, z_1) {1 \over \partial^2} T_{z z}({\vect{x}}_1, z_1),\\
 R_3 &=- {1 \over 4}\int {d^3 {\vect{x}}_1 dz_1 \over z_1^{2}}  \ \partial_j T_{z j}({\vect{x}}_1, z_1) \left({1 \over \partial^2}\right)^2 \partial_i T_{z i}({\vect{x}}_1, z_1),
\end{split}
 \end{align}
where ${1\over \partial^2 }$ denotes the inverse of $\partial_{x^i} \partial_{x^j} \delta^{ij}$.

Substituting for $\delta \phi$ from eq.\eqref{limdelphi} in the stress tensor eq.\eqref{defstressa} one can calculate both these contributions. 
The resulting answer is the sum of three terms shown in Fig 1.(a), 1.(b), 1.(c), which can be thought of as  corresponding
 to $S,T,$ and $U$ channel contributions respectively. In the S channel exchange, Fig. 1.(a), the  
momentum carried by the graviton  along the $x^i$ directions is, 
\be
\label{schnlk}
{\vect{k}}=\vk + \vkk.
\ee

The contributions of the $T,U$ channels can be obtained by replacing $\vkk \leftrightarrow \vkkk$,
and $\vkk \leftrightarrow \vkfour$ respectively. 

The $S$ channel contribution for $\widetilde {W}$ which we denote by $\widetilde{W}^S(\vk, \vkk, \vkkk,
\vkfour)$ turns out to be 
\be
\label{defwts}
\widetilde{W}^S(\vk,\vkk,\vkkk,\vkfour) =16 (2\pi)^3 \ \delta^3\bigg(\sum_{J=1}^4 {\vect{k}}_J\bigg) \bigg( \prod_{I=1}^4 \phi({\vect{k}}_I)\bigg) \widehat{W}^S(\vk,\vkk,\vkkk,\vkfour)
\ee

where 
\begin{align}
\label{hatW}
\begin{split}
 &\widehat{W}^S(\vk,\vkk,\vkkk,\vkfour) = - 2 \bigg[ \bigg\{\vk.\vkkk+\frac{\{(\vkk+\vk).\vk\} \{(\vkfour+\vkkk).\vkkk\}}{|\vk+\vkk|^2}\bigg\} \\ & \bigg\{\vkk.\vkfour+  \frac{\{(\vk+\vkk).\vkk\} \{(\vkkk+\vkfour).\vkfour\}}{|\vk+\vkk|^2}\bigg\} 
+  \bigg\{\vk.\vkfour+\frac{\{(\vkk+\vk).\vk\} \{(\vkfour+\vkkk).\vkfour\}}{|\vk+\vkk|^2}\bigg\}\\ & 
\bigg\{\vkk.\vkkk+\frac{\{(\vkk+\vk).\vkk\} \{(\vkfour+\vkkk).\vkkk\}} {|\vk+\vkk|^2}\bigg\}-
\bigg\{\vk.\vkk-\frac{\{(\vkk+\vk).\vk\} \{(\vk+\vkk).\vkk\}} {|\vk+\vkk|^2}\bigg\}\\ & 
\bigg\{\vkkk.\vkfour-\frac{\{(\vkkk+\vkfour).\vkfour\} \{(\vkfour+\vkkk).\vkkk\}}{|\vk+\vkk|^2}\bigg\} \bigg] \times\\ & 
\Bigg[ \bigg\{\frac{ k_{1} k_{2} (k_{1}+k_{2})^2 \left((k_{1}+k_{2})^2-k_{3}^2-k_{4}^2-4 k_{3} k_{4}\right)}{ (k_{1}+k_{2}-k_{3}-k_{4})^2 (k_{1}+k_{2}+k_{3}+k_{4})^2
   (k_{1}+k_{2}-|\vk+\vkk|) (k_{1}+k_{2}+|\vk+\vkk|)} \\ &\Big(-\frac{k_{1}+k_{2}}{2 k_{1}
   k_{2}}-\frac{k_{1}+k_{2}}{-(k_{1}+k_{2})^2+k_3^2+k_{4}^2+4 k_{3}
   k_{4}}+\frac{k_{1}+k_{2}}{|\vk+\vkk|^2-(k_{1}+k_2)^2}\\ &+\frac{1}{-k_{1}-k_{2}+k_{3}+k_{4}} -\frac{1}{k_{1}+k_{2}+k_{3}+k_{4}}+\frac{3}{2 (k_{1}+k_{2})}\Big)  + (1,2 \leftrightarrow 3,4) \bigg\}\\
&- \frac{|\vk+\vkk|^3 \left(-k_{1}^2-4 k_{2}
   k_{1}-k_{2}^2+|\vk+\vkk|^2\right) \left(-k_{3}^2-4
   k_{4} k_{3}-k_{4}^2+|\vk+\vkk|^2\right)}{2
   \left(-k_{1}^2-2 k_{2}
   k_{1}-k_{2}^2+|\vk+\vkk|^2\right)^2 \left(-k_{3}^2-2
   k_{4} k_{3}-k_{4}^2+|\vk+\vkk|^2\right)^2}\Bigg],
\end{split}
\end{align}

The $S$ channel contribution for $R$ is denoted by $R^S(\vk, \vkk, \vkkk, \vkfour)$ and is given by 
\be
\label{defrs}
R^S(\vk,\vkk,\vkkk,\vkfour) =16 (2\pi)^3 \ \delta^3\big(\sum_{J=1}^4 {\vect{k}}_J\big) \bigg[ \prod_{I=1}^4 \phi({\vect{k}}_I)\bigg]  
\widehat{R}^S(\vk,\vkk,\vkkk,\vkfour),
\ee
where 
\begin{equation} \label{remexp}
\begin{split}
\widehat{R}^S(\vk,\vkk,\vkkk,\vkfour) = &{A_1(\vk,\vkk,\vkkk,\vkfour) \over (k_1+k_2+k_3+k_4)}  + {A_2(\vk,\vkk,\vkkk,\vkfour) \over (k_1+k_2+k_3+k_4)^2} \\ & +{A_3(\vk,\vkk,\vkkk,\vkfour) \over (k_1+k_2+k_3+k_4)^3}
\end{split}
\end{equation}
with
\begin{align} \nonumber
  A_1(\vk,\vkk,\vkkk,\vkfour) = & \bigg[ \frac{\vkkk \cdot \vkfour \left(\vk \cdot \vkk \left(k_1^2+k_2^2\right)+2 k_1^2 k_2^2\right)}{8 |\vk+\vkk|^2} + \{1,2 \Leftrightarrow 3,4\}\bigg] \\ \nonumber &-\frac{k_1^2 \vkk \cdot \vkkk k_4^2+k_1^2 \vkk \cdot \vkfour k_3^2+\vk \cdot \vkkk k_2^2 k_4^2+\vk \cdot \vkfour k_2^2 k_3^2}{2 |\vk+\vkk|^2} \\ &-\frac{\left(\vk \cdot \vkk \left(k_1^2+k_2^2\right)+2 k_1^2 k_2^2\right) \left(\vkkk \cdot \vkfour \left(k_3^2+k_4^2\right)+2 k_3^2 k_4^2\right)}{8 |\vk+\vkk|^4}, \label{A1inR}  
  \end{align}
  \begin{align} \nonumber
  A_2(&\vk,\vkk,\vkkk,\vkfour) =  -\frac{1}{8 |\vk+\vkk|^4}\bigg[k_3 k_4 (k_3+k_4) \left(\vk \cdot \vkk \left(k_1^2+k_2^2\right)+2 k_1^2 k_2^2\right)\\  \nonumber & (k_3 k_4+\vkkk \cdot \vkfour) \nonumber  +k_1 k_2 (k_1+k_2) (k_1 k_2+\vk \cdot \vkk) \left(\vkkk \cdot \vkfour \left(k_3^2+k_4^2\right)+2 k_3^2 k_4^2\right)\bigg] \\&  
  -\frac{1}{2 |\vk+\vkk|^2}\bigg[k_1^2 \vkk \cdot \vkkk k_4^2 (k_2+k_3)+k_1^2 \vkk \cdot \vkfour k_3^2 (k_2+k_4) \nonumber \\ &+\vk \cdot \vkkk k_2^2 k_4^2 (k_1+k_3)+\vk \cdot \vkfour k_2^2 k_3^2 (k_1+k_4)\bigg] \nonumber  \\ & + \bigg[\frac{\vk \cdot \vkk}{8 |\vk+\vkk|^2} \big((k_1+k_2) \left(\vkkk \cdot \vkfour \left(k_3^2+k_4^2\right)+2 k_3^2 k_4^2\right) \nonumber \\
  & + k_3 k_4 (k_3+k_4) (k_3 k_4+\vkkk \cdot \vkfour)\big)  + \{1,2 \Leftrightarrow 3,4\} \bigg],\label{A2inR}
  \end{align} 
  \begin{align} \nonumber
  A_3(\vk&,\vkk,\vkkk,\vkfour) =  -\frac{k_1 k_2 k_3 k_4 (k_1+k_2) (k_3+k_4) (k_1 k_2+\vk \cdot \vkk) (k_3 k_4+\vkkk \cdot \vkfour)}{4 |\vk+\vkk|^4}\\ \nonumber & 
 -\frac{k_1 k_2 k_3 k_4 (k_1 \vkk \cdot \vkkk k_4+k_1 \vkk \cdot \vkfour k_3+\vk \cdot \vkkk k_2 k_4+\vk \cdot \vkfour k_2 k_3)}{|\vk+\vkk|^2}\\ & + \frac{1}{4 |\vk+\vkk|^2} \bigg[k_1 k_2 (k_1 k_2+\vk \cdot \vkk) \left(\vkkk \cdot \vkfour \left(k_3^2+k_4^2\right)+2 k_3^2 k_4^2\right) \nonumber \\ &+\vk \cdot \vkk k_3 k_4 (k_1+k_2) (k_3+k_4) (k_3 k_4+\vkkk \cdot \vkfour)+ \{1,2 \Leftrightarrow 3,4\}\bigg] \nonumber \\ & +\frac{3 k_1 k_2 k_3 k_4 (k_1 k_2+\vk \cdot \vkk) (k_3 k_4+\vkkk \cdot \vkfour)}{4 |\vk+\vkk|^2}. \label{A3inR},
\end{align}

Details leading to these results are given in appendix \ref{detailsadsos}.
The  full answer for $S_{\text{on-shell}}^{\text{AdS}}$ is obtained by adding the contributions of the $S, T, U$ channels. 
This gives, from eq.\eqref{onshellact},  
\begin{align}
\label{fullanswerS}
\begin{split}
S^{\text{AdS}}_{\text{on-shell}} = {M_{\text{pl}}^2 R_{\text{\text{AdS}}}^2 \over 4}  &\bigg[{1 \over 2} \big\{{\widetilde W}^S(\vk, \vkk, \vkkk, \vkfour) +
 {\widetilde W}^S(\vk, \vkkk, \vkk, \vkfour)  +
 {\widetilde W}^S(\vk, \vkfour, \vkkk, \vkk) \big\}\\
 &+ R^S(\vk, \vkk, \vkkk, \vkfour) +R^S(\vk, \vkkk, \vkk, \vkfour)  +
 R^S(\vk, \vkfour, \vkkk, \vkk) \bigg]
 \end{split}
\end{align}
where ${\widetilde W}^S(\vk, \vkk, \vkkk, \vkfour)$ is given in eq.\eqref{defwts} and $ R^S(\vk, \vkk, \vkkk, \vkfour)$ is given in eq.\eqref{defrs}.

\subsection{Analytic Continuation to de Sitter Space}
\label{dsres}
As we explained in section \ref{pfwf}, the on-shell action $S_{\text{on-shell}}^{\text{AdS}}
$ obtained above can be analytically continued to the de Sitter space on-shell action $S_{\text{on-shell}}^{dS}$. So, the AdS correlator that we have computed above continues directly to the coefficient function in the wave function of the Universe at late times on de Sitter space. More precisely, the result of eq.\eqref{fullanswerS} is just the Fourier transform of the coefficient function we are interested in by
\be
\label{valopos}
\langle O({\vect{x}}_1) O({\vect{x}}_2) O({\vect{x}}_3) O({\vect{x}}_4)\rangle =\int \prod_{I=1}^4 {d^3 k_I \over (2\pi)^3}
 e^{i ({\vect{k}}_I\cdot x_I)} \langle O(\vk) O(\vkk) O(\vkkk) O(\vkfour)\rangle 
\ee
where 
\be
\label{fourptwo}
\begin{split}
\langle O(\vk) O(\vkk)& O(\vkkk) O(\vkfour)\rangle =-4(2 \pi)^3 \delta^3(\sum_{i=1}^3
{\vect{k}}_i) \bigg[ {1\over2}\bigg\{\widehat{W}^S(\vk, \vkk, \vkkk, \vkfour)
\\&+\widehat{W}^S(\vk, \vkkk, \vkk, \vkfour)+\widehat{W}^S(\vk, \vkfour, \vkkk, \vkk)\bigg\} +\widehat{R}^S(\vk, \vkk, \vkkk, \vkfour) \\
&+\widehat{R}^S(\vk, \vkkk, \vkk, \vkfour) +\widehat{R}^S(\vk, \vkfour, \vkkk, \vkk)\bigg]
\end{split}
\ee
where ${\widehat W}^S(\vk, \vkk, \vkkk, \vkfour)$ is given in eq.\eqref{hatW} and $ \widehat{R}^S(\vk, \vkk, \vkkk, \vkfour)$ is given in eq.\eqref{remexp}. 
As was mentioned in subsection \ref{basicaspects}, once the coefficient function $\langle O({\vect{x}}_1) O({\vect{x}}_2) O({\vect{x}}_3) O({\vect{x}}_4)\rangle $ is obtained in eq.\eqref{valopos}, eq.\eqref{fourptwo}, we now know all the relevant terms in the wave function in eq.\eqref{wf2}.

\section{The Four Point Scalar Correlator in de Sitter Space}
\label{finaldscalc}
With the wave function eq.\eqref{wf2}, in our hand we can proceed to calculate the scalar four point correlator $\langle \delta \phi({\vect{x}}_1) \delta \phi({\vect{x}}_2) \delta \phi({\vect{x}}_3) \delta \phi({\vect{x}}_4)\rangle$ which was defined in eq.\eqref{fourptfi}.  As was discussed in subsection \ref{fgaugefix} we need to fix the gauge more completely for this purpose. We will work below first in gauge 2, described in subsection \ref{gauge2} and 
then at the end of the calculation transform the answer to be in gauge 1, section \ref{gauge1}. 

In gauge 2 the metric perturbation
 $\gamma_{ij}$ is transverse and traceless. Working in this gauge we follow the strategy outlined in subsection \ref{fgaugefix} and first integrate out the metric perturbation to obtain a probability distribution
 $P[\delta \phi]$ defined in eq.\eqref{defP}.  The functional integral over $\gamma_{ij}$ is quadratic.
Integrating it out gives, in momentum space,  
\be
\label{intmetra}
\gamma_{ij}(k)=-{1\over2 k^3}\int {d^3 k_1 \over (2\pi)^3}{d^3 k_2 \over (2\pi)^3}
\delta\phi(\vk)\delta\phi(\vkk) \langle O(-\vk)O(-\vkk)T^{l m}({\vect{k}})\rangle \widehat{P}_{l m i j}({\vect{k}})
\ee
where 
\be
\label{defPhat}
\begin{split}
\widehat{P}_{i j k l}({\vect{k}}) &= \widetilde{\cal T}_{i k}({\vect{k}}) \widetilde{\cal T}_{j l }({\vect{k}})+\widetilde{\cal T}_{i l}({\vect{k}}) \widetilde{\cal T}_{j k}({\vect{k}})-\widetilde{\cal T}_{ij}({\vect{k}}) \widetilde{\cal T}_{kl}({\vect{k}}),\\
\text{with} \ \widetilde{\cal T}_{i k}({\vect{k}}) &= \delta_{i k} -{k_i k_k \over k^2}.
\end{split}
\ee
Eq.\eqref{intmetra}
 determines the $\gamma_{ij}(k)$ in terms of the scalar perturbation $\delta \phi(k)$. 
Substituting back then leads to the expression, 
\begin{align}
\label{valpphi}
 \begin{split}
  P[\delta \phi ({\vect{k}})] = & \exp \bigg[{M_{Pl}^2 \over H^2} \bigg(-\int {d^3k_1 \over (2\pi)^3} 
\ {d^3k_2 \over (2\pi)^3} \ \delta \phi(\vk) \delta \phi(\vkk) \langle O(-\vk) 
O(-\vkk)\rangle  \\ 
  &+ \int \prod_{J=1}^4 \bigg\{{d^3k_J \over (2\pi)^3}\delta \phi({\vect{k}}_J)\bigg\}
  \bigg\{ {1 \over 12} \langle O(-\vk) O(-\vkk)O(-\vkkk) O(-\vkfour)\rangle \\
&+ {1 \over 8} \langle O(-\vk) O(-\vkk) T_{ij}(\vk+\vkk)\rangle' \langle O(-\vkkk) O(-\vkfour) T_{kl}(\vkkk+\vkfour)\rangle'  \\ & (2 \pi)^3  \delta^3 \big(\sum_{J=1}^4 {\vect{k}}_J \big)
\widehat{P}_{i j k l}(\vk+\vkk) {1\over |\vk+\vkk|^3}
 \bigg\} \bigg) \bigg]
 \end{split}
\end{align}
with $\widehat{P}_{i j k l}(\vk+\vkk)$ being defined in eq.\eqref{defPhat} and the prime in $\langle O(\vk) O(\vkk) T_{ij}(\vkkk)\rangle'$ signifies that a factor of $(2 \pi)^3  \delta^3 \big(\sum_{l=1}^4 {\vect{k}}_l \big)$ is being stripped off from the unprimed $\langle O(\vk) O(\vkk) T_{ij}(\vkkk)\rangle$, i.e. 

\be
\langle O(\vk) O(\vkk) T_{ij}(\vkkk)\rangle = (2 \pi)^3  \delta^3 \big(\sum_{J=1}^4 {\vect{k}}_J \big)\langle O(\vk) O(\vkk) T_{ij}(\vkkk)\rangle' .
\ee

We see that  in the exponent on the RHS of eq.\eqref{valpphi} there are two  terms which are quartic in $\delta \phi$,
the first is proportional to the $\langle O(\vk)O(\vkk)O(\vkkk)O(\vkfour) \rangle$ coefficient function, and the second  is an extra term which arises in the process of integration out $\gamma_{ij}$ to obtain the probability distribution,
$P[\delta \phi]$.    

The four  function can now be calculated from $P[\delta \phi({\vect{k}})]$ using eq.\eqref{expcorr}.
The answer consists of two terms which come from the two quartic terms mentioned above respectively and are 
straightforward to compute. We will refer to these two contributions with the subscript $``CF"$ and $``ET"$ respectively below.   
The $\langle O(\vk)O(\vkk)O(\vkkk)O(\vkfour) \rangle$ coefficient function in eq.\eqref{valpphi} gives the contribution,
\begin{align}
\label{valCF}
\begin{split}
\langle \delta \phi(\vk) &\delta\phi(\vkk) \delta \phi(\vkkk) \delta\phi(\vkfour) \rangle_{CF}= - 8 (2 \pi)^3 \delta^3\big(\sum_{J=1}^4 {\vect{k}}_J\big) {H^6 \over M_{Pl}^6}{1 \over \prod_{J=1}^4 (2 k_J^3)} \\ & \bigg[ {1\over2}\bigg\{\widehat{W}^S(\vk, \vkk, \vkkk, \vkfour) +\widehat{W}^S(\vk, \vkkk, \vkk, \vkfour)+\widehat{W}^S(\vk, \vkfour, \vkkk, \vkk)\bigg\} \\&+\widehat{R}^S(\vk, \vkk, \vkkk, \vkfour) +\widehat{R}^S(\vk, \vkkk, \vkk, \vkfour) +\widehat{R}^S(\vk, \vkfour, \vkkk, \vkk)\bigg]
\end{split}
\end{align}
where ${\widehat W}^S(\vk, \vkk, \vkkk, \vkfour)$ is given in eq.\eqref{hatW} and $ \widehat{R}^S(\vk, \vkk, \vkkk, \vkfour)$ is given in eq.\eqref{remexp}.

While the ET term which arises due to integration out $\gamma_{ij}$  gives, 
\begin{align}
\label{valGE}
\begin{split}
\langle \delta \phi(\vk)&\delta\phi(\vkk) \delta \phi(\vkkk) \delta\phi(\vkfour) \rangle_{ET}= 4 (2 \pi)^3 \delta^3\big(\sum_{J=1}^4 {\vect{k}}_J\big) {H^6 \over M_{Pl}^6}{1 \over \prod_{J=1}^4 (2 k_J^3)} \\ & \bigg[\widehat{G}^S(\vk, \vkk, \vkkk, \vkfour) +\widehat{G}^S(\vk, \vkkk, \vkk, \vkfour) +\widehat{G}^S(\vk, \vkfour, \vkkk, \vkk)\bigg]
\end{split}
\end{align}
where 
\be
\label{defGhat} 
\begin{split}
&\widehat{G}^S(\vk, \vkk, \vkkk, \vkfour) = {S(\widetilde{{\vect{k}}},\vk, \vkk)S(\widetilde{{\vect{k}}},\vkkk, \vkfour)  \over |\vk+\vkk|^3}\bigg[ \bigg\{\vk.\vkkk+\frac{\{(\vkk+\vk).\vk\} \{(\vkfour+\vkkk).\vkkk\}}{|\vk+\vkk|^2}\bigg\}\\ & \bigg\{\vkk.\vkfour+  \frac{\{(\vk+\vkk).\vkk\} \{(\vkkk+\vkfour).\vkfour\}}{|\vk+\vkk|^2}\bigg\} 
+  \bigg\{\vk.\vkfour+\frac{\{(\vkk+\vk).\vk\} \{(\vkfour+\vkkk).\vkfour\}}{|\vk+\vkk|^2}\bigg\}\\ & 
\bigg\{\vkk.\vkkk+\frac{\{(\vkk+\vk).\vkk\} \{(\vkfour+\vkkk).\vkkk\}} {|\vk+\vkk|^2}\bigg\}-
\bigg\{\vk.\vkk-\frac{\{(\vkk+\vk).\vk\} \{(\vk+\vkk).\vkk\}} {|\vk+\vkk|^2}\bigg\}\\ & 
\bigg\{\vkkk.\vkfour-\frac{\{(\vkkk+\vkfour).\vkfour\} \{(\vkfour+\vkkk).\vkkk\}}{|\vk+\vkk|^2}\bigg\} \bigg]
\end{split}
\ee with
\be
\label{singhat}
S( \widetilde{{\vect{k}}},\vk, \vkk) = (k_1+k_2+k_3) - {\sum_{i>j}k_i k_j \over (k_1+k_2+k_3)} - {k_1k_2k_3\over (k_1+k_2+k_3)^2} \Bigg|_{\widetilde{{\vect{k}}}=-(\vk+\vkk)}.
\ee

The full answer for the four point correlator in gauge 2 is then given by combining eq.\eqref{valCF} and eq.\eqref{valGE},
\be
\label{finalfourpta}
\begin{split}
\langle \delta \phi(\vk)\delta\phi(\vkk) \delta \phi(\vkkk) \delta\phi(\vkfour) \rangle=&\langle \delta \phi(\vk)\delta\phi(\vkk) \delta \phi(\vkkk) \delta\phi(\vkfour) \rangle_{CF} \\ &+\langle \delta \phi(\vk)\delta\phi(\vkk) \delta \phi(\vkkk) \delta\phi(\vkfour) \rangle_{ET}.
\end{split}
\ee
Let us end this subsection with one comment. We see from eq.\eqref{valpphi} that the  ET contribution is determined by 
the $\langle O O T_{ij} \rangle$ correlator. As discussed in \cite{MRT} this correlator is completely fixed by conformal invariance, so we see 
conformal symmetry completely fixes the   ET  contribution to the scalar $4$ point correlator.
\subsection{Final Result for the Scalar Four Point Function}
\label{finres4pt}
We can now convert the result to  gauge 1 defined in  \ref{gauge1} using the relation \footnote{
The relation in eq.\eqref{infper2} has  corrections in involving higher powers of $\zeta$
 which could lead to additional contributions in eq.\eqref{finalres} that  arise from the two point correlator of  $\delta \phi$. However these corrections are further suppressed in the slow-roll parameters as discussed in Appendix \ref{addterm}.} in eq.\eqref{infper2}.
 %(**XX**).} 
 
This gives 
\begin{align}
\label{finalres}
\begin{split}
 \langle \zeta(\vk) &\zeta(\vkk) \zeta(\vkkk) \zeta(\vkfour) \rangle  = (2 \pi)^3 \delta^3\big(\sum_{J=1}^4 {\vect{k}}_J\big) {H^6 \over M_{Pl}^6 \epsilon^2}{1 \over \prod_{J=1}^4 (2 k_J^3)} \\ & 
 \bigg[  \widehat{G}^S(\vk, \vkk, \vkkk, \vkfour) +\widehat{G}^S(\vk, \vkkk, \vkk, \vkfour)  +\widehat{G}^S(\vk, \vkfour, \vkkk, \vkk)\\ &
 - \widehat{W}^S(\vk, \vkk, \vkkk, \vkfour) -\widehat{W}^S(\vk, \vkkk, \vkk, \vkfour)-\widehat{W}^S(\vk, \vkfour, \vkkk, \vkk) \\&
 -2\bigg\{\widehat{R}^S(\vk, \vkk, \vkkk, \vkfour) +\widehat{R}^S(\vk, \vkkk, \vkk, \vkfour) +\widehat{R}^S(\vk, \vkfour, \vkkk, \vkk) \bigg\} \bigg]
 \end{split}
\end{align}
where ${\widehat W}^S(\vk, \vkk, \vkkk, \vkfour)$ is given in eq.\eqref{hatW}, $ \widehat{R}^S(\vk, \vkk, \vkkk, \vkfour)$ is given in eq.\eqref{remexp} and $\widehat{G}^S(\vk, \vkk, \vkkk, \vkfour)$ is given in eq.\eqref{defGhat}.

Eq.\eqref{finalres} is one of the main results of this paper.
The variables ${\vect{k}}_i,i=1,2,3,4$ refer
 to the spatial momenta carried by the perturbations,
The scalar perturbation $\zeta$  is defined in eq.\eqref{expgamma}, see also subsection \ref{gauge1}, 
  with $\zeta ({\vect{k}})$ being related to  $\zeta({\vect{x}})$ by a relation analogous to eq. \eqref{ftrel}. 
$H$ is the Hubble constant during inflation defined in eq.\eqref{scaleds},
eq.\eqref{relhv}.  Our 
conventions for $M_{Pl}$ are  given in eq.\eqref{action1},  eq. \eqref{defmpl}. And
 the slow-roll parameter $\epsilon$ is defined in 
eq.\eqref{defeps}, eq.\eqref{releps3}. 

The result, eq.\eqref{finalres}, was derived in the leading slow-roll approximation.
One way to incorporate corrections is to take $H$ in eq.\eqref{finalres} to be the Hubble parameter when the modes cross the horizon, at least for situations where all momenta,  ${\vect{k}}_i, i=1,\cdots 4$, are comparable in magnitude. Additional corrections 
 which depend on the slow-roll parameters,  $\epsilon, \eta$, eq.\eqref{defeps}, eq.\eqref{defeta}, will also arise. 

\paragraph{\bf Comparison with previous results \\}
Our result, eq.\eqref{finalres}, agrees with that obtained in \cite{Seery:2008ax}.
The result in \cite{Seery:2008ax} consists of two terms,  the CI term and the ET term, see eq.(4.7).
The CI term agrees with the $\hat{R}^S$ terms  in eq.\eqref{finalres} which arise 
due    the longitudinal graviton propagator, eq.\eqref{defrs}. The $\hat{W}$ terms in eq.\eqref{finalres} arise 
from the transverse graviton propagator, eq.\eqref{wittendiag}, while the $G^S$ terms arise from the extra contribution 
due to   integrating out the metric perturbation, eq.\eqref{defGhat}; these two together agree with the ET term in \cite{Seery:2008ax}. This comparison is carried out in a Mathematica file that is included with the source of the arXiv version of this paper \cite{mathematicaallchecks}.

\section{Tests of the Result and Behavior in Some Limits}
\label{Testsofresult}
We now turn to carrying out some tests on our result  for the four point function and examining its behavior 
in some limits. 
We will first verify that the result is consistent with the conformal invariance
of de Sitter space in subsection \ref{testconf}, and then  examine its behavior 
in various limits in subsection \ref{finreslimit} and show that this agrees with expectations. 

\subsection{Conformal Invariance}
\label{testconf}
Our calculation for the $4$ point function was carried out at leading order in the slow-roll approximation,
where the action governing the perturbations is that of a free scalar in de Sitter space, eq.\eqref{eqff}. 
Therefore the result must be consistent with the full symmetry group of $dS_4$ space which is $SO(1,4)$ as was discussed in section \ref{prelim}. 
In fact the wave function, eq.\eqref{wf2}, in this approximation  itself should be invariant under this symmetry,
as was discussed in section \ref{conformalinv}, see also, 
 \cite{Maldacena:2002vr}, \cite{Maldacena:2011nz} and \cite{MRT}. 

For the four point function we are discussing, as was discussed towards the end of section eq.\eqref{conformalinv} after 
eq.\eqref{opttr}, given the checks in the literature already in place only one remaining check needs to be carried out 
to establish the  invariance of all relevant terms in the wave function. This is to check the invariance of the 
 $\langle O(\vk) O(\vkk)O(\vkkk)O(\vkfour)\rangle$ coefficient defined in eq.\eqref{valopos} and eq.\eqref{fourptwo}. Conformal invariance of the coefficient function $\langle O(\vk) O(\vkk)O(\vkkk)O(\vkfour)\rangle$ gives rise to the equation
\begin{align}
\label{eqwi}
\begin{split}
& \langle\delta O(\vk) O(\vkk)O(\vkkk)O(\vkfour)\rangle +  \langle O(\vk) \delta O(\vkk)O(\vkkk) O(\vkfour)\rangle \\ & + \langle O(\vk) O(\vkk) \delta O(\vkkk) O(\vkfour)\rangle + \langle O(\vk) O(\vkk) O(\vkkk) \delta O(\vkfour)\rangle =0
\end{split}
\end{align}
where $\delta O({\vect{k}}) $ is given in Appendix \ref{appsubtrans}, eq.\eqref{osctmom}, and depends on ${\vect{b}}$ which is the parameter specifying the conformal transformation.

The coefficient function $\langle O(\vk) O(\vkk)O(\vkkk)O(\vkfour)\rangle$ contains an overall delta function which enforces momentum 
conservation, eq.\eqref{fourptwo}. 
As was argued in \cite{Maldacena:2011nz} all terms in eq.\eqref{eqwi} where the derivatives act on this
 delta function sum to zero, so the effect of the derivative operators acting on it  can be neglected.
We can also use  rotational invariance so take ${\vect{b}}$ to point along the $x^1$ direction. 
Our answer for $\langle O(\vk) O(\vkk)O(\vkkk)O(\vkfour)\rangle$ is given in eq.\eqref{fourptwo}. 
The complicated nature of the answer  makes it very difficult to check analytically whether  eq.\eqref{eqwi} is met. 
However, it is quite simple to check this numerically.
Using Mathematica \cite{mathematicaallchecks}, one finds that the LHS of eq.\eqref{eqwi} does indeed vanish with the four point function given in 
eq.\eqref{fourptwo}, there by showing that our result for $\langle O(\vk) O(\vkk)O(\vkkk)O(\vkfour)\rangle$ 
 does meet the requirement of conformal invariance. 
This then establishes that all terms in the wave function relevant  for the four point function calculation are invariant under
 conformal symmetry. 

A further subtlety having to do with gauge fixing, 
 arises in discussing the conformal invariance of correlation functions, as opposed to the wave function,
as discussed in section \ref{fgaugefix}. The relevant terms in the probability distribution $P[\delta \phi]$ were obtained in eq.\eqref{valpphi}. 
 The scalar four point function would be invariant if  $P[\delta \phi]$ 
 is invariant under the combined conformal transformation, eq.\eqref{cftrans1}, eq.\eqref{cftrans2} and compensating coordinate transformation, 
eq.\eqref{furthct}. From eq.\eqref{intmetra} and eq.\eqref{valcompv} we see that the compensating coordinate transformation in this case is 
specified by  
\be
\label{comcoor4pt}
v_i ({\vect{k}}) = -{3 b^k  \over k^5}\int {d^3 k_1 \over (2\pi)^3}{d^3 k_2 \over (2\pi)^3}
\delta\phi(\vk)\delta\phi(\vkk) \langle O(-\vk)O(-\vkk)T^{l m}({\vect{k}})\rangle \widehat{P}_{l m k i}({\vect{k}})
\ee
with $\widehat{P}_{l m k i}({\vect{k}})$ being given in eq.\eqref{defPhat}.

Thus the total change in $\delta \phi$, given in eq.\eqref{comdphi2} becomes, 
\be
\label{extchp}
\begin{split}
\delta (\delta \phi({\vect{k}}))= \delta^C (\delta \phi({\vect{k}}))+\delta^R (\delta \phi({\vect{k}}))
\end{split}
\ee
where $\delta^C (\delta \phi({\vect{k}}))$ is given in eq.\eqref{delphsctmom} of Appendix \ref{appsubtrans}. Using eq.\eqref{intmetra} in eq.\eqref{delphrmom}, $\delta^R (\delta \phi({\vect{k}}))$ becomes,
\begin{align}
\label{delRphin}
 \begin{split}
 \delta^R (\delta \phi({\vect{k}})) =& -3i k^i b^k  \int{d^3\vect{k_1} \over (2 \pi)^3} {d^3 \vect{k_2} \over (2 \pi)^3} {d^3 \vect{k_3} \over (2 \pi)^3}{\widehat{P}_{l m k i} ( {\vect{k}}-\vkk) \over |{\vect{k}}-\vkk|^5} \ \\ & \delta \phi( \vk) \delta \phi( \vkk) \delta\phi(\vkkk) \langle O(- \vk) O( -\vkk)T_{l m}(-{\vect{k}}+\vkk)\rangle.
 \end{split}
\end{align}
Note in particular that $v^i$ given in eq.\eqref{comcoor4pt} is quadratic in the scalar perturbation and as a result 
$\delta^R(\delta \phi)$ in eq.\eqref{delRphin} is cubic in $\delta \phi$.

The probability distribution, $P[\delta\phi]$ upto quartic order in $\delta \phi$
 is given in eq.\eqref{valpphi} and consists of 
quadratic terms and quartic terms.   In Appendix \ref{confinvprob} we show that, upto quartic order,  
\be
\label{condpphi}
P[\delta  \phi({\vect{k}})]=P[\delta \phi({\vect{k}}) + \delta \big(\delta \phi({\vect{k}})\big)]
\ee
where $\delta \big(\delta \phi({\vect{k}})\big)$ is given in eq.\eqref{extchp}, so that 
$P[\delta\phi]$ is invariant under $\delta(\delta\phi)$ upto this order. This invariance arises as follows.
The quadratic term in $P[\delta \phi]$ gives rise to a contribution going like $\delta \phi^4$,
since  $\delta^R(\delta \phi)$ is cubic in $\delta \phi$. This is canceled by a contribution coming from the quartic term due to $\delta^C(\delta \phi)$ which is linear in $\delta \phi$. 
Eq.\eqref{condpphi} establishes that the probability distribution function and thus also the four point scalar correlator calculated from it 
are  conformally invariant. 

\subsection{Flat Space Limit \label{subsecflatspace}}
We now describe another strong check on our computation of the four point correlator:  its ``flat space limit.'' This is the statement that, in a particular limit, this correlation function reduces to the flat space scattering amplitude of four minimally coupled scalars in four dimensions! More precisely, we use the flat space limit developed in \cite{Raju:2012zr,Raju:2012zs}, which involves
an analytic continuation of the momenta. In our context, the limit reads
\be
\lim_{\norm{k_1} + \norm{k_2} + \norm{k_3} +  \norm{k_4} \rightarrow 0} {\left(\norm{k_1} + \norm{k_2} + \norm{k_3} +  \norm{k_4} \right)^3\over \norm{k_1} \norm{k_2} \norm{k_3} \norm{k_4}}  \langle O(\vect{k_1})O(\vect{k_2}) O(\vect{k_3}) O(\vect{k_4}) \rangle'   = {\cal N} S_4(\vect{\tilde{k}_1}, \vect{\tilde{k}_2}, \vect{\tilde{k}_3}, \vect{\tilde{k}_4})
\ee
where $S_4$ is the four-point scattering amplitude of scalars minimally coupled to gravity, and the on-shell four-momenta of these scalars are related to the three-momenta of the correlators by $\vect{\tilde{k}_n} = (i \norm{k_n}, \vect{k_n})$. This means that we add an additional planar direction to the three boundary directions, and use $i \norm{k_n}$ for the this component of the four-dimensional momentum (which we call the $z$-component) and $\vect{k_n}$ for the other three components.  An intuitive way to understand this limit is as follows. The four-point flat space scattering amplitude conserves four-momentum, whereas the CFT correlator conserves
just three-momentum. The point where $\sum \norm{k_n} = 0$ corresponds to  the point in kinematic space, where the four-momentum is conserved. The claim is that the CFT correlator has a pole at this point, and the residue is just the flat space scattering amplitude.

Note that, consistent with our conventions, the prime  on the correlator indicates that a factor involving the delta function has been stripped off. Similarly, on the right hand side, $S_4$ is the flat-space scattering amplitude {\em without} the momentum conserving delta function. Here, ${\cal N}$ is an unimportant numerical factor (independent of all the momenta) that we will not keep track of, which just depends on the conventions we use to normalize the correlator and the amplitude.

\begin{figure}[!h]
\begin{center}
\includegraphics[height=0.3\textwidth]{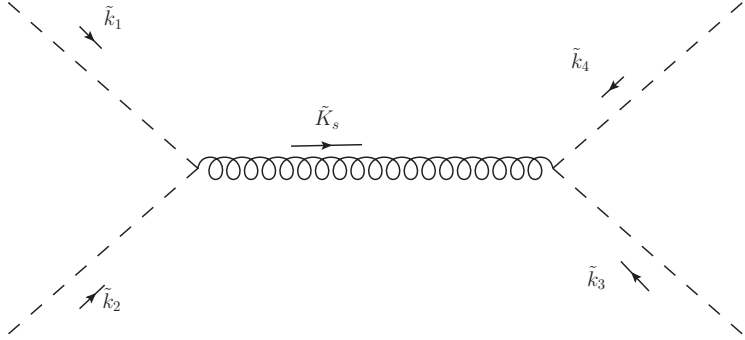}
\caption{S-channel Feynman diagram for scattering of minimally coupled scalars \label{flatspacefeynman}}
\end{center}
\end{figure}

Before we show this limit, let us briefly describe the flat space scattering amplitude of four minimally coupled scalars. In fact, the 
Feynman diagrams that contribute to this amplitude are very similar to the Witten diagrams of fig. \ref{threewittendiag}. The s-channel
diagram is shown in fig. \ref{flatspacefeynman}, and of course, the $t$ and $u$-channel diagrams also contribute to the amplitude.

So, the flat-space amplitude using the diagram in Fig. \ref{flatspacefeynman} evaluates to
\be
\label{flatspaceamp}
S_4 = T_{i j}^{\text{flat}}(\vect{\tilde{k}_1}, \vect{\tilde{k}_2}) G^{{\text{flat-grav}}}_{i j  k l} (\vect{\tilde{K}_s}) T_{k l}^{\text{flat}}(\vect{\tilde{k}_3}, \vect{\tilde{k}_4}) + \text{t+u-channels}
\ee
where $\vect{\tilde{K}_s} = \vect{\tilde{k}_1} + \vect{\tilde{k}_2}$ and 
\be
\label{flatspacecomponents}
\begin{split}
&T_{i, j}^{\text{flat}}(\vect{\tilde{k}_1}, \vect{\tilde{k}_2}) = {\tilde{k}_1}^i {\tilde{k}_2}^j - {1 \over 2} (\vect{\tilde{k}_1} \cdot \vect{\tilde{k}_2}) \delta^{i j},  \\
& G^{{\text{flat-grav}}}_{i j  k l} (\vect{\tilde{K}_s}) =   
{1 \over \vect{\tilde{K}_s}^2 -  i \epsilon}  {1 \over 2} \left[\left({\cal T}^{\text{flat}}_{i k} {\cal T}^{\text{flat}}_{j l} + {\cal T}^{\text{flat}}_{i l} {\cal T}^{\text{flat}}_{j k} - {\cal T}^{\text{flat}}_{i j} {\cal T}^{\text{flat}}_{k l} \right)\right], \\
&{\cal T}^{\text{flat}}_{i j} = \delta_{i j} + {\tilde{K_s}_i \tilde{K_s}_j \over (\tilde{K_s}^0)^2}
\end{split}
\ee
and we emphasize that in \eqref{flatspaceamp} the Latin indices are contracted only over the spatial directions. In \eqref{flatspacecomponents}, $\tilde{K_s}^0$ indicates the $z$-component of the exchanged momentum. 
The reader should compare this formula to our starting formula for the four-point correlator in \eqref{wittendiagorig}. In fact, if we consider the pole in the $p$-integral at $p = i (\norm{k}_1 + \norm{k}_2)$, in this integral, we see that we will get the right flat space limit. However, in our calculation it was convenient to divide the expression in \eqref{wittendiagorig} into two parts: a transverse graviton
contribution given in \eqref{wittendiag}, and a longitudinal graviton contribution that we wrote as a remainder comprised of three terms in \eqref{valR}. 

We can make the same split for the flat-space amplitude, by replacing the graviton propagator with a simplified version $\widetilde{G}^{\text{flat-grav}}_{i j k l}$, by replacing each ${\cal T}^{\text{flat}}_{i j}$ with
\be
\widetilde{\cal T}^{\text{flat}}_{i j} = \delta_{i j} - {\tilde{K_s}_i \tilde{K_s}_j \over (\vect{K_s})^2}.
\ee
and then writing
\be
S_4 = \tilde{S}_4 + 2 R^{\text{flat}}
\ee
where $\tilde{S}_4$ is given by an expression analogous to \eqref{flatspaceamp} 
\be
\tilde{S}_4 = T_{i j}^{\text{flat}}(\vect{\tilde{k}_1}, \vect{\tilde{k}_2}) \tilde{G}^{{\text{flat-grav}}}_{i j  k l} (\vect{\tilde{K}_s}) T_{k l}^{\text{flat}}(\vect{\tilde{k}_3}, \vect{\tilde{k}_4}) + \text{t+u-channels} = \tilde{k}_{1 i} \tilde{k}_{2 j} \widehat{P}^{i j k l}(\vect{\tilde{K}_s}) \tilde{k}_{3 k} \tilde{k}_{4 l}.
\ee
The second equality follows because projector in this simplified propagator now projects onto traceless tensors that are transverse to $\tilde{K}_s$, which we have denoted by $\widehat{P}^{i j k l}(\vect{\tilde{K}_s})$, as in \eqref{defPhat}.

Now, consider the expression in \eqref{hatW}. We see that the {\em third order} pole at $k_t = (\norm{k_1} + \norm{k_2} + \norm{k_3} + \norm{k_4})$ is given by
\be
\begin{split}
\widehat{W}^S &= {k_{1 i} k_{2 j} k_{3 k} k_{4 l}  P^{i j k l} \over k_t^3} \left[ \bigg\{\frac{ k_{1} k_{2} (k_{1}+k_{2})^2 \left((k_{1}+k_{2})^2-k_{3}^2-k_{4}^2-4 k_{3} k_{4}\right)}{ (k_{1}+k_{2}-k_{3}-k_{4})^2 ((k_{1}+k_{2})^2-|\vect{k_1}+\vect{k_2}|^2)} \right] + \Or[{1 \over k_t^2}] \\ 
&= {1 \over k_t^3} {k_1 k_2 k_3 k_4 \over (k_{1}+k_{2})^2-|\vect{k_1}+\vect{k_2}|^2}{k_{1 i} k_{2 j} k_{3 k} k_{4 l}  P^{i j k l}} + \Or[{1 \over k_t^2}] \\
&= {k_1 k_2 k_3 k_4 \over k_t^3} \tilde{S}_4  + \Or[{1 \over k_t^2}].
\end{split}
\ee
All that remains to establish the flat space limit is to show that
\be
\lim_{\norm{k_1} + \norm{k_2} + \norm{k_3} +  \norm{k_4} \rightarrow 0} {\left(\norm{k_1} + \norm{k_2} + \norm{k_3} +  \norm{k_4} \right)^3\over \norm{k_1} \norm{k_2} \norm{k_3} \norm{k_4}} \widehat{R}^S = R^{\text{flat}}.
\ee
The algebra leading to this is a little more involved but as we verify in the included Mathematica file \cite{mathematicaallchecks} this also remarkably
turns out to be true.

So, we find that our complicated answer for the four-point scalar correlator reduces in this limit precisely to the flat space scalar amplitude. This is a very non-trivial check on our answer.

\subsection{Other Limits}
\label{finreslimit}
In this subsection we consider our result for the four point function for various other limiting values 
of the external momenta. 
The first limit we consider in subsection \ref{label1}  is when one of the four momenta goes to zero. Without any loss of generality we can take this to be $\vkfour \rightarrow 0$. The second limit we consider in subsection \ref{limit2}
 is when 
two of the momenta, say $\vk, \vkk$ get large and nearly opposite to each other,
while the others,  $\vkkk, \vkfour$, stay fixed, so that,  
 $\vk\simeq-\vkk$, and $ k_1,k_2 \gg k_3,k_4$. In both cases we will see that the result agrees with what is 
expected from general considerations. Finally, in subsection \ref{limit3ccl} we consider the
counter-collinear  limit 
where the sum of two momenta vanish, say $|\vk+\vkk| \rightarrow 0$.
In this limit we find that the result has a characteristic third order
pole singularity resulting  in a characteristic divergence, as noted in \cite{Seery:2008ax}. 

\subsubsection{Limit I : $\vkfour\rightarrow 0$}
\label{label1}
This kind of a limit was first considered in \cite{Maldacena:2002vr} and is now referred to as a squeezed limit in the literature. 
It is convenient to think about the behavior of the four point function in this limit 
by analyzing what happens to the wave function eq.\eqref{wf2}. For purposes of the present discussion we can write the wave function, 
eq.\eqref{wf2},  as, 
\begin{align}
\label{wflim1}
\begin{split}
\psi[\delta \phi({\vect{k}})]=\exp\bigg[{M_{Pl}^2 \over H^2} \bigg(&-{1\over 2} \int {d^3 \vk \over (2 \pi)^3} {d^3 \vkk \over (2 \pi)^3} \delta \phi( \vk) 
\delta \phi( \vkk) \langle O(- \vk) O( -\vkk)\rangle  \\
&+{1\over 3!}\int {d^3 \vk \over (2 \pi)^3} {d^3 \vkk \over (2 \pi)^3} {d^3 \vkkk \over (2 \pi)^3} \delta\phi(\vk)\delta\phi(\vkk) \delta\phi(\vkkk) \\
& \langle O(-\vk) O(-\vkk) O(-\vkkk) \rangle \\
&+ {1\over 4!}\int {d^3 \vk \over (2 \pi)^3} {d^3 \vkk \over (2 \pi)^3} {d^3 \vkkk \over (2 \pi)^3} {d^3 \vkfour \over (2 \pi)^3}  \delta\phi(\vk)\delta\phi(\vkk) \delta\phi(\vkkk)\delta \phi(\vkfour) \\
& \langle O(-\vk) O(-\vkk) O(-\vkkk) O(-\vkfour) \rangle + \cdots \bigg)\bigg].
\end{split}
\end{align}
Here terms dependent on the tensor perturbations have not been shown explicitly since they are 
not relevant for the present discussion. We have also explicitly  included a three point function on the RHS.
This three point function vanishes in the slow-roll limit, \cite{Maldacena:2002vr}, but it is  important to 
 include it for  the general argument we give now, since the slow -roll limit for this general argument is a bit subtle. 

In the limit when $\vkfour\rightarrow 0$, $\delta\phi(\vkfour)$ becomes approximately constant 
and  its effect is to rescale the metric by taking $h_{ij}$,  eq.\eqref{defgij}, 
to be, eq.\eqref{defgij2}, eq.\eqref{expgamma}, 
\be
\label{regfv}
h_{ij}=e^{2Ht}(1+2 \zeta)\delta_{ij}
\ee
where 
$\zeta(\vkfour) $ is related to $\delta\phi(\vkfour)$ by, eq.\eqref{infper2},
\be
\label{relddp}
\delta \phi(\vkfour) =-\sqrt{2 \epsilon}\zeta(\vkfour).
\ee

The effect on the  wave function in this limit can   be obtained by first considering the wave 
function in the absence of the $\delta \phi(\vkfour)$ perturbation and then  rescaling the 
momenta to incorporate the dependence on $\delta \phi(\vkfour)$. 
The coefficient of the  term which is trilinear in $\delta \phi$ on the RHS of eq.\eqref{wflim1}
 is denoted by 
$\langle O(\vk O(\vkk) O(\vkkk) \rangle$. Under the rescaling which incorporates the effects of $\delta \phi(\vkfour)$
this coefficient will change as follows:
\be
\label{del3pt}
\delta  \langle O(\vk)O(\vkk)O(\vkkk) \rangle  \sim    {\delta \phi(\vkfour) \over \sqrt{\epsilon}} 
\bigg(\sum_{i=1}^3 {\vect{k}}_i \cdot \partial_{{\vect{k}}_i}\bigg) \langle O(\vk) O(\vkk) O(\vkkk) \rangle.
\ee
As a result the trilinear term now depends on four powers of $\delta \phi$ and gives a contribution to the four point function. 
We see that the resulting value of the coefficient of the term quartic in $\delta \phi$ is therefore 
\be
\label{quartlim}
\lim_{\vkfour\rightarrow 0} \langle O(\vk) O(\vkk) O(\vkkk) O(\vkfour)\rangle \sim {1\over \sqrt{\epsilon}} \bigg(\sum_{i=1}^3 {\vect{k}}_i \cdot \partial_{{\vect{k}}_i}\bigg) \langle O(\vk) O(\vkk) O(\vkkk) \rangle.
\ee

The three point function in this slow-roll  model of inflation  was calculated in \cite{Maldacena:2002vr} and the result is given in eq.(4.5),  eq.(4.6) of \cite{Maldacena:2002vr}. From this result it is easy to read-off the value of 
$\langle O(\vk) O(\vkk) O(\vkkk) \rangle $. One gets that 
\be
\label{tptmalda}
\langle O(\vk) O(\vkk) O(\vkkk) \rangle \propto (2\pi)^3 \delta(\sum_i k_i) {1\over \sqrt{\epsilon}} A
\ee
where 
\be
\label{valastar}
A= 2 {\ddot{\phi}_* \over \dot{\phi}_* H} \sum_{i=1}^3 k_i^3+ {\dot{\phi}_*^2 \over H^2} \bigg[ {1 \over2}\sum_{i=1}^3 k_i^3 +  {1 \over2} \sum_{i \neq j} k_i k_j^2 + 4 {\sum_{i > j} k_i^2 k_j^2 \over k_t}\bigg]
\ee
where the subscript $*$ means values of the corresponding object to be evaluated at the time of horizon crossing and $k_t = k_1+k_2+k_3$.
4
Substituting in eq.\eqref{quartlim} gives 
\be
\label{qlim2}
\lim_{\vkfour \to 0} \langle O(\vk)O(\vkk)O(\vkkk) O(\vkfour)\rangle \sim {1\over \epsilon} \bigg[\sum_{i=1}^3 {\vect{k}}_i \cdot \partial_{{\vect{k}}_i}\bigg]\bigg[\delta(\sum_i k_i) A\bigg].
\ee
Now, it is easy to see that due to the $\dot{\phi}_*, \ddot{\phi}_*$ dependent prefactors 
 $A$ is of order the slow-roll parameters $\epsilon, \eta$, eq.\eqref{defeta2}, eq.\eqref{releps3}. 
Thus in the limit where $\epsilon \sim \eta$ and both tend to zero the ${1\over \epsilon}$ prefactor
on the RHS will cancel the dependence in $A$ due to the prefactors. 
However, note that there is an additional suppression since $A$ is trilinear in the momenta 
and  therefore $\delta (\sum_i k) A$  will be invariant under rescaling all the momenta
to leading order in the slow-roll approximation,  
\be
\label{condlima}
\bigg[\sum_{i=1}^3 {\vect{k}}_i \cdot \partial_{{\vect{k}}_i}\bigg]\bigg[\delta(\sum_i k_i) A\bigg]=0.
\ee
As a result the RHS of eq.\eqref{qlim2} and thus the four point function will vanish in this limit 
in the leading slow-roll approximation. 
To subleading order in the slow-roll approximation the condition in eq.\eqref{condlima}
 will not be true any more since
the Hubble constant $H$ and $\dot{\phi}_*, \ddot{\phi}_*$  which appears in eq.\eqref{valastar} will also 
depend on $k$ and should be evaluated  to take the values they do when the modes cross the horizon. 

For our purpose it is enough to note that the behavior of the four point function in the leading
 slow-roll approximation is that it vanishes when $\vkfour \rightarrow 0$. 
It is easy to see that the result in eq.\eqref{fourptwo}
 does have this feature in agreement with the general 
analysis above. In fact expanding eq.\eqref{fourptwo}
 for small momentum we find it vanishes linearly with $k_4$.

\subsubsection{Limit II : $k_1,k_2$ get large }
\label{limit2}
Next, we consider a limit where two of the momenta, say $\vk, \vkk$, get large in magnitude 
and approximately cancel, so that their sum, $|\vk+ \vkk|$,  is held fixed.   The other two momenta, 
$\vkkk, \vkfour$, 
are held fixed in this limit.  
Note that this limit is a very natural one from the point of view of the CFT.
In position space in the CFT in this limit two operators come together, at the same spatial location
 and the behavior can be understood using the operator product expansion (OPE). 
We will see below that 
 our result for the four point function 
 reproduces the behavior expected from the OPE  in the CFT. 

It is convenient to start the analysis first from the CFT point of view and then compare with the 
four point function result we have obtained \footnote{One expects to justify the OPE from the bulk itself,
but we will not try to present a careful derivation along those lines here.}.
Consider the four point function of an operator $O$ of dimension $3$ in a CFT:
\be
\label{fptcft}
\langle O({\vect{x}}_1)O({\vect{x}}_2)O({\vect{x}}_3)O({\vect{x}}_4) \rangle.
\ee
The  momentum space correlator is 
\begin{align}
\label{ftlim2}
\begin{split}
\langle O(\vk)O(\vkk)O(\vkkk)O(\vkfour) \rangle &= \int  d^3{\vect{x}}_1 d^3{\vect{x}}_2 d^3{\vect{x}}_3 d^3{\vect{x}}_4 \ e^{-i(\vk \cdot {\vect{x}}_1+\vkk \cdot {\vect{x}}_2+\vkkk \cdot {\vect{x}}_3+\vkfour \cdot {\vect{x}}_4)} \\ & \langle O({\vect{x}}_1)O({\vect{x}}_2)O({\vect{x}}_3)O({\vect{x}}_4) \rangle  
\\ &=(2\pi)^3\delta^{(3)}(\sum_I{\vect{k_I}})\int d^3{\vect{y}}_2 d^3{\vect{y}}_3 d^3{\vect{y}}_4 \ 
e^{-i(\vkk \cdot {\vect{y}}_2+\vkkk \cdot {\vect{y}}_3+\vkfour \cdot {\vect{y}}_4)} \\
& \langle O({\vect{0}}) O({\vect{y}}_2)O({\vect{y}}_3)O({\vect{y}}_4) \rangle
\end{split}
\end{align}
where in the last line on the RHS above
\be
\label{defy}
{\vect{y}}_2={\vect{x}}_2-{\vect{x}}_1, \ {\vect{y}}_3= {\vect{x}}_3-{\vect{x}}_1, \ {\vect{y}}_4= {\vect{x}}_4-{\vect{x}}_1.
\ee

We are interested in the limit where  $\vkkk,\vkfour$ are held fixed while 
\be
\label{lim2a}
k_2 \rightarrow \infty \ \text{and} \ k_1=|-(\vkk+\vkkk+\vkfour) | \rightarrow \infty
\ee 
In position space in this limit ${\vect{x}}_1\rightarrow {\vect{x}}_2$ so that 
\be
\label{lim2b}
{\vect{y}}_2 \rightarrow 0.
\ee
The operator product expansion can be used when the condition in eq.\eqref{lim2b} is met, to expand
\be
\label{opea}
O(\vect{0})O({\vect{y_2}})= C_1\frac{y_2^i y_2^j}{y_2^5}T_{ij}({\vect{y_2}})+\ldots
\ee
where $C_1$ is a constant that depends on the normalization of $O$. 
In general there are  extra contact terms  which can also appear on the RHS of the OPE. We are ignoring such terms and considering the limit when $y_2$ is small but not vanishing. 
 
Using eq.\eqref{opea} in the RHS of eq.\eqref{ftlim2} we get 
\begin{align}
\label{lim2rela}
\begin{split}
\lim_{\vkk\rightarrow \infty} \langle O(\vk) O(\vkk) O(\vkkk) O(\vkfour)\rangle '  = \int & d^3 {\vect{y}}_2 \ d^3 {\vect{y}}_3 \ d^3 {\vect{y}}_4 \ C_1 
{y_2^i y_2^j\over y_2^5} \ e^{-i(\vkk \cdot {\vect{y}}_2+\vkkk \cdot {\vect{y}}_3+\vkfour \cdot {\vect{y}}_4)} \\ & \langle T_{ij}({\vect{y}}_2)  O({\vect{y}}_3) O({\vect{y}}_4)\rangle '
\end{split}
\end{align}
where by the $\lim$ on the LHS we mean more precisely the limit given in  eq.\eqref{lim2a} and 
the symbol $\langle O(\vk) O(\vkk) O(\vkkk) O(\vkfour)\rangle'$ with the prime superscript stands of  the four
 point correlator $\langle O(\vk) O(\vkk) O(\vkkk) O(\vkfour)\rangle$ without the factor of 
$(2\pi)^3 \delta^3(\sum_i{\vect{k}}_i)$. The variable  $\vk$ which appears in the argument on the LHS of eq.\eqref{lim2rela} is understood to take the value 
$\vk=-(\vkk+\vkkk+\vkfour)$. 

Now in the limit of interest when $k_2\rightarrow \infty$, the support for the integral on the RHS of 
eq.\eqref{lim2rela} comes from the region where $y_2\rightarrow 0$. Thus the integral in eq.\eqref{lim2rela} can be approximated to be 
\begin{align}
\label{intapp}
\begin{split}
&\int d^3 {\vect{y}}_2 \ d^3 {\vect{y}}_3 \ d^3 {\vect{y}}_4 \ C_1 
{y_2^i y_2^j\over y_2^5} \ e^{-i(\vkk \cdot {\vect{y}}_2+\vkkk \cdot {\vect{y}}_3+\vkfour \cdot {\vect{y}}_4)}\langle T_{ij}({\vect{y}}_2)  O({\vect{y}}_3) O({\vect{y}}_4)\rangle ' \\
& = C_1 D_1 {k_2^i k_2^j \over k_2^2} \int d^3 {\vect{y}}_3 \ d^3 {\vect{y}}_4  e^{-i(\vkkk \cdot {\vect{y}}_3+\vkfour \cdot {\vect{y}}_4)}\langle T_{ij}({\vect{0}})  O({\vect{y}}_3) O({\vect{y}}_4)\rangle ',
\end{split}
\end{align}
where the prefactor is due to
\be
\label{prefaclim2}
{k_2^i k_2^j\over k_2^2} = D_1 \int d^3y_2 {y_2^i y_2^j\over y_2^5} e^{i \vkk\cdot y_2}.
\ee 
Finally doing the integral in eq.\eqref{intapp} gives us,
\be
\label{inttoo}
\int d^3 {\vect{y}}_3 \ d^3 {\vect{y}}_4 \ e^{-i(\vkkk \cdot {\vect{y}}_3+\vkfour \cdot {\vect{y}}_4)}\langle T_{ij}({\vect{0}})  O({\vect{y}}_3) O({\vect{y}}_4)\rangle ' = \langle T_{ij}(-\vkkk-\vkfour) O(\vkkk) O(\vkfour)\rangle '
\ee
where the prime superscript again indicates the absence of the momentum conserving delta function,
and using eq.\eqref{lim2rela}, eq.\eqref{intapp},   and eq.\eqref{inttoo}, we get
\be
\label{finlim2eq}
\lim_{\vkk\rightarrow \infty} \langle O(\vk) O(\vkk) O(\vkkk) O(\vkfour)\rangle ' = 
C_1 D_1{k_2^i k_2^j \over k_2^2}  \langle T_{ij}(-\vkkk-\vkfour) O(\vkkk) O(\vkfour)\rangle '.
\ee
 
From eq.\eqref{finlim2eq}  we see that in this limit the behavior of the scalar four
 point function gets related to the three point $ \langle T_{ij}(-\vkkk-\vkfour) O(\vkkk) O(\vkfour)\rangle$ correlator.
For the slow-roll model we are analyzing this three point function was calculated in \cite{Maldacena:2002vr} and has been studied more generally in \cite{MRT}, see also \cite{Bzowski:2011ab}, \cite{Bzowski:2012ih}.
These results give the value of the $ \langle T_{ij}(-\vkkk-\vkfour) O(\vkkk) O(\vkfour)\rangle$ correlator after contracting with the polarization of the graviton, $e^{s,ij}$, to be,
\begin{align}
\label{too}
e^{s,ij} \langle T_{ij}(-\vkkk-\vkfour) O(\vkkk) O(\vkfour)\rangle ' = -2e^{s,ij}
k_{3i}k_{4j}S(\tilde{k},k_3,k_4)\,
\end{align}
where  the momentum
\be
\label{deftk}
{\vect{\tilde k}}=-(\vkkk+\vkfour),
\ee
 is the argument of $T_{ij}$.\footnote{The reader should not confuse this with the four-momentum that was introduced in subsection \ref{subsecflatspace}. Here $\vect{\tilde{k}}$ is a three-vector.}  Note that the polarization $e^{s,ij}$ is a traceless tensor perpendicular to 
${\vect{\tilde{k}}}$ and $S(\widetilde{k},k_3,k_4)$ is given in eq.\eqref{singhat}.

By choosing  $\vkk\,\bot\,(\vkkk+\vkfour)$ we use eq.\eqref{too} to 
obtain from eq.\eqref{finlim2eq}
 \be
\label{lim2int}
\lim_{\vkk\rightarrow \infty} \langle O(\vk) O(\vkk) O(\vkkk) O(\vkfour)\rangle' = C_1 D_1 {k_2^i k_2^j\over k_2^2 } k_{3i}k_{4j}S(\tilde{k},k_3,k_4).
\ee
The numerical constant $C_1D_1$ in eq.\eqref{lim2int} can be obtained independently looking at the correlator $\langle O O T_{ij} \rangle$. This correlator was completely fixed after contracting with the polarization of the graviton, $e^{s,ij}$, by conformal invariance in \cite{MRT}. Comparing the behavior of $\langle O O T_{ij} \rangle$ in the limit when two of the $O$'s come together in position space with the expectations from CFT one can obtain,
\be
\label{calcd}
C_1D_1 ={3 \over 2}.
\ee

For the correlator $\langle O(\vk) O(\vkk) O(\vkkk) O(\vkfour)\rangle$, to compare this expectation from CFT in eq.\eqref{lim2int} with our result  in eq.\eqref{fourptwo},
 it is convenient to parameterize $\vkk={\vect{a}} /\epsilon$ and then take the limit $\epsilon \rightarrow 0$, with $\vkkk,\vkfour$ held fixed and 
$\vk=-(\vkk+\vkkk+\vkfour)$. For comparison purposes we also  consider
the situation when $\vkk\,\bot\,(\vkkk+\vkfour)$. As discussed in appendix \ref{apptest1} in this limit and also for the cases when $\vkk$ is perpendicular to $\vkkk+\vkfour$, we find that eq.\eqref{fourptwo} becomes
\be
\label{lim2intexct}
\lim_{\vkk\rightarrow \infty} \langle O(\vk) O(\vkk) O(\vkkk) O(\vkfour)\rangle' = {3 \over 2} {k_2^i k_2^j\over k_2^2 } k_{3i}k_{4j}S(\tilde{k},k_3,k_4).
\ee
Comparing eq.\eqref{lim2int} with eq.\eqref{lim2intexct} it is obvious that, in this limit, our result for the four point function agrees precisely with the expectation from OPE in the CFT.
The agreement is  upto contact terms which have been neglected in our discussion based on  CFT considerations above anyways.
  
\subsubsection{Limit III: Counter-Collinear Limit}
\label{limit3ccl}
Finally, we consider a third limit in which the sum of two  momenta   
vanish while the magnitudes of all individual momenta, $k_i, i=1,\cdots 4$,
are non-vanishing. Below we consider the case where 
\be
\label{defk12}
{\vect{k}}_{12}\equiv \vk + \vkk \rightarrow 0.
\ee
Note that by momentum conservation it then follows that $(\vkkk+ \vkfour)$ also vanishes.
This limit is referred to as the counter-collinear limit in the literature. As we will see, in this limit our result, eq.\eqref{finalres},
 has a divergence which arises from a pole in the propagator of the  graviton which is exchanged to  give rise to
 the  term, eq.\eqref{valGE}. This divergence is a characteristic feature of the result and could  
 potentially be  observationally interesting. 
Towards the end of this subsection we will  see that the counter-collinear limit  can in fact be obtained as a 
special case of the limit considered in the previous subsection.

It is easy to see that in the limit eq.\eqref{defk12} the contribution of the CF term eq.\eqref{valCF} is  finite while 
that of the ET contribution term eq.\eqref{valGE}
 has a divergence arising from the $G^S(\vk, \vkk,\vkkk,\vkfour)$ term, leading to, 
\be
\label{ccge}
\begin{split}
\langle \delta \phi(\vk) \delta \phi(\vkk)  \delta \phi (\vkkk) \delta \phi (\vkfour) \rangle_{ET} \rightarrow & 4 (2 \pi)^3 \delta^3\big(\sum_J {\vect{k}}_J\big) {H^6 \over M_{Pl}^6} {1 \over \prod_{J=1}^4 (2k_J^3)} {9 \over 4}{ k_1^3 k_3^3 \over k_{12}^3} \\ & \sin^2(\theta_1)\sin^2(\theta_3) \cos(2 \chi_{12,34}).
\end{split}
 \ee
The RHS arises as follows. In this limit $k_1\simeq k_2$ and $k_3\simeq k_4$, and from eq.\eqref{singhat} 
\be
\label{limvalS}
S(k_1,k_2) \simeq {3 \over 2 } k_1,
\ee
 and similarly,  $S(k_3,k_4) \simeq {3 \over 2} k_3$. As explained in appendix \ref{detcontract}, eq.\eqref{valGE}
then gives rise to eq.\eqref{ccge}, where $\theta_1$ is the angle between $\vk$ and ${\vect{k}}_{12}$
$\theta_3$ is the angle between $\vkkk$ and ${\vect{k}}_{12}$, and $\chi_{12,34}$ is the angle  between the
 projections of  $\vk, \vkkk$ on the plane orthogonal to ${\vect{k}}_{12}$.

From eq.\eqref{ccge} it  then follows that in this limit 
\be
\label{finalzzcc}
\begin{split}
\langle \zeta(\vk) \zeta(\vkk)  \zeta (\vkkk) \zeta (\vkfour) \rangle_{ET} \rightarrow & 4 (2 \pi)^3 \delta^3\big(\sum_J {\vect{k}}_J\big){1 \over 4 \epsilon^2} {H^6 \over M_{Pl}^6} {1 \over \prod_{J=1}^4 (2k_J^3)} {9 \over 4}{ k_1^3 k_3^3 \over k_{12}^3} \\ & \sin^2(\theta_1)\sin^2(\theta_3) \cos(2 \chi_{12,34}).
\end{split}
\ee

Some comments are now in order. 
First, as was noted in subsection \ref{finaldscalc} after eq.\eqref{finalfourpta}
 the ET contribution term is completely fixed by conformal invariance and therefore
the $1/k_{12}^3$ divergence in eq.\eqref{finalzzcc} is also fixed by conformal symmetry and is  model independent. 
The model dependence in the result above could arise from the fact that the $CF$ term makes no contribution in the slow-roll case.
The behavior of the CF term (like that of the ET term, see below) in this limit depends on contact terms which arise in the OPE.
We have not studied these contact terms carefully and it could perhaps be that in other models the CF term also gives rise to a 
divergent contribution comparable   to eq.\eqref{finalzzcc}.   Of course departures from the result above can also arise in models where 
 conformal invariance is  not preserved. 
 %Second, the result, eq.\eqref{finalzzcc},  agrees with the counter-collinear limit of the result  obtained in \cite{Seery:2008ax}.
 %  In fact we have used the same notation for the angles $\chi_{12,34}$

Second, the  two factors $S(k_1,k_2), S(k_3,k_4)$ in eq.\eqref{defGhat} arise from the two factors of 
$\langle O O T_{ij} \rangle$ in the ET contribution to $P[\delta \phi]$,  eq.\eqref{valpphi}, since $\langle O O T_{ij}\rangle$ when contracted with a polarization tensor can be expressed in terms of $S$, eq.\eqref{ootinmrt} in Appendix \ref{coefffnnorm}. In   the three point function $\langle O O T_{ij} \rangle$ the limit where $\vk+\vkk$
vanishes is a squeezed limit. This limit  was investigated in \cite{MRT}, subsection 4.2, and it follows from eq.\eqref{limvalS} and eq.\eqref{ootinmrt} in Appendix \ref{coefffnnorm} that in this limit 
\be
\label{slimoot}
\langle O(\vk) O(\vkk) T_{ij}(\vkkk)\rangle'e^{s,ij}= -2 e^{s,ij}k_{1i}k_{2j} {3\over 2} k_1
\ee
and is a contact term since it is analytic in $\vkk$.  However it is easy to  see from eq.\eqref{ccge} that the contribution that  the
 product of two of these three point functions make to the four point scalar correlator  in the counter-collinear limit 
 is no longer  a contact term.  This example illustrates the importance of keeping track of contact terms carefully even
 for  the purpose of eventually evaluating non-contact terms in the correlation function.

Finally, we note that due to conformal invariance an equivalent way to phrase the counter-collinear limit 
is  to take the four momenta,  $\vk, \vkk, \vkkk, \vkfour$,
all  large while keeping the sum, $\vk+\vkk=-\vkkk+\vkfour$, fixed.  This makes it clear 
that the counter-collinear limit is a special case of the limit considered in the previous subsection.
However in our discussion of the previous section we did not keep track of contact terms. Here, in obtaining the leading 
divergent behavior it is important to keep these terms, as we have noted above.
 In fact without keeping the contact term contributions 
in $\langle O O T_{ij} \rangle$ the ET contribution would  vanish in this limit.

\section{Discussion}
\label{discussion}
In this paper we have calculated the primordial four point correlation 
function for scalar perturbations in the canonical model 
of slow-roll inflation, eq.\eqref{action1}. We worked to leading order in the slow-roll approximation where the calculations can be done in de Sitter space. Our final answer is given 
in eq.\eqref{finalres}. 
This answer agrees with the result obtained in \cite{Seery:2008ax}, which was obtained using quite different methods. 

The resulting answer is small, as can be seen from the prefactor in eq.\eqref{finalres}
 which goes like 
${H^6 \over M_{Pl}^6 \epsilon^2 }\sim P_{\zeta}^3 \epsilon$, where $P_{\zeta} \sim 10^{-10}$
 is the power in the scalar perturbation, eq.\eqref{defpzeta}.    
And it is  a  complicated function of the magnitudes of three 
independent momenta and three
angles. 

The smallness of the answer is expected, since it can be easily estimated without any detailed calculation by noting, for example,  that the coefficient of the $\langle O O O O \rangle$ term, 
eq.\eqref{wf2},  is not expected to vanish \footnote{
In contrast the $\langle OOO \rangle$ coefficient function  vanishes
 to leading order in the slow-roll approximation leading to $f_{NL}\sim \epsilon$.
This vanishing of $\langle OOO \rangle$ is expected from general considerations of CFT.}.
In discussions related to observations, it is conventional  to consider the four point correlator (also called the trispectrum) to be of 
the local form and parameterize it by two coefficients $\tau_{NL}, g_{NL}$. This local form arises by taking the perturbation to be of the type, 
\be
\label{lform}
\zeta=\zeta_g+ {1\over 2} \sqrt{\tau_{NL}}(\zeta_g^2-\langle\zeta_g^2\rangle )+{9\over 25} g_{NL} \zeta_g^3,
\ee
where $\zeta_g$ is a Gaussian field. 
The answer we get is not of this local type\footnote{We clarify that we are not excluding any local type terms from the answer for the four point function, but rather that the full answer is not of the form obtained by making the ansatz in eq.\eqref{lform}.} and so it is not possible to directly 
 compare 
our result with the experimental bounds quoted in the literature, see \cite{Ade:2013ydc}, \cite{Smidt:2010sv},
$\tau_{NL} < 2800$.
However, to get some feel for the situation, we note that the non-Gaussian term proportional to 
$\sqrt{\tau_{NL}}$ in eq.\eqref{lform} would give rise to a four point correlator of order
$\sim \tau_{NL} P_{\zeta}^3$. Thus, as a very rough estimate the 
slow-roll model gives rise to $\tau_{NL}\sim \epsilon \sim 10^{-2}$ which is indeed 
small and very far from the experimental bound mentioned above.

As mentioned in the introduction, one of the main motivations of this work was to use techniques drawn from 
the AdS/CFT correspondence for  calculating correlation functions of perturbations produced during inflation,
and to analyze how  the Ward identities of conformal invariance get implemented on these
 correlation functions.
The four point scalar correlator provides a concrete and interesting setting for these purposes. 

As the analysis above has hopefully brought out the calculation could be  done quite easily by 
continuing the result from AdS space. In fact  doing the calculation in this way naturally gives rise to the wave function. And the wave function is  well suited for studying how
the symmetries, including conformal invariance, are implemented, since the symmetries of the wave function are automatically symmetries of all correlators calculated from it. 
We found that the wave function, calculated upto the required order for the four point scalar correlator,
is conformally invariant and also invariant under spatial reparameterizations. The Ward identities for conformal invariance follow from this, 
and it also follows that the four point function satisfies these Ward identities, this is discussed further in
 the next paragraph. 
 Given the complicated nature of the result  
these Ward identities
serve as an important and highly non-trivial check on the  
result. An additional set of checks was also    provided by comparing  the behavior of the result in various 
limits to what is expected  from the operator product expansion in  a conformal field theory. 

Our analysis helped uncover an interesting  general subtlety with regard to
 the Ward identities of conformal invariance. This subtlety arises in de Sitter space, more generally inflationary backgrounds, 
and does not have an analogue in AdS space, and is a general feature for other  correlation functions as well.
 In the dS case one computes the wave function 
as a functional of the boundary values for the scalar and the tensor perturbations, 
 in contrast to the partition function in AdS space. As a result, calculating the 
correlation functions in dS space requires an additional step of integrating over all 
boundary values of the scalar and tensor perturbations. This last step is well defined only if we fix the gauge completely. 
 
The resulting correlation functions are then
only  invariant under a conformal transformation accompanied by a compensating  coordinate transformation 
that restores the gauge. Failure to include this additional coordinate transformation results in the wrong Ward identities. 
It is also worth emphasizing that due to these complications it is actually simpler to check for conformal invariance in the wave function,  before correlation functions are computed from it. 
The wave function is well defined without the additional gauge fixing mentioned above, and on general grounds can be argued to be 
invariant under both conformal transformations and general spatial reparameterizations. 
Once this is ensured the 
correlation functions calculated from it  automatically satisfy the required Ward identities.

Going beyond the canonical  slow-roll model we have considered,
 one might ask what constraints does conformal symmetry impose on the $4$ point 
correlator in general?  Our answer, eq.\eqref{finalres}, arises from a sum of two terms, see eq.\eqref{finalfourpta}.
The second contribution, the extra  term (ET), eq.\eqref{valGE}, is completely determined by conformal invariance and is model independent. This follows by noting that the boundary term is obtained from  the $\langle OOT_{ij} \rangle$
correlator\footnote{See eq.\eqref{valpphi} and related discussion leading to eq.\eqref{valGE} in section 
\ref{finaldscalc}.},
and the $\langle OOT_{ij} \rangle$ correlator in turn is completely fixed by conformal invariance, \emph{e.g.} as discussed in \cite{MRT}.
 The first contribution to the answer though, the $\langle OOOO \rangle$ dependent CF term, is more model dependent and is related to the 
$4$ point function of a dimension $3$ scalar operator $O$ in a CFT.  
In $3$ dimensional CFT there are $3$ cross-ratios for the $4$ point function in position space
which are conformally invariant, and any function of these cross ratios 
is allowed by conformal invariance. This results in  a rather weak constraint on the CF term.  

However, some model independent results can arise in various limits.
For example, in the counter-collinear limit considered in section \ref{limit3ccl}, 
 our full answer has a  characteristic pole and is dominated by the ET contribution. In contrast the 
contribution from the CF term is finite and sub-dominant. The difference in behavior can be traced to 
 contact terms in the OPE of two $O$ operators. While we have not studied these contact terms in enough detail, it seems to us reasonable that in  a large class of models  the ET term should continue to dominate in this limit and
 the resulting behavior of the correlator should then be model independent and be a robust prediction that follows 
just  from conformal invariance. 
A similar model independent result may also arise in another limit which was discussed in section section \ref{limit2},
in which two of the momenta grow large. 
In this limit again the behavior of $\langle OOOO \rangle$,
upto contact terms, is determined by the $\langle OOT_{ij} \rangle$ correlator, eq.\eqref{finlim2eq}, eq.\eqref{lim2int},  which  is model independent.
A better understanding of the extent of model independence in this limit   
also requires a deeper understanding of the relevant contact terms
which we leave for the future.

\acknowledgments

We are grateful to  Paolo Creminelli and Marko Simonovic for discussions and for 
generously sharing some Mathematica code. We are also grateful to Daniel Harlow for useful discussions. AG would like to thank Suratna Das and Shailesh Lal for useful discussions and TIFR Mumbai for hospitality during various stages of this work. NK acknowledges useful discussions with Satyabrata Sahu and also thanks A. Shukla for pointing out some typographical errors in the manuscript.  ST thanks the organizers and participants of the workshop at the KITP, Santa Barbara, on Primordial Cosmology, PRIMOCOSMO13, for helpful discussion and comments. NK and ST thank the organizers and participants of the  ICTS workshop on The Information Paradox, Entanglement And Black Holes for discussions. ST acknowledges support from the J. C. Bose fellowship, Government of India. SR is partially supported by a Ramanujan fellowship of the Department of Science and Technology  of the Government of India. We would all like to acknowledge our extensive debt to the people of India.

\appendix
\section{Two and Three Point Functions and Normalizations}
\label{coefffnnorm}
In this appendix we will summarize the two point functions of scalar and tensor perturbations and the 
scalar-scalar-tensor three point function and issues related to their normalizations. The wave function 
in momentum space, written in eq.\eqref{wf2} contains the relevant coefficient functions for our discussion. 
The label $s,s'$ corresponds to the polarizations of the graviton as shown in eq.\eqref{defgammas}. The polarization tensors $\epsilon^s_{ij}({\vect{k}})$, in eq.\eqref{defgammas} are transverse and traceless as already mentioned and are normalized according to,
\be
\label{normpoltens}
\epsilon^{s,ij} \epsilon^{s'}_{ij} = 2\delta^{s,s'}.
\ee
Similarly, we define for the stress tensor
\be
\label{polstrten}
T^s({\vect{k}}) = T_{ij}({\vect{k}}) \epsilon^{s,ij}(-{\vect{k}}).
\ee
In momentum space the coefficient functions are related to position spaces ones in eq.\eqref{wf2old} and can be written as,

\begin{align}
\label{deftwo}
\langle O(\vk) O(\vkk)\rangle =\int& d^3{\vect{x}} d^3{\vect{y}} e^{-i \vk \cdot {\vect{x}}} 
e^{-i \vkk \cdot {\vect{y}}} \langle O({\vect{x}}) O({\vect{y}}) \rangle.
\end{align}
With this convention all the other coefficient functions, $\langle T_{ij}({\vect{x}}) T_{kl}({\vect{y}}) \rangle$, $\langle O({\vect{x}}) O({\vect{y}}) T_{ij}({\vect{z}})\rangle$ and $\langle O({\vect{x}}) O({\vect{y}}) O({\vect{z}}) O({\vect{w}})\rangle$, will be related to their values in momentum space accordingly.

The coefficient functions, $\langle O(\vk)O(\vkk)\rangle $, $\langle T^s(\vk) T^{s'}(\vkk)\rangle$  are well known, in the literature. We write them here,
\be
\label{ootwopt}
\langle O(\vk)O(\vkk) \rangle =(2 \pi)^3 \delta^3(\vk+\vkk)  k_1^3
\ee
and
\be
\label{tttwopt}
\langle T^{s}(\vk) T^{s'}(\vkk) \rangle = (2 \pi)^3 \delta^3(\vk+\vkk) k_1^3 {\delta^{s s'}\over 2}.
\ee
From eq.\eqref{ootwopt} and the wave function eq.\eqref{wf2} we get 
\be
\label{defpzeta}
 \langle \zeta(\vk) \zeta(\vkk) \langle=(2\pi)^3 \delta (k_1+k_2) P_{\zeta}(\vk)
\ee
where
\be
\label{defpzeta2}
P_{\zeta}(\vk)={H^2 \over M_{pl}^2 }{1 \over \epsilon }{1 \over 4 k_1^3}.
\ee

For the slow-roll model of inflation being considered here the three point coefficient function $\langle O(\vk)O(\vkk)T^s(\vkkk)\rangle $ was computed in \cite{Maldacena:2002vr}. It was also obtained in \cite{MRT} from more general considerations, which is 
\begin{align}
\label{ootinmrt}
\begin{split}
\langle O(\vk) O(\vkk) T_{ij}(\vkkk)\rangle e^{s, ij}=&-2 (2\pi)^3 \delta(\sum_{J=1}^4{\vect{k}}_J) e^{s, ij} k_{1i}k_{2j} S(k_1,k_2,k_3). \\
\text{with \ } \ S(k_1,k_2,k_3)=&(k_1+k_2+k_3) - {\sum_{i>j}k_i k_j \over (k_1+k_2+k_3)} - {k_1k_2k_3\over (k_1+k_2+k_3)^2}.
\end{split}
\end{align}

\section{Ward Identities under Spatial and Time Reparameterization}
\label{appwardident}
In this appendix we will derive the Ward identities obeyed by the coefficient functions due to both spatial and time reparameterizations. They are also called the momentum and Hamiltonian constraints respectively. We will also discuss the transformation of the scalar and tensor perturbations under special conformal transformation (SCT) following invariance of wave function under SCT.
\subsection{Ward Identities under Spatial and Time Reparameterization}
We will consider the specific coefficient function $\langle T_{ij}({\vect{x}}) O({\vect{y_1}}) O({\vect{y_2}})\rangle$ and derive the Ward identities under spatial and time parameterization. For that we need to consider only two out of the four terms in the exponent, the first and the third term, on RHS of the wave function in eq.\eqref{wf2old},
\be
\label{wfapp1}
\begin{split}
\psi[\delta \phi, \gamma_{ij}]  =  \exp\bigg[{M_{Pl}^2 \over H^2} \bigg(-{1\over 2} \int & d^3x \sqrt{g({\vect{x}})} \ d^3y \sqrt{g({\vect{y}})}  \ \delta \phi({\vect{x}}) \delta \phi({\vect{y}}) \langle O({\vect{x}}) O({\vect{y}})\rangle  \\
 -{1\over 4} \int & d^3 x \sqrt{g({\vect{x}})} \ d^3 y \sqrt{g({\vect{y}})} \ d^3 z \sqrt{g({\vect{z}})}\\ 
& \delta \phi ({\vect{x}}) \delta \phi({\vect{y}}) \gamma_{ij}({\vect{z}}) \langle O({\vect{x}}) O({\vect{y}}) T^{ij}({\vect{z}})\rangle \bigg)\bigg].
\end{split}
\ee
In the leading order to the perturbations, relevant to our calculation, $\sqrt{g({\vect{x}})}$, defined in eq.\eqref{defgij2}, can be expanded as 
\be
\label{gexpapp}
\sqrt{g({\vect{x}})} = 1 + {1 \over2}\gamma_{ii}({\vect{x}}).
\ee

Under the spatial reparameterizations given in eq.\eqref{sprepara} the scalar and tensor perturbations, $\delta \phi \ \text{and} \ \gamma_{ij}$, transform as given in eq.\eqref{metrtsp} and eq.\eqref{scalartrsp} respectively. Following them we can obtain the change in $\sqrt{g({\vect{x}})}$ under the spatial reparameterizations
\be
\label{chsqg}
\sqrt{g({\vect{x}})} \rightarrow \sqrt{g({\vect{x}})} -\partial_i v_i ({\vect{x}}).
\ee

Using eq.\eqref{metrtsp}, eq.\eqref{scalartrsp} and eq.\eqref{chsqg} we can obtain the change in the two terms on the RHS of eq.\eqref{wfapp1}. They are, for the first term,
\begin{align}
\label{dels1}
\begin{split}
 &\delta^S\bigg[-{1\over 2} \int d^3x \sqrt{g({\vect{x}})} \ d^3y \sqrt{g({\vect{y}})}  \ \delta \phi({\vect{x}}) \delta \phi({\vect{y}}) \langle O({\vect{x}}) O({\vect{y}})\rangle\bigg] =  \\ & 
 {1\over 2}  \int d^3x d^3y \bigg[{\partial \over \partial x^i} \bigg\{v^i({\vect{x}}) \delta \phi({\vect{x}})\bigg\} \delta \phi({\vect{y}}) +\delta \phi({\vect{x}}) {\partial \over \partial y^i} \bigg\{v^i({\vect{y}}) \delta \phi({\vect{y}})\bigg\}  \bigg] \langle O({\vect{x}}) O({\vect{y}})\rangle
 \end{split}
\end{align}
and similarly for the second term
\begin{align}
\label{dels2}
\begin{split}
 &\delta^S\bigg[-{1\over 4} \int d^3 x \sqrt{g({\vect{x}})} \ d^3 y \sqrt{g({\vect{y}})} \ d^3 z \sqrt{g({\vect{z}})} \delta \phi ({\vect{x}}) \delta \phi({\vect{y}}) \gamma_{ij}({\vect{z}}) \langle O({\vect{x}}) O({\vect{y}}) T^{ij}({\vect{z}})\rangle\bigg]\\ &  =  
 {1\over 2}  \int d^3x \ d^3y \ d^3z {\partial v^i({\vect{z}}) \over \partial z^j} \delta \phi({\vect{x}}) \delta \phi({\vect{y}}) \langle O({\vect{x}}) O({\vect{y}}) T^{ij}({\vect{z}})\rangle.
 \end{split}
\end{align}

The invariance of the wave function under spatial reparameterizations as stated in eq.\eqref{wfinvsp}, translates to the requirement that the total change of the RHS of eq.\eqref{wfapp1} vanishes. Which in turn implies the sum of eq.\eqref{dels1} and eq.\eqref{dels2} vanishes. This requirement, after performing an integration by parts to move the derivatives in the RHS. of eq.\eqref{dels1} and eq.\eqref{dels2} to the coefficient functions, leads us to the desired Ward identity given in eq.\eqref{spreparaward}, which is the momentum constraint.

Under the time reparameterization in eq.\eqref{timerepara}, at late times when $e^{-Ht} \rightarrow 0$, the scalar perturbation, $\delta \phi$, does not change and the tensor perturbation, $\gamma_{ij}$, changes as given in eq.\eqref{mettrepara}. Also $\sqrt{g({\vect{x}})}$, in eq.\eqref{gexpapp}, changes by,
\be
\label{chsqgt}
\sqrt{g({\vect{x}})} \rightarrow \sqrt{g({\vect{x}})} + 3 H \epsilon({\vect{x}}).
\ee

Using these we can obtain the change in the two terms on the RHS of eq.\eqref{wfapp1} under time reparameterization, similarly as we did for the spatial reparameterization. They are, for the first term,
\begin{align}
\label{dels1a}
\begin{split}
 \delta^T\bigg[-{1\over 2} \int& d^3x \sqrt{g({\vect{x}})} \ d^3y \sqrt{g({\vect{y}})}  \ \delta \phi({\vect{x}}) \delta \phi({\vect{y}}) \langle O({\vect{x}}) O({\vect{y}})\rangle\bigg] =  
 -{1\over 2} \int d^3x \ d^3y \ d^3z \\ &\delta \phi({\vect{x}})\delta \phi({\vect{y}})  H \epsilon({\vect{z}})\bigg[ 3 \delta^3(z-x)\langle O({\vect{x}}) O({\vect{y}})\rangle + 3 \delta^3(z-y)\langle O({\vect{x}}) O({\vect{y}})\rangle  \bigg]
 \end{split}
\end{align}
and similarly for the second term,
\begin{align}
\label{dels2a}
\begin{split}
 &\delta^T\bigg[-{1\over 4} \int d^3 x \sqrt{g({\vect{x}})} \ d^3 y \sqrt{g({\vect{y}})} \ d^3 z \sqrt{g({\vect{z}})} \delta \phi ({\vect{x}}) \delta \phi({\vect{y}}) \gamma_{ij}({\vect{z}}) \langle O({\vect{x}}) O({\vect{y}}) T^{ij}({\vect{z}})\rangle\bigg]\\ &  =  
 -{1\over 2}  \int d^3x \ d^3y \ d^3z H \epsilon({\vect{z}})\delta \phi ({\vect{x}}) \delta \phi({\vect{y}})\langle O({\vect{x}}) O({\vect{y}}) T^{ii}({\vect{z}})\rangle.
 \end{split}
\end{align}

The invariance of the wave function under the time reparameterization as mentioned in eq.\eqref{wftrepara} then implies that the sum of eq.\eqref{dels1a} and eq.\eqref{dels2a} vanishes. Which leads us to the constraint on the coefficient function under time reparameterization as written in eq.\eqref{extreparaw}, also called the Hamiltonian constraint.
\subsection{Transformations of the Scalar and Tensor Perturbations under SCT}
\label{appsubtrans}
At late times, $e^{-H t} \rightarrow 0$, the special conformal transformation (SCT) takes the form as given in eq.\eqref{cftrans1} and eq.\eqref{cftrans2}.
Under SCT the scalar perturbation, $\delta \phi({\vect{x}})$, transforms like a scalar,
\be
\label{scsptr}
\delta \phi({\vect{x}}) \rightarrow \delta \phi'({\vect{x}})=\delta \phi(x^i -\delta x^i).
\ee
Therefore under SCT the change in scalar perturbation is,
\be \label{scsptr1}
\delta(\delta \phi) = \delta \phi'({\vect{x}})-\delta \phi({\vect{x}})= - (x^2 \eepsilon^i - 2 x^i \dotp[x, \eepsilon]) {\partial \over \partial x^i} \delta \phi.
\ee
Since $e^{2Ht} g_{m n}$ appears as metric component in eq.\eqref{admmetric}, eq.\eqref{gfversion1}, it should transform as a tensor under coordinate transformation
\be
e^{2Ht'} g'_{ij} ({\vect{x'}})= e^{2Ht} g_{m n} ({\vect{x}}) {\partial x^m \over \partial x'^i}{\partial x^n \over \partial x'^j}.
\ee
 From here one obtains the change in the tensor perturbation due to SCT,
\begin{align}
\label{gammachange}
 \begin{split}
  \gamma_{ij}({\vect{x}}) & \rightarrow \gamma'_{ij}({\vect{x}}) = \gamma_{ij}({\vect{x}}) + \delta \gamma_{ij}({\vect{x}}),\\
 \delta \gamma_{ij}({\vect{x}}) &= 2 {M^m}_j \gamma_{i m} + 2 {M^m}_i \gamma_{m j}-(x^2 \eepsilon^m - 2 x^m \dotp[x, \eepsilon]){\partial \gamma_{ij}({\vect{x}}) \over \partial x^m}, \\
  {M^m}_j &= x^m b^j- x^j b^m.
 \end{split} 
\end{align} 
The invariance of wave function under SCT implies that,
\begin{align}
\begin{split}
 \psi[\gamma_{ij}({\vect{x}})] = \psi[\gamma'_{ij}({\vect{x}})] =  \psi[\gamma_{ij}({\vect{x}}) + \delta \gamma_{ij}({\vect{x}})].
\end{split} 
\end{align}
Ignoring scalar perturbations one obtains,
\be
\int d^3 x \gamma_{ij}({\vect{x}}) T_{ij}({\vect{x}}) = \int  d^3 x \left(\gamma_{ij}({\vect{x}}) + \delta \gamma_{ij}({\vect{x}})\right)  T_{ij}({\vect{x}}).
\ee
After doing integration by parts one can move the derivatives acting on $\gamma_{ij}$ to $T_{ij}$ to obtain,
\be
\label{apptch}
\int d^3 x \gamma_{ij}({\vect{x}}) T_{ij}({\vect{x}}) = -\int d^3 x \gamma_{ij}({\vect{x}}) \left(T_{ij}({\vect{x}}) + \delta T_{ij}({\vect{x}})\right).
\ee
From eq.\eqref{apptch} one can obtain the change in $T_{ij}({\vect{x}})$ under SCT. Similar arguments based on invariance of the wave function, can lead us from eq.\eqref{scsptr1} to the change in $O({\vect{x}})$ under SCT. The position space expression for them are given in eq.\eqref{opstr} and eq.\eqref{opttr}. We can obtain the changes in $O({\vect{x}})$ and $T_{ij}({\vect{x}})$ under SCT in momentum space in a straight forward manner as given below
\begin{align}
\delta \phi({\vect{k}}) \rightarrow \delta \phi'({\vect{k}}) =& \delta \phi({\vect{k}}) +\delta \big(\delta \phi({\vect{k}})\big) \label{appa1},\\
O({\vect{k}}) \rightarrow O'({\vect{k}}) =& O({\vect{k}}) +\delta O({\vect{k}}) \label{appa2},\\
\gamma_{ij}({\vect{k}}) \rightarrow \gamma'_{ij}({\vect{k}}) =& \gamma_{ij}({\vect{k}}) + \delta \gamma_{ij}({\vect{k}})
\label{appa3},\\
T_{ij}({\vect{k}}) \rightarrow T'_{ij}({\vect{k}}) =& T_{ij}({\vect{k}}) + \delta T_{ij}({\vect{k}}) \label{appa4}
\end{align}
where
\begin{align}
\label{delphsctmom}
 \delta \big(\delta \phi({\vect{k}})\big) =& 6 ({\vect{b}}. {\bf \partial})\delta \phi({\vect{k}})+2 k_j\partial_{k_j} (\vect{\eepsilon}\cdot \vect{\partial_k})\delta \phi({\vect{k}})-\dotp[\eepsilon,k] \partial_{k^i} \partial_{k^i} \delta \phi({\vect{k}}),\\ \label{osctmom}
 \delta O({\vect{k}}) =& 2 k_j\partial_{k_j} (\vect{\eepsilon}\cdot \vect{\partial_k})O({\vect{k}})-\dotp[\eepsilon,k] \partial_{k^i} \partial_{k^i} O({\vect{k}}),\\ \label{gamsctmom}
 \delta \gamma_{ij}({\vect{k}})=&6 ({\vect{b}}. {\bf \partial})\gamma_{ij} + 2 \tilde{M}^l_i \gamma_{l j}+ 2 \tilde{M}^l_j \gamma_{i l} +2 k_l\partial_{k_l} (\vect{\eepsilon}\cdot \vect{\partial_k}) \gamma_{ij}-\dotp[\eepsilon,k] \partial_{k^l} \partial_{k^l}\gamma_{ij}, \\ \label{tsctmom}
 \delta T_{ij}({\vect{k}}) =& 2 \tilde{M}^l_i T_{l j}+ 2 \tilde{M}^l_j T_{i l} +2 k_l\partial_{k_l} (\vect{\eepsilon}\cdot \vect{\partial_k}) T_{ij}-\dotp[\eepsilon,k] \partial_{k^l} \partial_{k^l}T_{ij}, \\ \label{mtilmom}
 \tilde{M}^l_i  \equiv & \eepsilon^l \partial_{k^i}-\eepsilon^i\partial_{k^l}.
\end{align}

\section{More Details on Calculating the AdS Correlator}
\label{detailsadsos}
In this appendix we will discuss in some more detail the algebra leading us to the four point scalar correlator in AdS, written in eq.\eqref{fullanswerS}, which is the unknown coefficient in on-shell action in $AdS$ space, $S^{\text{AdS}}_{\text{on-shell}}$.

The basic technique to compute this correlator is simply to consider the Feynman-Witten diagrams in AdS that are shown in \ref{threewittendiag}. These diagrams are just like flat space Feynman diagrams, except that the propagators between bulk points are replaced by AdS Green functions, and the lines between the bulk and the boundary are contracted with regular solutions to the wave equation in AdS, which are called bulk to boundary propagators.

We start with our action in eq.\eqref{eucads} for a canonically coupled massless scalar field, $\delta \phi$, in $AdS_4$ space-time. The stress tensor for the scalar field, $T_{\mu\nu}$, acts as a source coupled to the metric perturbation $\delta g_{\mu\nu}$.  In the gauge given in eq.\eqref{gmetp}, the evaluation of the Witten diagram simplifies to the expression given in \eqref{wittendiagorig}, which we copy here for the reader's convenience
\be
\label{wittendiagcopied}
\begin{split}
&\int d z_1 d z_2 \sqrt{-g_1} \sqrt{-g_2} T_{i_1 j_1} (z_1)  g^{i_1 i_2} g^{j_1 j_2} G^{{\text{grav}}}_{i_2 j_2, k_2 l_2} (\vect{k}, z_1, z_2) g^{k_1 k_2} g^{l_1 l_2} T_{k_1 l_1}(z_2)  \\
&G^{{\text{grav}}}_{i j, k l}(\vect{k}, z_1, z_2)  =
\int \left[{
(z_1)^{-1 \over 2}  J_{{3 \over 2}}(p z_1) J_{{3 \over 2}} (p z_2) (z_2)^{-1 \over 2}  \over  
\left(\vect{k}^2 + p^2 - i \epsilon\right)} \right.   {1 \over 2} \left.\left({\cal T}_{i k} {\cal T}_{j l} + {\cal T}_{i l} {\cal T}_{j k} - 
{2 {\cal T}_{i j} {\cal T}_{k l}\over d-1} \right)\right] {-  d p^2
\over 2 },  
\end{split}
\end{equation}
Note that here, as opposed to \eqref{wittendiagorig}, we have suppressed all except for the radial coordinates.

In fact the Bessel functions, which appear above simplify greatly in $d=3$, so that we have
$J_{3/2}(z)$ which appears in eq.\eqref{tilwx1x2} refers to the Bessel function with index $3/2$. 
It has the form, 
\be
\label{valjth}
J_{3/2}(z)=\sqrt{\frac{2}{\pi}}\frac{1}{\sqrt{z}}\left(-\cos{z}+\frac{\sin{z}}{z}\right)=\sqrt{\frac{2}{\pi}}\frac{(1-i z)e^{i z}-(1+iz)e^{-iz}}{2iz\sqrt{z}}.
\ee

As we will see below, this makes the $z$-integrals involved in the evaluation of \eqref{wittendiagcopied} very simple. The integral over $p$ can also be done by residues. However, from an algebraic viewpoint, it is simply to replace ${\cal T}_{i j} = \delta_{i j} + k_i k_j/p^2$ with $\tilde{{\cal T}}_{i j} = \delta_{i j} - k_i k_j/k^2$, which is the transverse traceless projector onto the exchanged momentum. 

This replacement leads to an additional ``remainder'' term, which accounts for the contribution of the longitudinal modes of the graviton. It also reduces our propagator to the form given in eq.(4.14) of \cite{Liu:1998ty}, for $d=3$ which reduces to,
\begin{align}
 \label{liuonshell}
 \begin{split}
  I=&{1\over 4}{M_{Pl}^2 \over 2} \bigg[\int d^4x_1 d^4x_2 \sqrt{g(z_1)} \sqrt{g(z_2)} (z_1z_2)^2 t_{ij}(x_1)G(x_1,x_2) t_{ij}(x_2) \\ &
  - 2 \int d^4x_1 \sqrt{g(z_1)} z_1^2  T_{zj} {1 \over \partial^2} T_{zj}-\int d^4x_1 \sqrt{g(z_1)} z_1^3 \partial_j T_{zj}{1 \over \partial^2} T_{zz}  \\ &
  - {1 \over 2}\int d^4x_1 \sqrt{g(z_1)} z_1^2  \partial_j T_{zj}\left({1 \over \partial^2}\right)^2 \partial_i T_{zi}\bigg].
 \end{split}
\end{align}
Note that in writing the above equation we have corrected two typographical errors in eq.(4.14) of \cite{Liu:1998ty}. First, an overall factor of ${1\over4}$ and second, $z_1^3$ in place of $z_1^2$ in the third term on the RHS of eq.\eqref{liuonshell}.
In the first term on the RHS of eq.\eqref{liuonshell}, $t_{ij}=\widehat{P}_{ijkl} T_{kl}$, such that $\widehat{P}_{ijkl}$ is the transverse traceless projector in flat space, eq.\eqref{defPhat}, and $G(x_1,x_2)$ is the Green's function for a free massless scalar field in Euclidean $AdS_4$, obtained in appendix A, see eq.(A.3), of \cite{Liu:1998ty}. 
 
 It is straightforward to see that the first term on the RHS of eq.\eqref{liuonshell} becomes the contribution from the transverse graviton, $\widetilde{W}$ in eq.\eqref{wittendiag}, and also the three other terms on the RHS of eq.\eqref{liuonshell} which are the contributions from the longitudinal graviton become $R_1,R_2$ and $R_3$ in eq.\eqref{remainder1} respectively. 
 
Next we proceed to perform the integrations in eq.\eqref{wittendiag} and eq.\eqref{remainder1}. We start with $\widetilde{W}$ in eq.\eqref{wittendiag}. 

 After being fourier transformed to momentum space the stress tensor, appearing in eq.\eqref{wittendiag}, becomes
\be
\label{momtij}
T_{ij}({\vect{k}},z)=\int d^3 {\vect{x}} \ T_{ij}({\vect{x}},z) \ e^{-i \ {\vect{k}}\cdot {\vect{x}}}.
\ee
Using eq.\eqref{momtij}, $\widetilde{W}$, as in eq.\eqref{wittendiag}, takes the form
\be
\label{wittendiagmom}
\widetilde{W} =\int dz_1 dz_2 {d^3 {\vect{k}} \over (2 \pi)^3} T_{i_1 j_1}(-{\vect{k}}, z_1) \delta^{i_1 i_2} \delta^{j_1 j_2} \widetilde{G}_{i_2 j_2, k_2 l_2}({\vect{k}}, z_1, z_2) \delta^{k_1 k_2} \delta^{l_1 l_2} T_{k_1 l_1}({\vect{k}}, z_2)
\ee
in momentum space with
 \begin{equation}
 \label{grprop}
 \widetilde{G}_{i j, k l}({\vect{k}}, z_1, z_2)  =
 \int_0^{\infty} {dp^2\over2} \left[{ J_{{3 \over 2}}(p z_1) J_{{3 \over 2}} (p z_2) \over  
  \sqrt{z_1 z_2}\left({\vect{k}}^2 + p^2 \right)}   {1 \over 2} \left({\widetilde{\cal T}}_{i k} {\widetilde{\cal T}}_{j l} + {\widetilde{\cal T}}_{i l} {\widetilde{\cal T}}_{j k} - 
  {\widetilde{\cal T}}_{i j} {\widetilde{\cal T}}_{k l} \right) \right]\,.
 \end{equation}
The indices $i_1,j_1$ etc which appear in eq.\eqref{wittendiagmom} take values along the $x^i, i=1,2,3$ directions eq.\eqref{eadsr}. The $z$ components of the stress tensor do not appear because of our choice of gauge, eq.\eqref{gmetp}.

In eq.\eqref{wittendiagmom} the graviton propagator $\widetilde{G}_{i j k l}({\vect{k}}, z_1, z_2)$ is contracted against two factors of the stress tensors $T_{ij}({\vect{k}},z)$. The stress tensor for the scalar perturbation $\delta \phi$, acts like a source term for the metric perturbation $\delta g_{\mu \nu}$ as evident from the interaction vertex in  eq.\eqref{defint}. From the expression of stress tensor in eq.\eqref{defstressa} one can obtain,
\begin{align}
\label{eq:Tij}
T_{ij}(z,{\vect{x}})&=2(\partial_i\delta \phi)(\partial_j\delta\phi)-\delta_{ij}\left[(\partial_z\delta\phi)^2+{\eta^{mn}}(\partial_m\delta\phi)(\partial_{n}\delta\phi)\right].
\end{align}

The  two insertions of the stress tensor in eq.\eqref{wittendiagmom} correspond to two different values of the radial variable $z=z_1$ and $z=z_2$, which are integrated over. For the $S$-channel contribution one should substitute for $\delta \phi$ from eq.\eqref{delphidd} in eq.\eqref{eq:Tij} and keep only the bilinears of the form $\phi(\vk)\phi(\vkk)$ at $z=z_1$ and similarly $\phi(\vkkk)\phi(\vkfour)$ at $z=z_2$. For the $T$- channel and $U$- channel contributions one just needs to exchange two of the external momenta in the $S$-channel answer like $\vkk \leftrightarrow \vkkk$ and $\vkk \leftrightarrow \vkfour$ respectively.

In momentum space, the stress tensors, to be substituted for in eq.\eqref{wittendiagmom}, becomes
\begin{align}
\label{bnexplct}
\begin{split}
T_{ij}[\phi_1(z_1),\phi_2(z_1)]&=-4\left\{k_{1i}k_{2j}\phi_1\phi_2+\frac{1}{2}\eta_{ij}\left[(\partial_{z_1}\phi_1)(\partial_{z_1}\phi_2)-\vect{k_1}\cdot\vect{k_2}\,\phi_1\phi_2\right]\right\}\,,\\
T_{kl}[\phi_3(z_2),\phi_4(z_2)]&=-4\left\{k_{3k}k_{4l}\phi_3\phi_4+\frac{1}{2}\eta_{kl}\left[(\partial_{z_2}\phi_3)(\partial_{z_2}\phi_4)-\vect{k_3}\cdot\vect{k_4}\,\phi_3\phi_4\right]\right\}\,.
\end{split}
\end{align}
We have used $\partial_m\phi=-ik_m\phi$ and $g^{ij}=z^2\eta^{ij}$ and also the abbreviations
\begin{align}
\label{eq:abbrev}
\begin{split}
\phi_1\equiv \phi(\vk) (1+ k_1 z_1)e^{-k_1 z_1},\hspace{1cm} 
\phi_2\equiv \phi(\vkk) (1+ k_2 z_1)e^{-k_2 z_1},\\
\phi_3\equiv \phi(\vkkk) (1+ k_3 z_2)e^{-k_3 z_1},\hspace{1cm}
\phi_4\equiv \phi(\vkfour) (1+ k_4 z_2)e^{-k_4 z_2}.
\end{split}
\end{align}
It is important to note that only the first term on the RHS of eq.\eqref{bnexplct} in both $T_{ij}$ and $T_{kl}$ contribute to $\widetilde{W}$ in eq.\eqref{wittendiagmom}. This is because the second term in $T_{ij}$ on the RHS of eq.\eqref{bnexplct} carries $\eta_{ij}$ which when contracted with the transverse projector  $\left({\widetilde{\cal T}}_{i k} {\widetilde{\cal T}}_{j l} + {\widetilde{\cal T}}_{i l} {\widetilde{\cal T}}_{j k} -  {\widetilde{\cal T}}_{i j} {\widetilde{\cal T}}_{k l} \right)$ of the graviton propagator $\widetilde{G}_{i_2 j_2, k_2 l_2}({\vect{k}}, z_1, z_2)$ in eq.\eqref{grprop}, gives zero.
\be
\eta_{ij} \left({\widetilde{\cal T}}_{i k} {\widetilde{\cal T}}_{j l} + {\widetilde{\cal T}}_{i l} {\widetilde{\cal T}}_{j k} -  {\widetilde{\cal T}}_{i j} {\widetilde{\cal T}}_{k l} \right) =0 .
\ee

Therefore, the relevant terms in the stress tensors are,
\begin{align}
\label{bnexplct1}
\begin{split}
T_{ij}(z_1)&=-4k_{1i}k_{2j}\phi_1\phi_2, \\
T_{kl}(z_2)&=-4k_{3k}k_{4l}\phi_3\phi_4.
\end{split}
\end{align}

Finally substituting in eq.\eqref{wittendiagmom} for $T_{ij}$ from eq.\eqref{bnexplct1} with the $\phi_i$'s in eq.\eqref{eq:abbrev} and  the graviton propagator in eq.\eqref{grprop} we obtain $\widetilde{W}^S(\vk,\vkk,\vkkk,\vkfour)$ as,
\be
\label{defwtsdet}
\begin{split}
\widetilde{W}^S(\vk,\vkk,\vkkk,\vkfour) =&16 (2\pi)^3 \delta^3(\sum_i {\vect{k}}_i)\phi(\vk) \phi(\vkk) \phi(\vkkk) 
\phi(\vkfour) \\ & k^i_1 k^j_2 k^k_3 k^l_4\left({\widetilde{\cal T}}_{i k} {\widetilde{\cal T}}_{j l} + {\widetilde{\cal T}}_{i l} {\widetilde{\cal T}}_{j k} - 
 {\widetilde{\cal T}}_{i j} {\widetilde{\cal T}}_{k l} \right)  S(\norm{k_1}, \norm{k_2}, \norm{k_3}, \norm{k_4})
\end{split}
\ee
where
\begin{equation}
\begin{split}
S(k_1, k_2, k_3, k_4) = \int_0^{\infty} {d p^2 \over 2(p^2 + K_s^2)} &\int_0^{\infty} {d z_1 \over z_1^2} (1 + k_1 z_1) (1 + k_2 z_1) (z_1)^{3 \over 2} J_{3 \over 2}(p z_1) e^{-(k_1 + k_2) z_1} \\ & \int_0^{\infty} {d z_2 \over z_2^2} (1 + k_3 z_2) (1 + k_4 z_2) (z_2)^{3 \over 2} J_{3 \over 2}(p z_2) e^{-(k_3 + k_4) z_1},
\end{split}
\end{equation}
and $K_s$ which is the norm of the momentum of the graviton exchanged in the $S$ channel, is
\be
\label{defks}
{\vect{k}}_s=\vk+\vkk=-(\vkkk+\vkfour)
\ee
$S(k_1,k_2k_3,k_4)$ can be evaluated by explicitly carrying out the integrals. 
We first do the $z_1,z_2$ integrals by noting that,
\begin{equation}
\label{eq:z2int}
\begin{split}
 \int_0^\infty {d z_1 \over z_1^2} (1 + k_1 z_1) (1 + k_2 z_1) (z_1)^{3 \over 2} J_{3 \over 2}(p z_1) e^{-(k_1 + k_2) z_1} = \sqrt\frac{2}{\pi}\frac{p^{3/2} \left(k_1^2+4 k_2
   k_1+k_2^2+p^2\right)}{ \left((k_1+k_2)^2+p^2\right)^2 }
\end{split}
\end{equation}
This gives, 
\begin{equation}\label{eq:Ss-pre-int}
S(k_1, k_2, k_3, k_4) = {\int_0}^{\infty} d p \frac{2}{\pi} \frac{p^4 \left(k_{1}^2+4 k_{2}
   k_{1}+k_{2}^2+p^2\right) \left(k_{3}^2+4 k_{4}
   k_{3}+k_{4}^2+p^2\right)}{
   \left((k_{1}+k_{2})^2+p^2\right)^2
   \left((k_{3}+k_{4})^2+p^2\right)^2
   \left(K_s^2+p^2\right) },
\end{equation}
The $p$ integral is now easy to do, by noting that the integrand is an even function of $p$, and then doing the integral by the method of residues.  This is a significant advantage of massless fields in momentum space in AdS$_{d+1}$, where $d$ is odd: exchange interactions can be evaluated algebraically! 
This leads to
\begin{equation}\label{eq:Ss-post-int}
\begin{split}
S =-2 &\Bigg[\frac{ k_{1} k_{2} (k_{1}+k_{2})^2 \left((k_{1}+k_{2})^2-k_{3}^2-k_{4}^2-4 k_{3} k_{4}\right)}{(k_{1}+k_{2}-k_{3}-k_{4})^2 (k_{1}+k_{2}+k_{3}+k_{4})^2
   (k_{1}+k_{2}-K_s) (k_{1}+k_{2}+K_s)}  \\ &\Big(-\frac{k_{1}+k_{2}}{2 k_{1}
   k_{2}}-\frac{k_{1}+k_{2}}{-(k_{1}+k_{2})^2+k_3^2+k_{4}^2+4 k_{3}
   k_{4}}+\frac{k_{1}+k_{2}}{K_s^2-(k_{1}+k_2)^2}\\ &+\frac{1}{-k_{1}-k_{2}+k_{3}+k_{4}} -\frac{1}{k_{1}+k_{2}+k_{3}+k_{4}}+\frac{3}{2 (k_{1}+k_{2})}\Big) + (1,2 \leftrightarrow 3,4) \\
&- \frac{K_s^3 \left(-k_{1}^2-4 k_{2}
   k_{1}-k_{2}^2+K_s^2\right) \left(-k_{3}^2-4
   k_{4} k_{3}-k_{4}^2+K_s^2\right)}{2
   \left(-k_{1}^2-2 k_{2}
   k_{1}-k_{2}^2+K_s^2\right)^2 \left(-k_{3}^2-2
   k_{4} k_{3}-k_{4}^2+K_s^2\right)^2} \Bigg].
\end{split}
\end{equation}

This value of $S$ from eq.\eqref{eq:Ss-post-int} along with the index contraction
{\small
\begin{align}
\label{eq:s-contraction}
\begin{split}
& k^i_1 k^j_2 k^k_3 k^l_4 (\widetilde{\cal T}_{ik} \widetilde{\cal T}_{jl}+\widetilde{\cal T}_{il} \widetilde{\cal T}_{jk}-\widetilde{\cal T}_{ij} \widetilde{\cal T}_{kl})= \bigg\{\vk.\vkkk+\frac{\{(\vkk+\vk).\vk\} \{(\vkfour+\vkkk).\vkkk\}}{|\vk+\vkk|^2}\bigg\} \\ & \bigg\{\vkk.\vkfour+  \frac{\{(\vk+\vkk).\vkk\} \{(\vkkk+\vkfour).\vkfour\}}{|\vk+\vkk|^2}\bigg\} 
+  \bigg\{\vk.\vkfour+\frac{\{(\vkk+\vk).\vk\} \{(\vkfour+\vkkk).\vkfour\}}{|\vk+\vkk|^2}\bigg\}\\ & 
\bigg\{\vkk.\vkkk+\frac{\{(\vkk+\vk).\vkk\} \{(\vkfour+\vkkk).\vkkk\}} {|\vk+\vkk|^2}\bigg\}-
\bigg\{\vk.\vkk-\frac{\{(\vkk+\vk).\vk\} \{(\vk+\vkk).\vkk\}} {|\vk+\vkk|^2}\bigg\}\\ & 
\bigg\{\vkkk.\vkfour-\frac{\{(\vkkk+\vkfour).\vkfour\} \{(\vkfour+\vkkk).\vkkk\}}{|\vk+\vkk|^2}\bigg\},
\end{split}
\end{align}}
can now  be substituted in eq.\eqref{defwtsdet} to obtain eq.\eqref{hatW}.

Once the transverse graviton contribution, $\widetilde{W}$, is calculated, we are left with the longitudinal contributions for the graviton eq.\eqref{valR}. The momentum space expressions for eq.\eqref{remainder1} becomes
\begin{align}
 \label{remainder2}
 \begin{split}
  R_1 &=  \int {dz_1 \over z_1^{2}} {d^3 {\vect{k}} \over (2 \pi)^3} T_{zj}({\vect{k}}, z_1) {1 \over k^2} T_{zj}(-{\vect{k}}, z_1),\\
  R_2 &= {1 \over 2} i \int {dz_1 \over z_1} {d^3 {\vect{k}} \over (2 \pi)^3} k_j T_{zj}({\vect{k}}, z_1){1 \over k^2} T_{zz}(-{\vect{k}}, z_1),\\
  R_3 &= -{1 \over 4}\int {dz_1 \over z_1^{2}} {d^3 {\vect{k}} \over (2 \pi)^3} k_j T_{zj}({\vect{k}}, z_1){1 \over k^4} k_i T_{zi}(-{\vect{k}}, z_1).
 \end{split}
\end{align}

Using the relevant components of the stress tensors obtained from eq.\eqref{defstressa}, as given below, 
\begin{align}\label{t0j1}
 T_{zj}(\phi_1, \phi_2)&= i  z_1 e^{-(k_1+k_2)z_1}\bigg[k_1^2k_{2j}(1+k_2 z_1)+k_2^2k_{1j}(1+k_1 z_1)\bigg],\\ \label{t001} \nonumber
 T_{zz} (\phi_1, \phi_2)&=e^{-(k_1+k_2)z_1}\bigg[\vk.\vkk+(\vk.\vkk) (k_1+k_2)z_1 \\ &+ k_1k_2(k_1k_2+\vk.\vkk)z_1^2  \bigg],\\ \label{delt0j1} \nonumber
 k_j T_{zj}(\phi_1, \phi_2)&= z_1 e^{-(k_1+k_2)z_1}\bigg[2k_1^2k_2^2+(\vk.\vkk)(k_1^2+k_2^2)\\&+k_1k_2(k_1+k_2)(k_1k_2+\vk.\vkk)z_1 \bigg]
\end{align}
in eq.\eqref{remainder2} one obtains eq.({\ref{defrs}}). Adding eq.\eqref{hatW} and eq.\eqref{defrs} one finally obtains $S^{\text{AdS}}_{\text{on-shell}}$ as given in eq.\eqref{fullanswerS}.

\section{Possibility of Additional Contributions in Changing Gauge}
 \label{addterm}
Once we obtained the four point function $\langle \delta \phi(\vk)\delta\phi(\vkk) \delta \phi(\vkkk) \delta\phi(\vkfour) \rangle$, eq.\eqref{finalfourpta}, in gauge 2, we finally moved to gauge 1 using the relation eq.\eqref{infper2}. It was mentioned in subsection \ref{finres4pt}, in the footnote, that there will be additional higher order terms in $\zeta$ on RHS of eq.\eqref{infper2}  which might lead to additional contribution in $\langle \zeta(\vk)\zeta(\vkk) \zeta(\vkkk) \zeta(\vkfour) \rangle$ coming from the two point function $\langle \delta \phi(\vk)\delta\phi(\vkk)\rangle$. In this appendix we will first compute the additional higher order terms in eq.\eqref{infper2} and then argue that the possible additional contributions to eq\eqref{infper2} are further suppressed in the slow-roll parameters.

For the discussion of contribution due to higher order terms in $\zeta$, we will not consider any tensor perturbations and define the scalar perturbation, $\zeta$, in the metric in the following way, from eq.\eqref{fmetric},
\be
\label{defhijsc}
h_{ij}=e^{2(\rho(t)+\zeta)}  \delta_{ij}.
\ee
The scalar perturbation in the inflaton is defined in eq.\eqref{delphi}. As was discussed in subsection \ref{gauge2}, to go from gauge 2 to gauge 1 one needs to do a time reparameterization, which to leading order in $\zeta$ was given in eq.\eqref{tre}. Considering the time coordinate in gauge 2 being $\tilde{t}$ and that in gauge 1 being $t$, we write this infinitesimal time reparameterization as,
\be
\label{trepn}
\tilde{t}= t+T.
\ee
The scalar perturbation in the inflaton in gauge 2 is given eq.\eqref{delphi}, the scalar perturbation in the metric, $\zeta$, is 
zero in this gauge and the metric takes the form,
\be
\label{meting2}
ds^2=-d\tilde{t}^2+e^{2\rho(\tilde{t})} \delta_{ij}dx^i dx^j.
\ee
Notice that for $\rho(\tilde{t})=H \tilde{t}$ we get eq.\eqref{defgij}, here we allow for a more general time dependence. 

In gauge 1 there is no scalar perturbation in the inflaton, $\delta \phi=0$ and the metric, now containing a scalar perturbation, becomes,
\be
\label{meting1}
ds^2=-dt^2+e^{2(\rho(t)+\zeta)} \delta_{ij}dx^i dx^j.
\ee

Using the time reparameterization, eq.\eqref{trepn}, in eq.\eqref{delphi} and demanding that $\delta \phi$ vanishes as we go to gauge 1 yields the relation, upto cubic order in $T$ \footnote{Here we neglected a term like $\partial_t \delta \phi$, because at late times, upon horizon crossing $\delta \phi$ becomes constant.},
\be
\label{reltprn1}
\delta \phi=-T \partial_t \bar{\phi}(t) -{T^2 \over 2} \partial_t^2 \bar{\phi}(t)-{T^3 \over 6} \partial_t^3 \bar{\phi}(t).
\ee
Using the time parameterization, eq.\eqref{trepn}, in the metric of gauge 2, eq.\eqref{meting2}, and then comparing with the metric in gauge 1, eq.(\eqref{meting1}), we obtain $\zeta$, upto cubic order in $T$ ,
\be
\label{reltprn3}
\zeta = T \partial_t \rho+ {T^2 \over 2} \partial_t^2\rho+{T^3 \over 6} \partial_t^3\rho.
\ee
Inverting this equation we obtain $T$ in terms of $\zeta$,
\be
\label{reltprn4}
T= {\zeta \over \partial_t \rho}- {\zeta^2 \over 2 (\partial_t \rho)^3} \partial_t^2 \rho-{\zeta^3 \over 6 (\partial_t \rho)^4} \partial_t^3 \rho.
\ee
Substituting for $T$ from eq.\eqref{reltprn4} in eq.\eqref{reltprn1}, we obtain the correction to eq.\eqref{infper2} due to higher order terms in $\zeta$,
\be
\label{reltprn5}
\begin{split}
\delta \phi=&-{\zeta \over \partial_t \rho} \partial_t \bar{\phi}+{1\over2} {\zeta^2 \over (\partial_t \rho)^2}(\partial_t \bar{\phi})^2\bigg({\partial_t^2 \rho \over \partial_t \rho \ \partial_t \bar{\phi}} -  {\partial_t^2 \bar{\phi} \over (\partial_t \bar{\phi})^2} \bigg) \\ &
-{1\over2}{\zeta^3 \over (\partial_t \rho)^3}(\partial_t \bar{\phi})^3\bigg(-{1\over3}{\partial_t^3 \rho \over \partial_t \rho \ (\partial_t \bar{\phi})^2} -{\partial_t^2 \rho \over \partial_t \rho}{\partial_t^2 \bar{\phi} \over(\partial_t \bar{\phi})^3 }+{1\over3} {\partial_t^3 \bar{\phi} \over (\partial_t \bar{\phi})^3 }\bigg).
\end{split}
\ee

The linear term in $\zeta$ on the RHS of eq.\eqref{reltprn5} gives the leading contribution to the four point correlator of curvature perturbations 
 $\langle \zeta(\vk)\zeta(\vkk) \zeta(\vkkk) \zeta(\vkfour) \rangle$ as obtained in eq.\eqref{finalres}.
This arises due to the quartic coefficient term $\langle O(\vk)O(\vkk) O(\vkkk) O(\vkfour) \rangle$ in eq.\eqref{wf2}.
Additional contributions to the $\langle \zeta(\vk)\zeta(\vkk) \zeta(\vkkk) \zeta(\vkfour) \rangle$ correlator  arise from the higher terms on the RHS of eq.({\ref{reltprn5}).
For example the term going like $\zeta^2$ on the RHS of eq.\eqref{reltprn5} when inserted twice in the quadratic term of the wave 
function, which goes like ${M_{Pl}^2 \over H^2}\int - \delta \phi \delta \phi \langle OO \rangle$, eq.\eqref{wf2},
 can give a contribution of this type.    
However,  due to the extra factor $\big({\partial_t^2 \rho \over \partial_t \rho \ \partial_t \bar{\phi}} -  {\partial_t^2 \bar{\phi} \over (\partial_t \bar{\phi})^2} \big)$ that multiples the $\zeta^2$ term on the RHS of eq.\eqref{reltprn5}
such a contribution will   be of order $\epsilon$ and therefore will be suppressed in the slow-roll parameters. 

Similarly, it is easy to see that the other terms in the expansion in 
eq.\eqref{reltprn5} also give subleading contributions to the four-point correlator. 

\section{Conformal Transformations and  Compensating Reparameterizations}
\label{confinvprob}

Following our discussion in section \ref{testconf}, in this Appendix we will show that the probability distribution $P[\delta \phi]$, defined in eq.(\ref{defP}),  is invariant under the combined conformal transformation, eq.\eqref{cftrans1}, eq.\eqref{cftrans2}, and 
the compensating coordinate reparameterization, eq.\eqref{furthct} upto quartic order in $\delta \phi$. 
This will, in turn, prove the invariance of the scalalr four point function  under the combined transformations mentioned above.

The probability distribution was calculated upto quartic order in $\delta \phi$  in eq.\eqref{valpphi}. 
Let us begin by writing the  different terms in $P[\delta \phi]$ schematically as,
\begin{align}
\label{probdenst}
 \begin{split}
  P[\delta \phi] = \exp\bigg[&-\int \delta\phi\delta\phi\langle OO \rangle + {1\over8} \int \delta\phi\delta\phi\delta\phi\delta\phi {\widehat{P}_{ijkl}({\vect{k}}_{12}) \over k_{12}^3}\langle OOT_{ij} \rangle \langle OOT_{kl} \rangle\\
  &+{1\over12}\int \delta\phi\delta\phi\delta\phi\delta\phi \langle OOOO \rangle \bigg]
 \end{split}
 \end{align}
 with, $k_{12}=|\vk+\vkk|$.  
In particular the  second term in the RHS of eq.\eqref{probdenst}  is responsible for the ET contribution and we will refer to it loosely as the ET term below.
 
The invariance of $P[\delta \phi]$ under the combined transformations works out in a rather non-trivial way as follows.
The first and third terms in the RHS of eq.\eqref{probdenst} which are quadratic and proportional to $\langle OOOO \rangle$ respectively, 
are both invariant under a  conformal transformation. But the  ET term in the RHS of eq.\eqref{probdenst} 
 is not. However, its transformation  under a conformal transformation is exactly canceled by the  transformation 
of   the first term in the RHS of eq.\eqref{probdenst} under the compensating 
reparameterization. 
Under a conformal transformation $\delta \phi$ transforms linearly, therefore the resulting change of the ET term under a conformal transformation  is quartic
 in $\delta \phi$. Now as we see from eq.(\ref{delRphin}) under the compensation reparameterization $\delta \phi$ transforms by a  term which is cubic
in $\delta \phi$.  As a result  the change under the compensating reparameterization
 of the first term in the RHS of eq.\eqref{probdenst}, which is quadratic in $\delta \phi$ to begin with,
  is also quartic in $\delta \phi$.
We will show in this appendix that these two terms exactly cancel.  
Additional contributions under the compensating reparameterization arise from the second and third terms in 
eq.\eqref{probdenst} but these are of order $(\delta \phi)^6$ and not directly of concern for us, since we are only keeping terms upto quartic order in $\delta \phi$. 
We expect that once all terms of the required order are kept $P[\delta \phi]$ should be invariant to all orders in $\delta \phi$. 

We turn now to establishing this argument in more detail. First, we will calculate the change in the first term in the 
RHS of eq.\eqref{probdenst} under the compensating reparameterization. Next, we will calculate the change in the ET term in eq.\eqref{probdenst} under a conformal transformation and show that they cancel each other.

\subsection{Change in $\int \delta \phi \delta \phi \langle OO \rangle$ under the Compensating  Reparameterization}

The change of $\int \delta \phi \delta \phi \langle OO \rangle$ under the coordinate reparameterization $x^i \rightarrow x^i + v^i(x)$ with $v^i ({\vect{x}}) =  -{6 b^k \gamma_{kj}({\vect{x}})\over \partial^2}$ can be written in momentum space as,
\begin{align}
\label{delRoo}
 &\delta^R\bigg[-\int \delta \phi \delta \phi \langle OO \rangle\bigg] = I_1 +I_2, \\
 I_1 = -&\int {d^3 \vk \over (2 \pi)^3} {d^3 \vkk \over (2 \pi)^3} \delta^R\big(\delta \phi( \vk) \big)
\delta \phi( \vkk) \langle O(- \vk) O( -\vkk)\rangle, \\
 I_2 =-&\int {d^3 \vk \over (2 \pi)^3} {d^3 \vkk \over (2 \pi)^3} \delta \phi( \vk)
\delta^R\big(\delta \phi( \vkk) \big) \langle O(- \vk) O( -\vkk)\rangle.
\end{align}
 Using $\delta^R\big(\delta \phi( {\vect{k}}) \big)$ from eq.\eqref{delRphin}, one gets 
 \begin{align}
 \begin{split}
 I_1 =  3b^m &\int {d^3 k_1 \over (2 \pi)^3} {d^3 k_2 \over (2 \pi)^3} {d^3 k_3 \over (2 \pi)^3}{d^3 k_4 \over (2 \pi)^3} {d^3 k_5 \over (2 \pi)^3} \delta \phi( \vk-{\vect{k}}_5) \delta \phi( \vkk) \delta\phi(\vkkk) \delta \phi( \vkfour) \\ 
 & k_1^i \ {\widehat{P}_{imkl}( {\vect{k}}_5)\over k_5^8}\langle O(- \vk) O( -\vkk)\rangle \ \langle O(- \vkkk) O( -\vkfour)T_{kl}(-{\vect{k}}_5)\rangle.
 \end{split}
 \end{align}
With the change of variable 
\be
\vk = {\bf q}_1+{\vect{k}}_5
\ee
and then again relabeling ${\bf q}_1 \rightarrow \vk$ one arrives at,
\begin{align}
 \begin{split}
 I_1 = & 3b^m \int {d^3 k_1 \over (2 \pi)^3} {d^3 k_2 \over (2 \pi)^3} {d^3 k_3 \over (2 \pi)^3}{d^3 k_4 \over (2 \pi)^3} {d^3 k_5 \over (2 \pi)^3} \delta \phi( \vk) \delta \phi( \vkk) \delta\phi(\vkkk) \delta \phi( \vkfour) \\ 
 & { \widehat{P}_{imkl}( {\vect{k}}_5) \over 
 k_5^8} (k_1^i+k_5^i) \ \langle O(- \vk- {\vect{k}}_5) O( -\vkk)\rangle \langle O(- \vkkk) O( -\vkfour)T_{kl}(-{\vect{k}}_5)\rangle.
 \end{split}
 \end{align}
 Similar procedure for $I_2$ gives,
 \begin{align}
 \begin{split}
 I_2 =&  3b^m \int {d^3 k_1 \over (2 \pi)^3} {d^3 k_2 \over (2 \pi)^3} {d^3 k_3 \over (2 \pi)^3}{d^3 k_4 \over (2 \pi)^3} {d^3 k_5 \over (2 \pi)^3} \delta \phi( \vk) \delta \phi( \vkk) \delta\phi(\vkkk) \delta \phi( \vkfour) \\ 
 & { \widehat{P}_{imkl}( {\vect{k}}_5) \over 
 k_5^8} (k_2^i+k_5^i) \ \langle O(- \vk) O( -\vkk- {\vect{k}}_5)\rangle  \langle O(- \vkkk) O( -\vkfour)T_{kl}(-{\vect{k}}_5)\rangle.
 \end{split}
 \end{align}
Substituting for $I_1$ and $I_2$ in eq.\eqref{delRoo} one obtains,
\begin{align}
\label{delRdpdp}
\begin{split}
 \delta^R\bigg[-&\int \delta \phi \delta \phi \langle OO \rangle\bigg] = 3b^m \int \prod_{i=1}^4 \bigg({d^3 k_i \over (2 \pi)^3}\delta \phi( {\vect{k}}_i)\bigg) {d^3 k_5 \over (2 \pi)^3}{ \widehat{P}_{imkl}( {\vect{k}}_5) \over k_5^8}   \\ 
 & \big\{(k_1^i+k_5^i) \ \langle O(- \vk- {\vect{k}}_5) O( -\vkk)\rangle+(k_2^i+k_5^i) \ \langle O(- \vk) O( -\vkk- {\vect{k}}_5)\rangle \big\}\\ 
 & \langle O(- \vkkk) O( -\vkfour)T_{kl}(-{\vect{k}}_5)\rangle.
 \end{split}
\end{align}
We further use the Ward identity for the conservation of stress tensor,
\begin{align}
\begin{split}
\partial_{x_3^j} \langle O({\vect{x}}_1) O({\vect{x}}_2)T_{ij}({\vect{x}}_3) \rangle &= \delta^3({\vect{x}}_3-{\vect{x}}_1) \langle\partial_{x_1^j} O({\vect{x}}_1) O({\vect{x}}_2)\rangle\\& +\delta^3({\vect{x}}_3-{\vect{x}}_2) \langle O({\vect{x}}_1) \partial_{x_2^j}O({\vect{x}}_2)\rangle
\end{split}
\end{align}
which in momentum space is of the form,
\begin{align}
\label{wiconsvst}
\begin{split}
(k_3^i+k_1^i) \ \langle O( \vk+\vkkk) O(\vkk)\rangle&+(k_3^i+ k_2^i) \ \langle O( \vk) O(\vkk+ \vkkk)\rangle \\
&= k_3^j \langle O(\vk) O(\vkk)T_{ij}(\vkkk) \rangle.
\end{split}
\end{align}
Using eq.\eqref{wiconsvst}, one can simplify eq.\eqref{delRdpdp} further to obtain,
\begin{align}
\label{delRdpdp1}
\begin{split}
 \delta^R\bigg[-\int \delta \phi \delta \phi \langle OO \rangle\bigg]& = 3b^m \int\prod_{i=1}^4  \bigg[{d^3 k_i \over (2 \pi)^3}\delta \phi( {\vect{k}}_i) \bigg] {d^3 k_5 \over (2 \pi)^3}{ \widehat{P}_{imkl}( {\vect{k}}_5) \over k_5^8} \\ 
 & k_5^j \ \langle O(- \vk) O( -\vkk) T_{ij}(-{\vect{k}}_5)\rangle \ \langle O(- \vkkk) O( -\vkfour)T_{kl}(-{\vect{k}}_5)\rangle.
 \end{split}
\end{align}

\subsection{Change in ET Contribution Term under Conformal Transformation}
The  ET term in the RHS of eq.\eqref{probdenst} in more detail is given  by 
\begin{align}
\begin{split}
&P[\delta \phi]_{ET}= \exp\bigg[ \int \prod_{J=1}^4 \bigg\{{d^3k_J \over (2\pi)^3}\delta \phi({\vect{k}}_J)\bigg\} \ I \bigg], \ \text{with}\\
\label{defI}
&I = \int {d^3k_5 \over (2\pi)^3}{d^3k_6 \over (2\pi)^3} {\langle T_{ij} ({\vect{k}}_5) T_{kl}({\vect{k}}_6)\rangle \over k_5^3  \ k_6^3}\langle O(\vk)O(\vkk)T_{ij}({\vect{k}}_5)\rangle \ \langle O(\vkkk)O(\vkfour)T_{kl}({\vect{k}}_6)\rangle.
\end{split}
\end{align}
Where we have used the relation,
\be
\label{projtt}
\widehat{P}_{ijkl} ={8 \langle T_{ij}( {\vect{k}}_5) T_{kl}( -{\vect{k}}_5)\rangle ' \over k_5^3}
\ee
to write $\widehat{P}_{ijkl}$ appearing in eq.\eqref{valpphi} in terms of $ \langle T_{ij}( {\vect{k}}_5) T_{kl}( -{\vect{k}}_5)\rangle '$.
In the above expression one can integrate over $k_5$ and $k_6$ using the momentum conserving delta function in $\langle T_{ij} ({\vect{k}}_5) T_{kl}({\vect{k}}_6)\rangle$ and also in the coefficient functions $\langle O(\vk)O(\vkk)T_{ij}({\vect{k}}_5)\rangle$ and $\langle O(\vkk)O(\vkkk)T_{ij}({\vect{k}}_6)\rangle$ to get  the ET  term in eq.\eqref{valpphi}. 
Under a conformal transformation $\delta \phi$ transforms as given in eq.(\ref{delphsctmom}).
After integrating by parts we obtain,
\begin{align}
\label{confdpI}
 \delta^C\bigg[\int \prod_{J=1}^4 \bigg\{{d^3k_J \over (2\pi)^3}\delta \phi({\vect{k}}_J)\bigg\} \ I\bigg]=- \int \prod_{J=1}^4 \bigg\{{d^3k_J \over (2\pi)^3}\delta \phi({\vect{k}}_J)\bigg\}\ \delta^C \big(I\big).
\end{align}
Where the resulting change in $I$ is given by 
\begin{align}
\label{confI}
\begin{split}
 \delta^C \big(I\big) =&\int {d^3k_5 \over (2\pi)^3}{d^3k_6 \over (2\pi)^3}  {\langle T_{ij} ({\vect{k}}_5) T_{kl}({\vect{k}}_6)\rangle \over k_5^3  \ k_6^3} \\ & \bigg[\langle \delta O(\vk)O(\vkk)T_{ij}({\vect{k}}_5)\rangle \ \langle O(\vkkk)O(\vkfour)T_{kl}({\vect{k}}_6)\rangle  \\ 
 & + \langle O(\vk)\delta O(\vkk)T_{ij}({\vect{k}}_5)\rangle \ \langle O(\vkkk)O(\vkfour)T_{kl}({\vect{k}}_6)\rangle \\
 &+\langle O(\vk)O(\vkk)T_{ij}({\vect{k}}_5)\rangle \ \langle \delta O(\vkkk)O(\vkfour)T_{kl}({\vect{k}}_6)\rangle \\ 
 &+\langle O(\vk)O(\vkk)T_{ij}({\vect{k}}_5)\rangle \ \langle O(\vkkk)\delta O(\vkfour)T_{kl}({\vect{k}}_6)\rangle \bigg]
 \end{split}
\end{align}
and $\delta O({\vect{k}})$ is given in eq.\eqref{osctmom}.

Now we will use the fact that $\langle O(\vkkk)O(\vkfour)T_{kl}({\vect{k}}_6)\rangle$ is invariant under special conformal transformation, which will allow us to write,
\be
\label{invsctoot}
\langle \delta O(\vk)O(\vkk)T_{ij}({\vect{k}}_5)\rangle + \langle O(\vk)\delta O(\vkk)T_{ij}({\vect{k}}_5)\rangle = - \langle  O(\vk)O(\vkk)\delta T_{ij}({\vect{k}}_5)\rangle.
\ee
Using eq.\eqref{invsctoot} back in eq.\eqref{confI},
\begin{align}
\begin{split}
 &\delta^C \big(I\big) =-\int {d^3k_5 \over (2\pi)^3}{d^3k_6 \over (2\pi)^3} {\langle T_{ij} ({\vect{k}}_5) T_{kl}({\vect{k}}_6)\rangle \over k_5^3  \ k_6^3} \bigg[\langle O(\vk)O(\vkk)\delta T_{ij}({\vect{k}}_5)\rangle \ \\
 & \langle O(\vkkk)O(\vkfour)T_{kl}({\vect{k}}_6)\rangle  +\langle O(\vk)O(\vkk)T_{ij}({\vect{k}}_5)\rangle \ \langle O(\vkkk) O(\vkfour)\delta T_{kl}({\vect{k}}_6)\rangle \bigg].
 \end{split}
\end{align}
Next we use the expression for the change of $T_{ij}({\vect{k}})$ as given in eq.\eqref{tsctmom} in the above expression and 
integrate by parts to move the derivatives acting on $T_{ij}({\vect{k}})$ to other terms.
After using the fact that $\langle T_{ij} ({\vect{k}}_5) T_{kl}({\vect{k}}_6)\rangle$ is also invariant under conformal transformation,
 we are left with the terms which arise when the differential operators acts on the factors of $1/k_5^3$ and $1/k_6^3$.
This gives,
\begin{align}
\label{confI1}
\begin{split}
 \delta^C \big(I\big) = \int & {d^3k_5 \over (2\pi)^3}{d^3k_6 \over (2\pi)^3} {12 \over k_5^3  \ k_6^3}\bigg[{1\over k_5^2}(b_mk_{5i}-b_i k_{5m})\langle T_{mj} ({\vect{k}}_5) T_{kl}({\vect{k}}_6)\rangle \\ 
 &+{1\over k_6^2}(b_mk_{6k}-b_k k_{6m})\langle T_{ij} ({\vect{k}}_5) T_{ml}({\vect{k}}_6)\rangle\bigg] \\ & \langle O(\vk)O(\vkk)T_{ij}({\vect{k}}_5)\rangle \ \langle O(\vkkk) O(\vkfour) T_{kl}({\vect{k}}_6)\rangle.
\end{split}
\end{align}
Further we can use the fact that $\langle T_{ij} ({\vect{k}}_5) T_{ml}({\vect{k}}_6)\rangle$ satisfies the relation
\be
k_{5m} \langle T_{mj} ({\vect{k}}_5) T_{kl}({\vect{k}}_6)\rangle =0
\ee
which follows for example from eq.(\ref{projtt}) and the fact that  $\widehat{P}_{ijkl}$ is transverse and traceless. 
This yields, 
\begin{align}
\label{confI2}
\begin{split}
 \delta^C \big(I\big) = \int  &{d^3k_5 \over (2\pi)^3}{d^3k_6 \over (2\pi)^3} {24 \over k_5^3  \ k_6^3} {b_mk_{5i}\over k_5^2}\langle T_{mj} ({\vect{k}}_5) T_{kl}({\vect{k}}_6)\rangle \\ & \langle O(\vk)O(\vkk)T_{ij}({\vect{k}}_5)\rangle \ \langle O(\vkkk) O(\vkfour) T_{kl}({\vect{k}}_6)\rangle.
\end{split}
\end{align}
In obtaining the above equation we have also used the fact that $I$ is symmetric under the exchange 
\begin{align}
 \begin{split}
 \text{the external momenta} \  \vk & \Leftrightarrow \vkkk,\ 
  \vkk  \Leftrightarrow \vkfour, \
  {\vect{k}}_5  \Leftrightarrow {\vect{k}}_6 \\
  \text{and for the indices} \ \{i,j\} & \Leftrightarrow \{k,l\}.
 \end{split}
\end{align}
Finally, using eq(\ref{projtt}) gives, 
\begin{align}
\label{confI3}
\begin{split}
 \delta^C \big(I\big) = 3 b_m \int {d^3k_5 \over (2\pi)^3} {\widehat{P}_{imkl}( {\vect{k}}_5) \over  k_5^8} k_{5j} \langle  O(\vk)O(\vkk)T_{ij}({\vect{k}}_5)\rangle  \langle O(\vkkk) O(\vkfour) T_{kl}(-{\vect{k}}_5)\rangle.
\end{split}
\end{align}
Using  $\delta^C \big(I\big)$ from eq.\eqref{confI3} in eq.\eqref{confdpI} and comparing with eq.\eqref{delRdpdp1}, we see that they exactly cancel each other so that their sum vanishes. 
This proves that $P[\delta \phi]$ is invariant upto quartic order in $\delta \phi$. 

\section{More Details on Different Limits of the Final Result}
\label{apptest}
As was mentioned in subsection \ref{finreslimit}, in this appendix we will provide more details on the discussion of different limits of our final result in the following subsections.
\subsection{Details on Deriving eq.\eqref{lim2int} for Limit II in Subsection \ref{limit2}}
\label{apptest1}
Let us start examining the behavior of $\langle O(\vk) O(\vkk) O(\vkkk) O(\vkfour)\rangle'$ in the limit $\vkk\rightarrow \infty$. We do so by parameterizing $\vkk={\vect{a}} /\epsilon$ and then take the limit $\epsilon \rightarrow 0$, with $\vkkk,\vkfour$ held fixed and 
$\vk=-(\vkk+\vkkk+\vkfour)$. One then obtains,
\begin{align}
\label{expf}
\lim_{\epsilon \rightarrow 0}\langle O(\vk) O(\vkk) O(\vkkk) O(\vkfour)\rangle'= \frac{1}{\epsilon} W^{div.} + W^{const.} + \mathcal{O}(\epsilon)\, .
\end{align}
Here  $W^{div.}$ is the coefficient of  a term which is divergent as $\epsilon \rightarrow 0$,
and  $W^{\rm const.}$ is  a term which is $\epsilon$ independent. The presence of a 
 divergent term might at first seem to contradict eq.\eqref{lim2int}. 
  However it turns out that the divergent piece is entirely a contact term analytic in the momenta.
\begin{align}
W^{div.}=W_s^{div.}+W_t^{div.} +W_u^{div.} 
\end{align}
where the contributions from the individual channels are
\begin{align}
{1 \over \epsilon}W_s^{div.}&=-\frac{5}{8 k_2} \dotp[k_2,k_3]\dotp[k_2,k_4] + \frac{k_2}{4} \vkkk \cdot \vkfour, \\
{1 \over \epsilon}W_t^{div.}={1 \over \epsilon}W_u^{div.}&=-\frac{1}{8 k_2} \dotp[k_2,k_3]\dotp[k_2,k_4] - \frac{k_2}{4} \vkkk \cdot \vkfour\,.
\end{align}
These terms are clearly analytic functions of $\vkkk, \vkfour$. 
Such analytic terms in position space give rise to contact terms, which are proportional to 
delta functions or their derivatives for one or more of the arguments. We had mentioned in 
our discussion after eq.\eqref{opea} that we are neglecting such contact terms in the OPE, it is 
therefore no contradiction that they are appearing in an expansion of the full answer in eq.\eqref{expf}
 above,
but did not appear in our discussion based on the OPE in eq.\eqref{lim2int}.

Neglecting these divergent pieces we get  in the limit $\epsilon \rightarrow 0$ that 
\be
\label{newf}
\lim_{\epsilon \rightarrow 0}\langle O(\vk) O(\vkk) O(\vkkk) O(\vkfour)\rangle'=W^{const}.
\ee
We get contributions to $W^{const}$ from all the three channels,
\be
\label{wconst_stu}
W^{const}= W^{const}_S+W^{const}_T+W^{const}_U
\ee
where,
\be
\label{wconst_tu}
W^{const}_T+W^{const}_U= \frac{(\vkk \cdot \vkkk) (k_3^2-k_4^2)}{8 k_2}.
\ee
It is obvious from RHS of eq.\eqref{wconst_tu} that the contribution to $W^{const}$ from $T$ and $U$ channels are analytic functions of $\vkkk, \vkfour$. Therefore, effectively they don't contribute  to $W^{const}$ for the same reason described earlier, and the contribution to $W^{const.}$ comes only from the $S$- channel, which is
\begin{align}
\label{wsconst}
W^{const}_S =  W^{const.}_{S(1)}+ W^{const.}_{S(2)},
\end{align}
with
\begin{align}
\label{wsconst_12}
\begin{split}
W^{const}_{S(1)} &= {3\over8}\frac{(\vkk \cdot \vkkk)(\vkk \cdot \vkfour)}{ k_2^2} \bigg((\tilde{k}+k_3+k_4)-\frac{\tilde{k}k_3+k_3k_4+k_4\tilde{k}}{(\tilde{k}+k_3+k_4)}-\frac{\tilde{k}
k_3k_4}{(\tilde{k}+k_3+k_4)^2}\bigg), \\
W^{const.}_{S(2)} &= \frac{3}{64} \tilde{k} \left(-\tilde{k}^2+k_3^2+k_4^2\right)+\frac{3 k_3^2 k_4^2}{8 (\tilde{k}+k_3+k_4)} + \frac{5(\vkk \cdot \vkkk) (k_3^2-k_4^2)}{16 k_2}.
\end{split}
\end{align}
It is obvious from eq.\eqref{wsconst_12} that the term $W^{const.}_{S(2)}$ does not contribute to eq.\eqref{newf}, because the first two terms on LHS of $W^{const.}_{S(2)}$ does not depend on $\vkk$ and the last term is analytic in the momenta $\vkkk, \vkfour$, therefore when fourier transformed back to position space they produce delta functions or derivatives of them.

Finally we get,
\be 
\label{lim2final}
\begin{split}
\lim_{\epsilon \rightarrow 0}\langle O(\vk) O(\vkk) O(\vkkk) O(\vkfour)\rangle' \sim & \frac{(\vkk \cdot \vkkk)(\vkk \cdot \vkfour)}{ k_2^2} \bigg((\tilde{k}+k_3+k_4) \\ &-\frac{\tilde{k}k_3+k_3k_4+k_4\tilde{k}}{(\tilde{k}+k_3+k_4)}-\frac{\tilde{k}
k_3k_4}{(\tilde{k}+k_3+k_4)^2}\bigg).
\end{split}
\ee
Therefore we have confirmed eq\eqref{lim2int}.

\subsection{More Details on Obtaining eq.\eqref{ccge}}
\label{detcontract}
In this subsection we will explain in more detail the contribution of the ET contribution term to the scalar four point correlator $\langle \delta \phi(\vk) \delta \phi(\vkk)  \delta \phi (\vkkk) \delta \phi (\vkfour) \rangle_{ET}$ in counter-collinear limit eq.\eqref{defk12}. It is obvious from eq.\eqref{valGE} and eq.\eqref{defGhat} that in this limit the dominant contribution in $\langle \delta \phi(\vk) \delta \phi(\vkk)  \delta \phi (\vkkk) \delta \phi (\vkfour) \rangle_{ET}$ comes from the S-channel term, $\widehat{G}^S(\vk,\vkk,\vkkk,\vkfour)$. This term diverges as ${1 \over k_{12}^3}$ in the counter-collinear limit. In the RHS of eq.\eqref{defGhat} the term within the parenthesis is due to the four external momenta corresponding to the four perturbations being contracted with the transverse traceless projector $\widehat{P}_{ijkl}$, eq.\eqref{defPhat}, as follows,
\be
\label{contk4p}
\begin{split}
 &k_1^i k_2^j k_3^k k_4^l \widehat{P}_{ijkl} (\vk+\vkk) = \\
 &\bigg[ \bigg\{\vk.\vkkk+\frac{\{(\vkk+\vk).\vk\} \{(\vkfour+\vkkk).\vkkk\}}{|\vk+\vkk|^2}\bigg\} \bigg\{\vkk.\vkfour+  \frac{\{(\vk+\vkk).\vkk\} \{(\vkkk+\vkfour).\vkfour\}}{|\vk+\vkk|^2}\bigg\} \\ &
+  \bigg\{\vk.\vkfour+\frac{\{(\vkk+\vk).\vk\} \{(\vkfour+\vkkk).\vkfour\}}{|\vk+\vkk|^2}\bigg\} 
\bigg\{\vkk.\vkkk+\frac{\{(\vkk+\vk).\vkk\} \{(\vkfour+\vkkk).\vkkk\}} {|\vk+\vkk|^2}\bigg\}\\ &-
\bigg\{\vk.\vkk-\frac{\{(\vkk+\vk).\vk\} \{(\vk+\vkk).\vkk\}} {|\vk+\vkk|^2}\bigg\} 
\bigg\{\vkkk.\vkfour-\frac{\{(\vkkk+\vkfour).\vkfour\} \{(\vkfour+\vkkk).\vkkk\}}{|\vk+\vkk|^2}\bigg\} \bigg].
\end{split}
\ee
One can also use the polarization tensors $\epsilon^s_{ij}$, defined in eq.\eqref{defgammas}, to write the RHS of eq.\eqref{contk4p} in an alternate way. In a spherical coordinate system having $\{ {\vect{e}}, \bar{\vect{e}},\widehat{{\vect{k}}}_{12}\}$ as basis (denoting $\widehat{{\vect{k}}}={{\vect{k}} \over k}$) one can obtain the relation 
\be
\label{polsumproj}
\sum_{s} \epsilon^s_{ij} \epsilon^s_{kl} = \widehat{P}_{ijkl}.
\ee
Let us define $\theta_i$ being the angle between ${\vect{k}}_i$ and ${\vect{k}}_{12}$, whereas $\phi_i$ being the angle between ${\vect{k}}_i$ and ${\vect{e}}$. As was shown in \cite{Seery:2008ax}, see eq.(2.32), using eq.\eqref{polsumproj}, one can then get,
\be
\label{contk4p1}
k_1^i k_2^j k_3^k k_4^l \widehat{P}_{ijkl} (\vk+\vkk) = k_1^2 k_3^2 \sin^2(\theta_1)\sin^2(\theta_3) \cos(2 \chi_{12,34}).
\ee
Using eq.\eqref{contk4p}, eq.\eqref{contk4p1} and eq.\eqref{singhat} in eq.\eqref{defGhat}, one obtains the form of the ET contribution term $\langle \delta \phi(\vk) \delta \phi(\vkk)  \delta \phi (\vkkk) \delta \phi (\vkfour) \rangle_{ET}$ in the counter-collinear limit as in eq.\eqref{ccge} in a straight forward way.

\subsection{More Details on the Check of the Relative Coefficient between CF and ET Terms}
\label{apptest3}
In subsection \ref{limit3ccl} towards the end we discussed the counter-collinear limit in an alternative but equivalent way compared to what already exists in literature. We took all the individual momenta i.e. ${\vect{k}}_i, \ i=1,2,3,4$ to diverge keeping $\vk+\vkk=-(\vkkk+\vkfour)$ fixed. 
This way of interpreting the counter-collinear limit provides us a further check to fix the relative coefficient between the CF term in eq.\eqref{valCF} and the ET term in eq.\eqref{valGE}. Here we will discuss in some detail.

We implement this alternate way of counter-collinear limit in two steps. First we take $\vk,\vk\rightarrow \infty$ keeping ${\vect{k}}_{12}$ fixed and then we take $\vkkk,\vkfour\rightarrow \infty$ keeping $\vkkk+\vkkk$ fixed. After the first limit the $\langle O(\vk) O(\vkk) O (\vkkk)O (\vkfour) \rangle$, given in eq.\eqref{fourptwo}, becomes,
\be
\label{cclcf1}
\begin{split}
\langle O(\vk) O(\vkk)& O (\vkkk)O (\vkfour) \rangle\rightarrow  4 (2\pi)^3 \delta^3\big(\sum_J {\vect{k}}_J\big) {3\over8} {(\vkk .\vkkk )(\vkk .\vkfour)\over k_2^2} S(k_3,k_4) 
\end{split}
\ee
with $S(k_3,k_4)$ being given in eq.\eqref{singhat}.
Next we take the limit, i.e. $\vkkk,\vkfour\rightarrow \infty$ keeping $\vkkk+\vkkk$ fixed and in this limit the leading non-analytic behavior of $S(k_3,k_4)$ goes as,
\be
\label{snonan}
S(k_3,k_4) \sim -{3\over8} {k_{34}^3 \over k_3^2}.
\ee
Using eq.\eqref{snonan}, one obtains the limiting behavior of the CF term contribution to scalar 4 point correlator, eq.\eqref{valCF} in the counter-collinear limit as,
\be
\label{cclcf}
\langle \delta \phi(\vk) \delta \phi(\vkk)  \delta \phi (\vkkk) \delta \phi (\vkfour) \rangle_{CF} = -8(2\pi)^3 \delta^3\big(\sum_J {\vect{k}}_J\big) \bigg({3\over8}\bigg)^2 {(\vkk .\vkkk )(\vkk .\vkfour)\over k_2^2}{k_{34}^3 \over k_3^2}.
\ee

For the ET contribution to the scalar four point correlator, eq.\eqref{valGE}, we have two factors of $S$, and as it was mentioned in the previous subsection, the term within the parenthesis in the RHS of eq.\eqref{defGhat} is $k_1^i k_2^j k_3^k k_4^l \widehat{P}_{ijkl} (\vk+\vkk)$, eq.\eqref{contk4p}. In the sequence of steps for the counter-collinear limit we are concerned with, this term goes as,
\be
\label{cclk4p}
k_1^i k_2^j k_3^k k_4^l \widehat{P}_{ijkl} (\vk+\vkk) \sim -2 (\vkk .\vkkk )(\vkk .\vkfour).
\ee
Using eq.\eqref{snonan} and eq.\eqref{cclk4p} in eq.\eqref{valGE}, we obtain the form of the contribution of the ET term in 4 point scalar correlator as,
\be
\label{cclge}
\langle \delta \phi(\vk) \delta \phi(\vkk)  \delta \phi (\vkkk) \delta \phi (\vkfour) \rangle_{ET} = -8(2\pi)^3 \delta^3\big(\sum_J {\vect{k}}_J\big) \bigg({3\over8}\bigg)^2 {(\vkk .\vkkk )(\vkk .\vkfour)\over k_2^2}{k_{34}^3 \over k_3^2}.
\ee
Now comparing eq.\eqref{cclcf}and eq.\eqref{cclge} we conclude that, once we take the counter-collinear limit in the sequential order prescribed above, the leading behavior of both the CF and ET term matches perfectly including coefficients and thus provides a further check on the relative coefficient of these two contributions.

\bibliographystyle{jhepmod}
\bibliography{references}

\end{document}